%% file: survey.tex
\documentclass[acmlarge]{acmart}

\AtBeginDocument{%
  \providecommand\BibTeX{{%
    \normalfont B\kern-0.5em{\scshape i\kern-0.25em b}\kern-0.8em\TeX}}}

\setcopyright{acmlicensed}
\copyrightyear{2025}
\acmYear{2025}
\acmDOI{XXXXXXX.XXXXXXX}

\usepackage{tabularx}
\usepackage{xcolor}
\usepackage[normalem]{ulem}  
\usepackage[flushleft]{threeparttable}
\usepackage{array,booktabs,makecell}
\usepackage{longtable}
\usepackage{multirow,array}
\usepackage{hyperref}   
\usepackage{pdflscape}  
\usepackage{colortbl}

\definecolor{light-gray}{gray}{0.96}
\definecolor{LightCyan}{rgb}{0.88,1,1}
\usepackage[skins]{tcolorbox}
\usepackage{wrapfig}
\usepackage{verbatim}
\usepackage{times}
\usepackage{color}
\usepackage{comment}
\usepackage{epsfig}
\usepackage{url}
\usepackage{xspace}
\usepackage{pgfgantt}
\usepackage{booktabs}
\usepackage{float}
\usepackage{stfloats}
\usepackage[ruled]{algorithm2e}

\usepackage{enumitem} 
 
\newcommand{\defn}[1]{\textbf{\emph{#1}}}
\newcommand{\poly}[1]{{\mathrm{poly}({#1})}}
\newcommand{\uar}[1]{uniform at random\xspace}
\newcommand{\whp}[1]{with high probability\xspace}

\newif\ifcomments
\commentsfalse

\begin{document}

\title{A Survey of Recent Advancements in Secure Peer-to-Peer Networks}

\author{Raj Patel}
\authornote{Both authors contributed equally to this research.}
\affiliation{%
  \institution{The University of Alabama}
  \city{Tuscaloosa}
  \country{USA}}
\email{rpatel38@ua.edu}

\author{Umesh Biswas}\authornotemark[1]
\affiliation{%
  \institution{Mississippi State University}
  \city{Mississippi State}
  \country{USA}}
\email{ucb5@msstate.edu}

\author{Surya Kodipaka}
\affiliation{%
  \institution{Mississippi State University}
  \city{Mississippi State}
  \country{USA}}
\email{sk2008@msstate.edu}

\author{Will Carroll}
\affiliation{%
  \institution{Mississippi State University}
  \city{Mississippi State}
  \country{USA}}
\email{woc17@msstate.edu}

\author{Preston Peranich}
\affiliation{%
  \institution{Mississippi State University}
  \city{Mississippi State}
  \country{USA}}
\email{plp122@msstate.edu}

\author{Maxwell Young}
\affiliation{%
  \institution{Mississippi State University}
  \city{Mississippi State}
  \country{USA}}
\email{young@cse.msstate.edu}

\renewcommand{\shortauthors}{Patel et al.}

\begin{abstract}
Peer-to-peer (P2P) networks are a cornerstone of modern computing, and their security is an active area of research. Many defenses with strong security guarantees have been proposed; however, the most-recent survey is over a decade old. This paper delivers an updated review of recent theoretical advances that address classic threats, such as the Sybil and routing attacks, while highlighting how emerging trends---such as machine learning, social networks, and dynamic systems---pose new challenges and drive novel solutions. We evaluate the strengths and weaknesses of these solutions and suggest directions for future research.\medskip
\end{abstract}

\ccsdesc[500]{General and reference~Surveys and overviews}
\ccsdesc[500]{Theory of computation~Distributed algorithms}
\ccsdesc[500]{Security and privacy~Formal methods and theory of security}

\keywords{Survey, distributed algorithms, theoretical results, secure peer-to-peer networks, Byzantine attacks, Sybil attacks, routing attacks}


\maketitle

\section{Introduction}

Since the debut of Napster in 1999 \cite{honigsberg2001evolution}, \defn{peer-to-peer (P2P)} networks have drawn steady attention from the research community. Many early results focused on designing topologies for  content storage and retrieval, with provable guarantees on common metrics such as latency, state, and message complexity.  However, due to a lack of admission control and any trusted authority, P2P networks are vulnerable to a wide range of attacks, and  attention soon turned to security challenges. In the decade that followed, a flurry of results on secure P2P networks appeared. Specifically, the research community demonstrated that, despite formidable attacks, correct content storage and retrieval can be preserved.

In 2011, 
Urdaneta, Pierre,  and Van Steen \cite{urdaneta2011survey} published a comprehensive survey of these results. However, in the years since, the networking landscape has changed significantly. Key advancements include the application of machine learning (ML) to network security, the ubiquity of social networks, the emergence of highly dynamic ad-hoc networks such as the Internet of Things, and the growing adoption of resource burning as a security tool. Given these developments, we feel that a ``sequel'' survey is warranted.

Here, our primary goal is to summarize results on secure P2P networks that have been published subsequent to the survey by Urdaneta et al. \cite{urdaneta2011survey}. We provide context, especially for newer contributions that build on older results; however, we avoid retreading ground that has already been surveyed. As a secondary aim, throughout our survey, we draw attention to instances where network security concerns have shifted over the past decade, and how this area of research has been influenced by contemporary technologies. Finally, as a tertiary contribution, based on developments in the literature, we discuss  open problems that we believe provide promising avenues for future work.

\subsection{Definitions and Terminology}\label{sec:definitions-terminology}

\noindent We begin by reviewing core definitions and terminology that are common to the literature on P2P systems. These concepts are essential for understanding the performance and security guarantees of the results we survey.\medskip

\noindent\textbf{Global System Knowledge.} The participants in a P2P system are (1) \defn{good peers} that obey a prescribed protocol, and (2) \defn{bad peers} that may deviate arbitrarily from a protocol. Since it is often helpful to view a P2P network as a graph (see our discussion in Section \ref{sec:overview-p2p}), we often refer to peers as  \defn{nodes}.

Throughout this survey, {\boldmath{$n$}} denotes the total number of peers currently in the network. In much of the literature, $n$ is unknown {\it a priori} to any good peers, although there are methods for estimating $n$ to within a small constant factor (e.g., \cite{king:choosing,king2007choosing}). Many results assume a fairly \defn{static setting}, where  $n$ does not change by more than a constant factor over time,  while others account for a \defn{dynamic setting} where $n$ may change significantly. Unsurprisingly, by comparison to the static setting, designing and analyzing algorithms in the dynamic setting is challenging.

The topology of the network is also largely unknown to good peers. For example, where data is stored is not known in advance. Moreover, any peers contacted in the process of  retrieving data are also unknown in advance. Due to joins and departures, membership in the system may also change over time, and thus any view of the network held by a peer may quickly become stale.
\smallskip

\noindent\textbf{Performance Metrics.} A ubiquitous metric is \defn{latency}, which in practice typically refers to the time required to route data from one peer to another. However, in the theoretical literature, latency often refers to the maximum number of communication links traversed from any source peer to any destination peer (that, for example, may hold content of interest to the source peer). Many of the P2P networks we survey have low latency in the sense that the maximum number of links traversed between any pair of peers is $O(\poly{\log n})$.

Another key metric is \defn{state complexity}, which represents the amount of information each peer must maintain. This includes information such as routing tables and peer identifiers, metadata on content, and cryptographic credentials. Typically, state is quantified by the number of objects tracked; for example, the number: of entries in a routing table, data items stored, and cryptographic keys held by a peer. Again, state complexity is typically $O(\poly{\log n})$.

Finally, the \defn{message complexity} measures the total number of messages exchanged among peers during the execution of a protocol, such as routing in order to retrieve content, or for having a peer join the network. This metric reflects the communication overhead and, again, is often $O(\poly{\log n})$.

We highlight that, in many theoretical results,  $O(\poly{\log n})$ is considered to be \defn{scalable} in the sense that this quantity is (roughly) exponentially small in the system size $n$. \medskip

\noindent\textbf{Adversary.} In the literature, all bad peers are typically assumed to be under the control of a single \defn{adversary}, since this captures perfect collusion and cooperation between the bad peers. The term \defn{Byzantine} is sometimes used to describe an adversary, originating from extensive research in secure distributed computing (e.g., \cite{Lamport,pease,castro:practical,castro:byzantine}). 

Generally, the adversary has full knowledge of the network topology, and can send spurious messages to any good peer. In settings where network resources (e.g., computation or bandwidth) play an important role, the adversary is often assumed to control a constant fraction of such resources (this is discussed further in Section \ref{resource-burning-methods}).

\subsection{Overview of P2P Networks}\label{sec:overview-p2p}

To set the stage, we provide a brief overview of P2P systems and related security challenges. Generally, P2P networks provide a decentralized environment, whereby all participants may act as both a client and a server \cite{pourebrahimi2005survey}.  This allows participants to share resources with each other, such as computing capacity, disk storage, and content. To support this sharing, P2P networks provide content lookup and retrieval functionality, which hinges on efficient routing. 

In analyzing routing, many P2P constructions are explicit about the topology of the network; these are \defn{structured} networks. In particular, we can view the network as a graph $G=(V,E)$, where the elements of the vertex set $V$ correspond to peers, and edges in $E$ correspond to communication links in an overlay. Here, when a new peer joins the system, there are rules about how it selects its neighbors; conversely, when a peer departs the system, there are rules for how the graph is restructured due to the peers absence.  

Arguably, the most popular example of a structured network is a  \defn{distributed hash table (DHT)}. Informally, a DHT solves the same problem as standard hash table with one caveat: the stored data and computing load is distributed over multiple peers. In their simplest form, DHTs need to provide a single \texttt{lookup}($k$) function, which returns a value associated with the \emph{key} $k$. This key is often an identifier (e.g., title or name) for some content, and the \emph{value} is typically the IP address of the peer that holds this content. The efficiency of a \texttt{lookup()} request is usually measured by the number of edges traversed until the content is located; in practice, this loosely corresponds to latency of a request. However, we note that there is significant work on this aspect alone (e.g., addressing geographical constraints \cite{fujita:proximity-aware,cramer:proximity}, reliability \cite{damiani2002reputation}, parallel searches \cite{stutzbach2006improving}).   

The literature on DHTs is vast, with popular constructions such as Chord \cite{stoica2001chord}, Viceroy \cite{malkhi2002viceroy}, CAN \cite{ratnasamy2001scalable}, Pastry \cite{rowstron2001pastry}, Kademlia \cite{maymounkov}, Koorde \cite{kaashoek2003koorde}, Cycloid \cite{shen:cycloid}, and D2B \cite{fraigniaud:d2b}. We recommend the surveys by Lua et al. \cite{P2P:network-overlay-survey}, Hassanzadeh-Nazarabadi et al. \cite{hassanzadeh2021dht}, and Zhang et al.  \cite{dht:TheoryPlatformAndApplication} for comprehensive review of these prior results. 

Beyond DHTs, there are many other structured P2P networks; for example,  BATON \cite{jagadish:baton}, skip graphs \cite{aspnes2003skip}, skip net \cite{HJSTW,harvey:deterministic}, rainbow skip graphs \cite{goodrich:rainbow}, skip-webs \cite{skip-webs:arge}, and the hyperring \cite{awerbuch:hyperring}. Many of these results arise from decentralizing other familiar data structures; for example, skip graphs are a decentralized version of the well-known skip list \cite{pugh1990skip}. These constructions offer various enhancements, some allowing for more complex lookup functionality (e.g., range queries), or additional robustness to non-adversarial faults (e.g., crash failures). We refer the interested reader to the survey by Malatras \cite{malatras:survey}, who discusses many of these  constructions.

In contrast to the above, \defn{unstructured} P2P networks, are far more flexible in terms of the allowed topology. Often, peers are allowed to select their neighbors randomly. We highlight that there are a number of constructions, e.g., Freenet \cite{clarke2001freenet}, Napster, and Gnutella \cite{saroiu2003measuring}, FastTrack \cite{liang2006fasttrack}, Gia \cite{chawathe2003making}, Phenix \cite{wouhaybi2004phenix}, UMM \cite{ripeanu2010search}, and Bittorent \cite{cohen2003incentives}; for the interested reader, we recommend the survey by Malatras \cite{malatras:survey} for an in-depth discussion. Here, however, we are concerned with surveying results that offer strong security guarantees, and from this perspective, the literature predominantly
focuses on structured P2P networks.




\subsection{Overview of Security Challenges}\label{sec:overview-security}

Having reviewed P2P systems, we now analyze the ongoing security challenges in P2P networks.  To begin, it is worth considering why obtaining security guarantees can be  challenging. For instance, if a peer deviates from protocol, how can this be reliably detected? In a traditional client-server setting, a server might unilaterally diagnose  a misbehaving client. However, in the P2P setting, there is no trusted authority analogous to a server; therefore, mechanisms designed to identify malicious peers might be exploited by an adversary to wrongfully accuse legitimate peers. 

Furthermore, even if a malicious peer is detected, mitigating its impact on the system remains difficult. In a client-server model, servers can authenticate clients and block access using a blacklist. However, P2P networks are open (or permissionless), allowing peers to join or leave freely without credentials. Consequently, a malicious peer can easily rejoin using a new digital identity, making it challenging to effectively exclude bad actors.

More generally, although defensive strategies have advanced---as we explore in this survey---the fundamental security challenges addressed in the literature remain largely unchanged since the review by Urdaneta et al.  \cite{urdaneta2011survey}, which we now review.

A \defn{Sybil attack} occurs when a malicious actor creates multiple fake \defn{identifiers (IDs)} in the network to gain disproportionate influence or control over the system \cite{douceur,detweiler:pseudospoofing}. By overwhelming the network with these fraudulent IDs, the adversary can compromise security measures that depend on consensus mechanisms. Throughout, in this context, we will often refer to bad peers as \defn{Sybil peers} or \defn{Sybil IDs}. While Sybil attacks may have been largely a theoretical concern, they become a serious security threat in practice \cite{bitcoin-sybil,Tran:2009:SOC:1558977.1558979,wang,ma:detecting,sridhar:content,prunster:total}.

The Sybil attack is  effective in large part because creating IDs can be cheap. The generation of many IDs also allows the adversary to  intentionally induce high churn to disrupt the network. Each time a peer joins or departs, links must be forged or torn down in order to accommodate the change in network topology. By repeatedly inserting and removing malicious peers, an attacker can generate excessive churn, commonly known as a \defn{join-leave} attack. Similarly, the adversary may use its Sybil IDs to issue many (spurious) lookup requests in order to consume network resources.  As noted earlier, the ease of generating new peer IDs complicates efforts to counter these threats.

\begin{figure*}[t!]
    \includegraphics[scale=0.542]{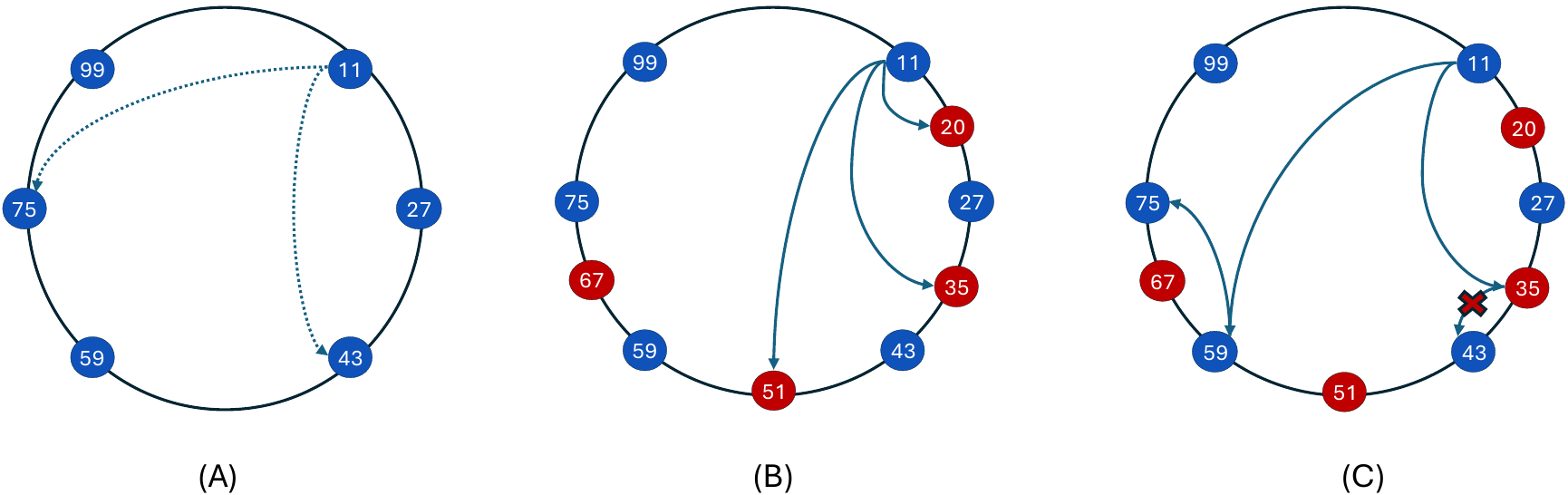}
    \caption{Good peers are colored blue, while bad peers are colored red. (Left) A DHT with only good peers. (Center) Peer with ID 11 suffers an eclipse attack. (Right) An example of routing attack. }
    \label{fig:Attacks_in_P2P_systems}
    \Description{The figure shows three scenarios in a Chord DHT peer-to-peer network. In (A), all nodes are honest and routing is normal. In (B), some finger links from honest nodes point to bad nodes, allowing attackers to control those paths. In (C), some finger links reach bad nodes while others reach honest nodes, exposing honest nodes to Sybil attacks.}
\end{figure*}

Routing is necessary for looking up content. In the case of structured networks, routing is also needed to move data from incoming peers to  specific network locations to enable efficient lookups. For instance, in a DHT, each peer maintains a set of neighbors, as depicted in Figure \ref{fig:Attacks_in_P2P_systems}(A). When executing a lookup, a good peer selects one of these neighbors to forward its request; for example, this is shown in Figure \ref{fig:Attacks_in_P2P_systems}(B), where the peer with ID $11$ forwards its request to the peers with IDs $59$ and then $75$. 

However, a single malicious or unresponsive peer along the route can disrupt the process by altering forwarded content or failing to cooperate, causing the operation to fail. This behavior constitutes a \defn{routing attack}.  This is illustrated in Figure \ref{fig:Attacks_in_P2P_systems}(B), where a lookup initiated by the good peer with ID $11$ passes through the bad peer with ID $35$, who corrupts or drops the lookup request.

The related \defn{eclipse attack} occurs when an attacker or adversary isolates a targeted victim's (node of a network) communication with the rest of the network in a P2P network system, effectively controlling or manipulating all kinds of communication for the victim node or nodes ~\cite{singh2006eclipse}. In Figure \ref{fig:Attacks_in_P2P_systems}(C), the good peer with ID $11$ is eclipsed, since all of its neighbors to which it links under the construction rules of the DHT are bad peers.

Instead of corrupting routing, the adversary may target the availability of content in the P2P network.  More specifically, bad peers may deliberately sabotage the availability of content by mistitling it, or replacing it with malware; alternatively,  the quality of stored content can be intentionally degraded---such malicious behavior is often referred to as \defn{poisoning or pollution attacks}, respectively \cite{christin2005content,liang:poison,liang:pollution}. In P2P applications that exist to provide access to popular media, this type of malicious behavior can be especially disruptive.


\subsection{Survey Scope and Organization}

Over a decade ago, Urdaneta et al. \cite{urdaneta2011survey} gave a comprehensive review of results on security techniques for DHTs. Here, we aim to provide a follow-up survey of more-recent literature, with an expanded focus on general P2P systems (rather than just DHTs).

Over a decade ago, Urdaneta et al. \cite{urdaneta2011survey} reviewed security techniques for DHTs. This study extends their work by surveying recent literature, broadening the scope to encompass P2P systems beyond DHTs. We largely focus on theoretical results and well-established analytical frameworks that provide strong security guarantees; however, we also include heuristic approaches that exploit subtle patterns of Sybil behavior to enhance security, provided they are grounded in rigorous evaluation or analysis. 

Outside of this scope, we note there is an expansive literature on related topics, such as reputation systems (e.g., \cite{hoffman2009survey,kohnen_a,kohnen_b,gracia,liu}) and anomaly detection (e.g., \cite{chandola2009anomaly,ahmed2016survey}), which can complement the security guarantees given by the papers reviewed here. Additionally, there has been a surge in research addressing blockchain technology, and we refer the interested reader to the following surveys  \cite{guo2022survey,monrat2019survey,casino}. P2P networks and (open) blockchains, face similar security challenges due to the lack of centralized control. Often, blockchain technology depends on a P2P network for disseminating information; therefore, compromising the P2P network can jeopardize the corresponding blockchain applications (e.g., see \cite{heilman,croman2016scaling}). Thus, while there is a relationship between permissionless blockchains and P2P systems, the former relies on the security of the latter, which is the focus of our survey.


The remainder of our survey is structured as follows. In Section \ref{sec:Sybil-Attacks}, we explore defenses against the Sybil attack, along with related adversarial behavior such as join-leave attacks, that make use of contemporary methods; specifically, the use of: social-network information (Subsection \ref{sec:graph-based}), ML (Subsection \ref{ml-based-methods}), and resource burning (Subsection \ref{resource-burning-methods}). In Section \ref{DA_Routing_Storage}, we review the literature on routing attacks (Section~\ref{DA_Routing_Storage}) and associated misbehavior (e.g., eclipse and poisoning/pollution attacks). In particular, we review approaches that leverage redundant routing (Subsection \ref{DA_Routing_Storage}); group-based approaches (Subsection \ref{group_based_approaches}), and constructions that tolerate high churn (Subsection \ref{high_churn}). We also provide tables that summarize each paper we review, allowing for a quick comparison of the results. Finally, at the end of each section, we provide a discussion, along with potential directions for future work.

\section{Defenses Against the Sybil Attack}\label{sec:Sybil-Attacks}

Recall from Section \ref{sec:overview-security} that, since P2P systems typically have little-to-no administrative control, an adversary may create multiple IDs and use them to disrupt critical functionality.  This malicious behavior, known as the Sybil attack, was first anticipated two decades ago~\cite{douceur}. In the interim, this attack has evolved from a theoretical concern to a credible threat, with several real-world instances involving social networks~\cite{shane:facebook,benevenuto,Yang:2011:USN:2068816.2068841} and P2P-based applications~\cite{wang,bitcoin-sybil}.  As a result, the research community has proposed methods for detecting and mitigating Sybil attacks, and here we survey this literature.

\begin{figure}
    \centering
    \includegraphics[width=0.75\linewidth]{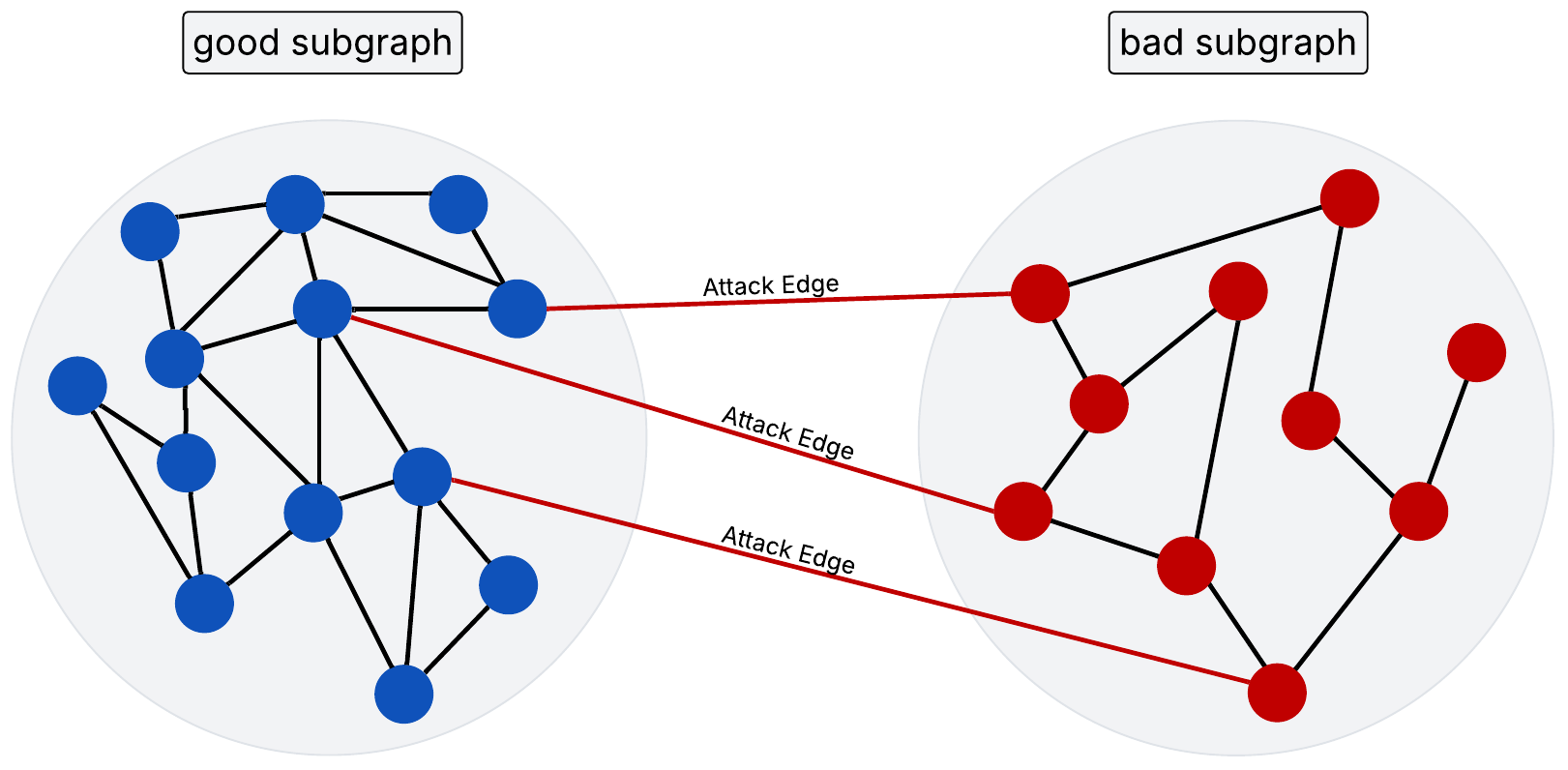}
    \caption{A depiction of an good subgraph (blue) and a bad subgraph connected by a small number of attack edges.}
    \label{fig:leveraging_social_network}
    \Description{The figure shows a social network split into two parts: a good subgraph and a bad subgraph. Only a few attack edges connect the good subgraph to the bad subgraph, representing a low number of attack edges between honest users and Sybil nodes.}
\end{figure}

\subsection{Leveraging Social Networks}\label{sec:graph-based}




The popularity of social networks has led to a vast literature on defenses that leverage their topological characteristics. In the context of viewing the network as a graph, a key property in this section is a fast \defn{mixing time}. The mixing time of a graph is defined as the number of steps required for the distribution of a random walk's position to become ``close'' to the graph’s stationary distribution, irrespective of the initial starting vertex.  A closely-related concept is \defn{expansion}. Informally, a graph has good expansion if, for every subset $S \subset V$ with $|S| \leq n/2$, the number of distinct neighbors of $S$, denoted $\Gamma(S)$, is at least a constant times $|S|$; that is, $|\Gamma(S)| \geq c|S|$ for some constant $c > 1$. For an overview of expansion properties and related notions, see the survey by Kowalski \cite{kowalski2019introduction}. Graphs with strong expansion properties support fast-mixing random walks, which are essential for several defenses we examine. 

Urdaneta et al. \cite{urdaneta2011survey} survey several results from this area, including the pioneering work by Yu et al. \cite{yu_sybilguard}, which presents a defense against Sybil attacks named SybilGuard. Here, we assume that the P2P network structure incorporates or is overlaid with a social trust graph. This graph might be informed by real-world social relationships (e.g., friend-to-friend networks), protocol-level trust mechanisms (e.g., manual authentication or key exchanges), or implicit trust inferred from long-term interactions (e.g., nodes that have reliably shared resources). Thus, edges represent a notion of social trust, and there should be few edges between bad and good nodes; such edges are referred to as \defn{attack edges}; see Figure \ref{fig:leveraging_social_network}. If we cut the attack edges, then the graph decomposes into two subgraphs: the \defn{good subgraph} containing all good nodes, and the \defn{bad subgraph} containing all bad nodes (i.e., the IDs controlled by the Sybil adversary).

Under SybilGuard, each good node performs a deterministic analog to a random walk, called a \defn{random route}: every good node maps each incoming edge to an outgoing edge randomly at first, and then this mapping is maintained throughout the execution. A random route possesses the properties of  \emph{convergence} and \emph{back-traceability}. \emph{Convergence} ensures that two random routes entering along the same edge $e$ to a good node will always exit along the same edge $e'$ (it is not necessarily the case that $e=e'$).  \emph{Back-traceability}  ensures that if a random route exits a node on edge $e$, then we can identify the edge $e'$ that it entered this node on. 

Each random route has length $O(\sqrt{n}\log{}n)$, where $n$ is the number of good nodes in the graph. By a birthday paradox-like argument, two good nodes are likely have their random routes intersect within the good subgraph. Such an intersection allows these two nodes to trust each other. A bad node may not follow protocol, but it is nonetheless constrained by the scarcity of attack edges. Specifically, Yu et al. \cite{yu_sybilguard} prove that the probability of a good node entering the bad subgraph is $o(1)$; that is, bad nodes are unlikely to be trusted. Subsequent results (e.g., \cite{yu_sybillimit,lesniewski-laas:whanau,danezis2009sybilinfer}) share common ground with this approach, and are addressed by Urdaneta et al. \cite{urdaneta2011survey}. \smallskip

{\it What new defenses have emerged in the interim?}  Cao et al. \cite{cao} propose a defense named SybilRank, which exploits the difference in mixing time between regions composed of good nodes and regions composed of bad nodes. Combining the Louvain method \cite{louvain} to  identify communities in the social network, along with truncated power iteration to approximate trust propagation by simulating many short random walks \cite{langville2004deeper}, SybilRank assigns a trust value to nodes. Simulation results imply that SybilRank can outperform  SybilLimit\cite{yu_sybillimit} and SybilInfer\cite{danezis2009sybilinfer}.  Additionally, the authors conduct a real-world experiment using the social network Tuenti \cite{tuenti}, where SybilRank yielded an 18-fold increase in the efficiency of identifying Sybil users as compared to Tuenti’s abuse-report-based approach.

Another notable system is Wh\~{a}nau, proposed by 
Lesniewski-Laas et al. \cite{lesniewski-laas:whanau}, which is a  Sybil-resistant DHT that uses random walks to construct correct routing tables in the face of a Sybil adversary. Wh\~{a}nau is able to tolerate $O(n/\log n)$ attack edges, and guarantees successful routing with high probability. Simulation results involving the social networks Flickr, LiveJournal, YouTube, and DBLP suggests that most routing requests succeed when the attack edges are less than $10$\% of good nodes. 

Making use of statistical methodology, Danezis and Mittal \cite{danezis2009sybilinfer} introduce SybilInfer. For a given start and endpoint of a random walk, called a \defn{trace}, the authors design a Bayesian inference approach to determine the likelihood that a trace involved good nodes. Through simulation results using real-world and synthetic data, the authors show that SybilInfer can outperform SybilLimit and SybilGuard.

Wei et al. \cite{wei:sybildefender} propose SybilDefender, a centralized defense mechanism that identifies Sybil nodes using random walks from known good nodes to reach the stationary distribution and select judge nodes, which then classify others based on visit frequency. Nodes resembling the good subgraph in walk behavior are deemed good; otherwise, they are flagged as Sybils. A complementary component, Sybil Community Detection, initiates walks from known bad nodes to efficiently identify clusters of Sybil identities. To further reduce attack edges, the authors introduce relationship rating and activity rating, leveraging user-rated relationships and interaction frequency, respectively. SybilDefender is evaluated using data sampled from Orkut and Facebook, and the results favor SybilDefender over SybilLimit, in terms of accuracy and run time by up to two orders of magnitude.

Tran et al. \cite{GateKeeper} propose GateKeeper, a defense that assumes the fast-mixing properties of random expander graphs. GateKeeper uses $m$ random nodes as sources of ``tickets'', which (loosely) are a form of proof that allow nodes admission to the network. Informally, good nodes should obtain many tickets, given that they belong to a fast-mixing subgraph, while bad nodes should receive far fewer. Nodes must obtain some threshold number of tickets in order to be admitted.  GateKeeper offers strong theoretical guarantees, restricting bad nodes to $O(\log k)$ per attack edge (where $k$ is the number of attack edges).

Given the reliance on random walks in literature, we highlight the survey by Alvisi et al. \cite{alvisi2013sok}, which addresses the structural properties of social graphs in Sybil defense protocols based on random walks to detect Sybil nodes. The characteristics of social graphs, such as conductance and mixing time of random walks, which help in differentiating between good and bad (i.e., Sybil) regions within the network are referred to as \defn{structural properties}. The authors address previous assumptions about these networks being made of two separate regions (good and Sybil), which are incorrect, as good regions are actually groups of loosely connected nodes. The authors highlight the importance of careful selection of community selection algorithms to avoid mis-classifying adversarial nodes as good. They suggest that Sybil defense should aim to securely whitelist a set of trustworthy nodes rather than attempting universal Sybil detection.

Finally, we highlight an earlier result, not surveyed by Urdaneta et al. \cite{urdaneta2011survey}, by 
Scheideler and Schmid \cite{scheideler2009distributed}. Here, the authors design and analyze a distributed (min) heap called SHELL, with efficient methods for peer joins/departures and routing. Specifically, each join operation can be performed in $O(\log n)$ time and requires $O(\log^2 n)$ updates to the network topology; each deletions requires constant time with, again, $O(\log^2 n)$ updates. Routing can be accomplished in $O(\log n)$ hops, allowing for fast lookups. A major novelty of the work is that any two peers, $u$ and $v$ can route along paths only involving peers that joined the system prior to $u$ and $v$. This yields some advantages with respect to tolerating churn and the Sybil attack. In particular, peers that frequently join and depart the system will not interfere with these ``older'' routes. Informally, this is accomplished by setting the key value of each peer in the heap to be its join time; these ``time stamps'' are assumed to be verifiable. Given this assignment of key values, peers that contribute to churn are located near the bottom of the heap, while older (more stable) peers are positioned near the top of the heap. Thus, routing among these older peers involves only the top portion of the heap. This structural property also yields robustness to the Sybil attack. Suppose that Sybil attack occurs at some time $t$ after join times of  $u$ and $v$. All of the bad peers introduced by the Sybil adversary will have key values larger than those of $u$ and $v$; therefore, again, $u$ and $v$ may route using only those older peers, thus being unaffected by the attack.

\subsection{Machine Learning Methods}\label{ml-based-methods}

In the previous section, we surveyed defenses that leveraged the fast-mixing properties of networks in order to mitigate Sybil attacks. While this property may be available for some P2P systems, there is evidence to suggest that not all overlays do \cite{viswanath2012exploring, boshmaf_botnet}.  What defenses are possible then?

In this section, we review defenses based on ML methods, which often avoid the same degree of  reliance on purely structural properties. The application of these techniques to various aspects of network security issues has been surveyed (e.g., \cite{shaukat2020survey,buczak2015survey}); here, we focus on results with a close connection to P2P systems.

Early work by Haribabu et al. \cite{haribabu2010detecting}  employs a multi-layered feed-forward neural network approach to detect Sybil behavior in P2P networks. Aspects such as the number of queries sent, number of neighbors, and amount of content contributed are used as input features for a neural network. The network is trained using supervised learning with backpropagation to identify potentially malicious peers. Suspected peers are assigned a CAPTCHA, and their responses determine whether they are classified as Sybil attackers or legitimate users.

A result by Cai and Rojas-Cessa \cite{cai2014containing} addresses Sybil attacks on P2P trust management systems. The authors propose a framework with three mechanisms: a local trust table for tracking suspicious interactions, a $k$-means clustering algorithm for grouping peers by trust, and a cryptographic transaction verification method to prevent false reports. 

Gong et al.~\cite{gong_sybilbelief} discuss the vulnerability of P2P systems to the Sybil attack and propose a defense, SybilBelief, which they claim overcomes a shortcoming in previous approaches. Unlike the supervised learning used by the method of Haribabu et al \cite{haribabu2010detecting},  SybilBelief is designed by recasting the social network as a Markov Random Field (MRF)~\cite{clifford1990markov}, and then using Loopy Belief Propagation (LBP)~\cite{murphy2013loopy} to arrive at a probability that a node is good or Sybil. The authors argue that this approach is a more robust and scalable approach.

Subsequent to SybilBelief, Gao et al. \cite{gao_sybilframe} propose SybilFrame,  which the authors claim to improve upon SybilBelief. This improvement is achieved by combining local node classification---which is informed by structural properties of the graph and crowdsourcing---and MRF with LBP. An interesting aspect of SybilFrame is that it finds application in settings where trust between participants is weak and the number of attack edges can be high; this is in contrast to much of the prior work, which assumes relatively few attack edges are forged. Another related result is by Gao et al. \cite{gao_sybilfuse} who propose SybilFuse, which is a defense that combines node and edge classifiers in order to compute local trust scores for both nodes and edges,  respectively. The use of this in P2P systems is proposed by Gupta et al. \cite{gupta2023bankrupting} (discussed in Section \ref{resource-burning-methods}).

Another classification-based result that is related to secure P2P systems is by Boshmaf et al.\cite{boshmaf_botnet}. The authors argue that a sophisticated Sybil adversary may not impact the mixing time of the network, thus thwarting many methods discussed in Section \ref{sec:graph-based}. For example, in a structured P2P system, the adversary may have its Sybil IDs adopt the network topology; therefore, there is no structural difference between the good and bad subgraphs. Boshmaf et al.\cite{boshmaf_botnet} discuss these ideas and their implications, specifically with respect to DHTs.

\subsection{Resource-Burning Approaches}\label{resource-burning-methods}

\defn{Resource burning (RB)} is the verifiable expenditure of a  resource, and its application to security problems goes back over two decades (e.g., \cite{dwork:pricing,dwork2003memory}). What is a resource? Commonly, this refers to computational power,  bandwidth, computer memory, or even human attention. Puzzles or challenges of tunable difficulty can be designed around such a resource; for instance, computational puzzles have a long history in the literature \cite{laurie-proof,liu:proof,dwork:pricing}. Another common puzzle type is the  {\it completely automated public Turing test to tell computers and humans apart (CAPTCHA)}, where the resource is human effort~\cite{von2003captcha,moradi2015captcha,baird2005scattertype}. Challenges based on bandwidth require the user to send a sufficient number of bits within a bounded amount of time \cite{walfish:ddos}. Additional puzzle constructions exist: {\it Proof of Space-Time}, which requires allocation of storage capacity~\cite{moran2019simple}; and {\it Proof of useful-work}, which consumes CPU cycles to solve real-world scientific or engineering problems~\cite{shoker2017sustainable, ball2018proofs}.

With the adoption of proof-of-work (PoW) in systems such as BitCoin~\cite{nakamoto:bitcoin}, FileCoin~\cite{filecoin}, and (inital versions of) Ethereum~\cite{ethereum}, RB has become a ubiquitous security tool. Ultimately, burning such resources translates into a monetary cost.  As we will see,  RB-based defenses can have a ``rate-limiting'' effect against a resource-limited adversary by imposing a cost on producing and maintaining Sybil IDs in the system.

In the context of DHTs, a cost analysis by Urue\~{n}a et al. \cite{uruena} involves estimating the total number of Sybil IDs needed to thwart access to specific content \cite{jennings,stoica2001chord}.  Given this estimate, the authors argue that an attacker must control a large number of IDs in order to successfully deny access to targeted content. Therefore, a small amount of RB for each ID generated should be sufficient to inflict a prohibitive cost on the adversary, thus preventing the majority of attacks on content storage and retrieval.

Several results assign puzzles to peers that must be solved within a certain amount of time; otherwise, the associated peer is essentially blacklisted by other peers.  One such result pertaining to DHTs is by
Li et al. \cite{li2012sybilcontrol}, who propose a defense called \textsc{SybilControl\xspace}. This defense employs computational puzzles to restrict the number of IDs that a Sybil adversary can generate. For any ID \(u\), its neighbors issue individual puzzles, which can be combined into a single puzzle for \(u\) to solve. A verifiable solution to this puzzle must be returned to \(u\)’s neighbors within a short timeframe; failure to do so results in these neighbors excluding \(u\) from the system. Simulation results using the Chord DHT show that SybilControl enables successful routing with latency comparable to the original Chord system, even in the presence of a powerful Sybil adversary.

A related result by Gupta et al. \cite{gupta_proof} leverages PoW to protect general distributed systems, including peer-to-peer networks. Each ID that wishes to join must solve a puzzle, which imposes a cost for entrance into the system. Furthermore, each time the total system size changes by a constant factor,  all IDs must solve a puzzle in order to keep participating; this is referred to as a \defn{purge}, and it imposes a \defn{purge cost}. A committee of IDs is responsible for issuing puzzles and verifying responses. Committee membership is decided by selecting IDs uniformly at random, which guarantees (with high probability) that the committee contains a good majority; therefore, the committee can act correctly by consensus. By forcing purges, the fraction of bad IDs in the system is limited to a minority. The authors show how to carefully trade off between entrance and purge costs such that the cumulative cost imposed on the good IDs is $O(T + g)$, where $T$ is the computational cost to the adversary, and $g$ is the number of good IDs that have joined the system so far. Importantly, if $T$ is small with respect to $g$, then the good IDs pay a resource cost that is proportional to their number, which seems reasonable. Otherwise, when $T$ is large to $g$ (i.e., the system is under substantial attack), the good IDs are not paying a cost that is asymptotically greater than the adversary.

Continuing in this line of work,  Gupta et al. \cite{gupta2019peace,gupta2023bankrupting} propose a RB-based defense, where the difficulty of the entrance puzzle is tunable. As in~\cite{gupta_proof}, purges are conducted  whenever the system size changes by a constant factor.  By carefully tuning the  hardness of the entrance puzzles, the cost to the algorithm is $O(\sqrt{\mathcal{T}\mathcal{J}} + \mathcal{J})$, where now $\mathcal{T}$ is the RB rate of the adversary (i.e., the total resources burned by the adversary divided by the execution time), and $\mathcal{J}$ is the join rate of good IDs (i.e., the total number of good IDs divided by the execution time).  Notably, this result provides an advantageous defense in the sense that attackers will spend asymptotically more resources than the good IDs. Through simulation results using real-world network data, the authors conclude that their algorithms are competitive with the contemporary PoW-based Sybil defense, \textsc{SybilControl\xspace}, and can be significantly more efficient under attack.


G\"{u}nther and Pietrzak  \cite{gunther:putting} explore the use of Proof of Space (PoSp) in DHTs, where peers must prove wasted disk space to limit the number of Sybil peers. This approach is suitable for storage-heavy applications, making attacks costly by requiring adversaries to contribute significant disk space to control a substantial portion of peers in the DHT.

\begin{figure}
    \centering
    \includegraphics[width=\linewidth]{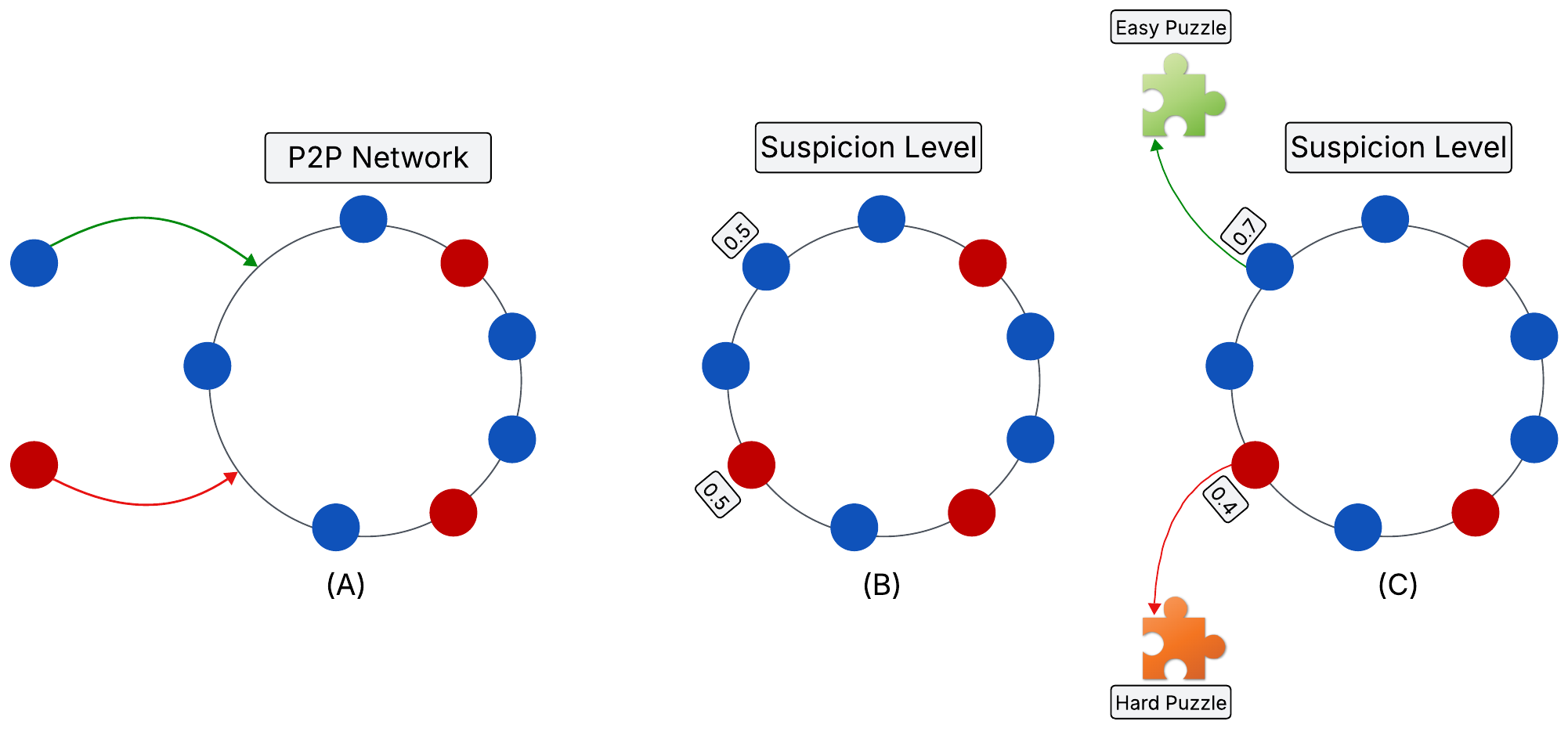}
    \caption{ (Left) When new peers join the system, they both receive an equal $0.5$ trustworthiness score (Center). As other factors contribute, these scores changes, which (Right) impacts the difficult level of puzzles they must solve in order to remain in the system.}
    \label{fig:RB_approach}
    \Description{The figure shows nodes evaluated by a suspicion level. Nodes with low suspicion are treated as honest and solve easy puzzles. Nodes with high suspicion are considered potential Sybils and must solve harder puzzles, making it costly for Sybil nodes to persist in the network.}
\end{figure}

In addition to a time constraint on receiving a valid puzzle solution, some approaches use additional information to identify peers suspected of being malicious and assigns them increasingly hard puzzles; this is illustrated in Figure \ref{fig:RB_approach}. Li et al. \cite{Liu2020PoWSybil} presents a defense against the Sybil attack that requires every peer to periodically solve PoW puzzles, issued to all identities simultaneously at fixed intervals; this is similar to the notion of a purge used in \cite{gupta_proof}. If a peer misses a verification deadline, it is marked as suspicious and must solve a harder puzzle in the next round, increasing the cost for attackers. Similarly, Da Costa Cordeiro et al. \cite{Cordeiro2012AdaptivePuzzle} 
 uses computational puzzles to limit identity creation. Given an identity request, the source peer is assigned a score based on the ratio of identities granted to its associated users (e.g., 
 peers operating from the same IP address range, or located in the same geographic region) compared to the average across other peers. Those peers with frequent identity requests must solve more difficult puzzles.  In both works, the analysis and experimental results demonstrate that the approach can significantly  reduce the effectiveness of Sybil attacks while maintaining low overhead for good peers.

\subsection{Summary of Results}

The following table summarizes the results surveyed in Section \ref{sec:Sybil-Attacks}.

\input{Tables/sybil-defense-table}

\subsection{Discussion and Future Work}\label{discussion-sybil-attack}

It is worthwhile to compare the general approaches surveyed in Sections \ref{sec:graph-based}, \ref{ml-based-methods}, and \ref{resource-burning-methods}. Given the continuing popularity of social networks, defenses based on their topological properties allow us to exploit data on human interactions in order to enhance security. However, almost all of these defenses rely on two assumptions. First, the good subgraph must be fast mixing, which may not hold true \cite{alvisi2013sok}; thus, methods of improving the mixing time would benefit these defenses, which is explored by Mohaisen and Hollenbeck \cite{mohaisen2014improving}. Second, forming attack edges is difficult; however, this may not always hold in real-world scenarios where users accept connections without scrutiny, as experimental evidence suggests \cite{bilge:all}.

ML and classification-based approaches---such as SybilBelief, SybilFrame, and SybilFuse---avoid reliance on the topological assumptions discussed above. These methods leverage techniques like Markov Random Fields and Loopy Belief Propagation to assign probabilities to nodes, providing scalability and resilience to noisy data \cite{gong_sybilbelief}. However, they are not without flaws; sophisticated adversaries can mimic network topologies to evade detection, particularly in structured systems like DHTs \cite{boshmaf_botnet}. Additionally, although substantial empirical evidence demonstrates the effectiveness of these approaches, they lack the rigorous mathematical guarantees established by theoretical results in the literature.

Finally, as discussed, RB has become a popular security technique in practice. Even for challenging adversarial models, provable security guarantees can be demonstrated, such as bounds on the number of Sybil peers, and the relative amount of RB performed by the adversary versus the good peers. However, many RB-based defenses are {\it proactive}, always imposing an RB cost, even in the absence of an attack. For example, in the simulations performed with \textsc{SybilControl\xspace}~\cite{li2012sybilcontrol},  each ID $u$ is issued a puzzle by its neighbors every $5$ seconds, regardless of the number of Sybil IDs present in the system.  In contrast, the result by Urue\~{n}a et al. \cite{uruena} does not suffer as much from this issue, since their proposed defense attaches a small RB cost to each ID created; thus, good IDs that remain in the system are not charged in perpetuity. This situation is remedied to an extent by the results in~\cite{gupta2019peace,gupta_proof}, where the amount of RB imposed on IDs is {\it reactively} tuned such that the good IDs have an RB cost that is asymptotically equal to or smaller than the adversary's cost. Nonetheless, all defenses incur some additional overhead, even when there is no attack.

In terms of future work, there are several directions that seem promising. First, in the context of defenses based on social-network topology, it may be useful to consider a network model that captures a cost to the adversary for establishing each edge between bad nodes and good nodes.  For example, in practice, large language models (LLMs) may be able to fool good participants into forming strong social connections with Sybil IDs; however, such an attack may incur a monetary cost or require substantial time to execute. 

Conversely, what about the cost to defenders? Integrating ML-based techniques with Sybil detection algorithms has shown progress, as we discussed above. However, there may be costs associated with these ML-driven approaches used to detect attack edges or Sybil nodes. For example, training and retraining can have a substantial cost in terms of computation and time.  More meaningful defenses may be designed by accounting for these costs in the model. Additionally, such costs may be reduced by leveraging information gleamed from the graph structure in order to aid ML (e.g., \cite{JETHAVA2022:hybrid-approach-sybil-detection}, \cite{Yue2020SybilTrustGCN}).

On a related note,  it might be intriguing to design and analyze a defense where key structural invariants---such as low diameter, the prevention of eclipse attacks, maintaining good expansion---can be preserved at a cost, despite an adversary adding or removing edges, also with associated costs. There is some general prior work in this direction (e.g., \cite{tochner2020gametheoretic,manshaei:game-theory}). Specific to expansion, work by Pandurangan et al. \cite{pandurangan2016dex} shows how this can be efficiently maintained, despite adversarial node deletions and additions. However, this result seems unsuitable for social networks given that edges represent trusted relationships between the endpoints, which cannot easily be altered. Is there a way to maintain expansion over time, while accounting for such constraints in social networks? Developing defenses that maintain key properties, and  where the adversary's costs significantly exceed those of good nodes, would be compelling.


\section{Defenses for Routing Attacks} \label{DA_Routing_Storage}

Routing attacks encompass various malicious actions, such as when a malicious peer, tasked with facilitating a route, becomes unresponsive or deliberately drops or corrupts the data it is supposed to forward. An eclipse attack (recall Section \ref{sec:overview-security}) can be viewed as an attack on routing, where an adversary attempts to isolate a set of nodes and, thus, control what information can be routed to them. Here, we survey results that propose defenses against these attacks.
 
\subsection{Redundant Routing}\label{redundant_routing}
 
Redundant routing in P2P systems involves sending multiple parallel messages along diverse and (largely) independent network paths to limit the impact of bad nodes. By leveraging path diversity, we can reduce the likelihood that all routes are compromised or controlled by the adversary, thus enhancing security. In the following, we examine several  works that incorporate redundant routing as a core defense mechanism.

Li et al. \cite{li} propose modifications to the Kademlia DHT \cite{maymounkov} that make it more robust to eclipse attacks. In  order to check that content is published correctly, a random peer is selected to perform a lookup, and if the correct content is received, publication is assumed to be successful. Otherwise,  a more involved  protocol is used that involves exploring  nodes  close to the corresponding content's key value, in the hopes that any faults along the original lookup path can be bypassed. Additionally, the correctness of a peer $u$ (and its  data) is verified through the use of ``witnesses'', which are other peers that vouch for $u$. These approaches are used under the assumption that at least $95$\% of the peers are correct. The authors report that their approach performs well, using a network of $10,000$ peers. More than $1,000$ distinct public IP addresses are needed to successfully achieve an eclipse attack. We note that, in practice, the largest  Kademlia-based system is the BitTorrent Mainline DHT, which serves between $10$ million and $25$ million users on a daily basis \cite{wang}. The result by Li et al. \cite{li} implies that an adversary would need an enormous number of IP addresses to successfully attack a system of this size.

Germanus et al. \cite{germanus_susceptibility} consider an  adversary who uses knowledge of the network's topology to   eclipse good nodes with minimal resources. The authors illustrate the feasibility of attacks in several structured P2P systems, such as Chord, Kademlia, and Pastry.  Subsequent work by Germanus et al. \cite{germanus_mitigating} proposes methods aimed at providing routing that is resilient to eclipse attacks. Simulations suggest that these heuristics offer promising performance in Kademlia-based DHTs, although the system size is limited to approxmiately $10K$, and high churn hampers the effectiveness of their mitigation strategies. In a follow-up paper Germanus et al. \cite{germanus_pass} propose a mitigation strategy that scales to   hundreds of thousands of peers and allows for 90\% to 100\% of lookups to succeed, even in the presence of high churn.

In closely-related work, Heilman et al. \cite{heilman} use of Monte Carlo simulations to quantify the resources needed by an adversary to perform eclipse attacks on the Bitcoin network. In general, these simulations show that a successful attack requires a large number of addresses in a number of different IP address blocks. However, the authors point out that organizations do exist with sufficient IP addresses to launch an attack; additionally, there exist botnets with over $29,000$ hosts that could satisfy these requirements. Finally, the paper discusses  potential countermeasures to these attacks, three of which were subsequently implemented by the Bitcoin developers as of $2015$.  

In the context of the Chord DHT,  Rottondi et al. \cite{rottondi} develop a simulation to investigate the effects of eclipse attacks, and they argue that attacks can be mitigated by  modifications to the routing protocol. Specifically, the authors call for the addition of extra peers to each node's finger table that allow for redundant routing; these extra peers form an \emph{auxiliary node list}. Creation of this list can be done in either a centralized or distributed manner; the former requires a trusted third party with full topological information to set up the lists, while the latter has a peer add to its auxiliary list any source node whose request it facilitates. For a system size up to $1,000$ peers, simulation results illustrate that the use of the proposed counter measure can mitigate the influence of an eclipse attack, with the centralized implementation performing best.  

Kapadia and Triandopoulos \cite{kapadia2008halo} also examine routing attacks in the Chord DHT. The authors propose  a ``High-Assurance LOcate" (or HALO) protocol for DHTs that leverages the use of redundant routes to tolerate Byzantine faults (recall Section \ref{sec:definitions-terminology}). Here, the threat model assumes that each peer is Byzantine independently with uniform probability; that is, this is a random fault model. Using the Chord DHT with $n$ peers,  HALO operates by identifying the $\Theta(\log n)$ immediate predecessors of a target peer $v$; these are known as the ``knuckles'' of $v$. Intuitively, \whp~there will be at least one knuckle that is correct, given that faults occur randomly.\footnote{An event occurs with probability at least $1-1/n^c$ for some tunable constant $c\geq 1$ is said to occur ``with high probability''.} A query is routed using the regular Chord lookup procedure on each knuckle, after which the query succeeds given that at least one knuckle is correct. Simulation results indicate that HALO can provide secure routing (i.e., without failure) when up to 22\% of the network peers are Byzantine. Even when 25\% of the peers are Byzantine, the percentage of lookup failures is low at most 3\%.

\subsection{Group-Based Approaches}\label{group_based_approaches}

\begin{figure*}[t!]
    \includegraphics[scale=0.38]{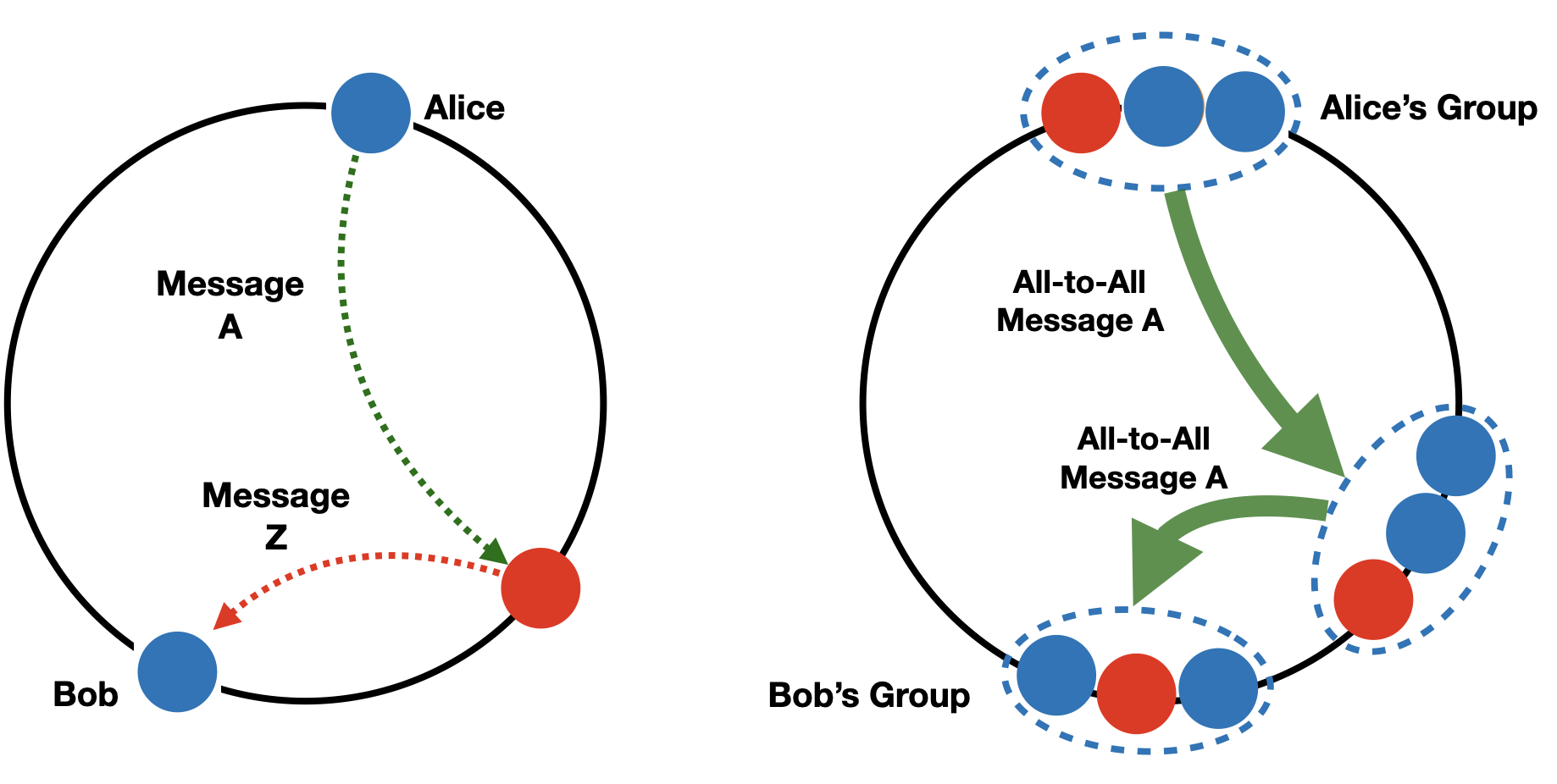}
   \caption{(Left) Alice attempts to route a message $A$ to Bob, but the message is corrupted by an intermediate bad peer. (Right) The use of groups to ensure Alice's message is securely routed despite bad peers.}\label{fig:P2P}\vspace{-5pt}
   \Description{The left side shows Alice’s message to Bob being corrupted by a bad peer. The right side shows groups, where messages are sent between all group members, preventing bad peers from disrupting the communication.}
\end{figure*}

The use of small sets of peers, often called \defn{groups}\footnote{Sometimes groups are referred to by different names, such as clusters \cite{guerraoui2013highly} or quorums \cite{saia2008reducing}.}, is  a very common solution to attacks on routing, including the eclipse attack. So long as each group has a majority of good peers, secure routing can occur via all-to-all communication between groups. Of all the messages a fixed peer receives,  the message in the majority will be correct, which the peer then acts on to facilitate a lookup request; this is illustrated in Figure \ref{fig:P2P}.  Thus, groups rather than peers become the ``atomic'' unit of the P2P network, and majority filtering provides scalable  resilience to malicious behavior.

We wish to highlight, that the use of groups also mitigates pollution/poisoning (recall Section \ref{sec:overview-security}). Each peer can store replicas of the data stored by other members of its group. In this way, content contributed by good nodes cannot be deleted or corrupted by bad nodes. Additionally, if paired with a defense that limits the Sybil IDs to a minority (e.g.,\cite{gupta_proof,gupta2019peace,gupta2023bankrupting}), groups can lessen the impact of join-leave attacks and spurious lookups. Each group may control the number of join events they permit per interval of time. While this does not prevent bad peers from joining, it does limit their join rate, and thus their contribution to churn. Similarly, each peer $u$ may be allocated an upper limit on the number of lookups it can perform per time interval. This limit can be enforced by the corresponding group, since if their messages do not accompany $u$'s, then the good peers in the receiving group will simply ignore $u$'s request.


Urdaneta et al. \cite{urdaneta2011survey} describe several earlier results on the use of groups. This approach was initiated by Awerbuch and Scheideler \cite{awerbuch_scheideler:group}, which achieves provable security so long as the fraction of bad peers is $O (1/\log{N})$, the  join rate of bad peers remains $\mathcal{O}(1/\log^2{N})$, and the join and departure rate of honest peers stays bounded at $\mathcal{O}(1/\log{N})$. Later results by Awerbuch and Scheideler \cite{awerbuch2006towards,awerbuch:towards2}---again, surveyed by Urdaneta et al. \cite{urdaneta2011survey}---achieve stronger guarantees, being able to tolerate a constant fraction of bad peers, and having no restrictions on the join and departure rates. 
More specifically, Awerbuch and Scheideler design an algorithm called the {\it cuckoo rule}, which works as follows. If a new peer $v$ wants to join the system, then $v$ is placed at a random point in the normalized ID space, $[0, 1)$. For a constant $k > 0$, all of the peers in a region of size $k/n$ are evicted to uniformly and independently chosen random points in the ID space. Awerbuch and Scheideler show that this rule maintains two important conditions: the Balancing Condition, and the Majority Condition. For every interval of $I$ in the normalized ID space, $[0,1)$, of size at least $(c \log n) / n$ for a constant $c > 0$: the Balancing Condition is met if $I$ contains $\Theta(|I| n)$ peers, and the majority condition is met if $I$ contains a majority of honest peers.


Subsequent to the survey by Urdaneta et al. \cite{urdaneta2011survey}, 
Sen and Freedman \cite{sen2012commensal} demonstrated that a revision to the cuckoo rule can improve performance. Simulation results of the original cuckoo rule indicate that, over time, peers become distributed non-uniformly in the ID space; this has a compounding effect whereby the number of peers per $k$-region exhibits high variance. This results in the need for a larger group size to be used (e.g., $256$ peers per group when $n=2048$), which impacts practical performance. To address this issue, the authors propose a new {\it commensal cuckoo rule} that is executed each time a new peer joins at some location $x\in [0,1)$. Informally, in the corresponding $k$-region, rather than moving all peers to locations chosen \uar~in the ID space, the new rule moves only $k\cdot r$ randomly chosen peers, where $r$ is the ratio of the current group size to the average group size.  Simulating system sizes of up to approximately $8000$ peers, the authors find that, in comparison to the original cuckoo rule, their new rule provides a significant increase in the robustness to Byzantine peers using groups of size $64$.


Random number generation is important to the cuckoo rule. As we have seen, a joining peer is typically assigned a group \uar~, which mitigates (to some degree) the ability of the adversary to obtain a bad majority in any group. The original cuckoo rule requires $O(\log^2 n)$ random bits per join operation. Therefore, efficient random number generation is an important ingredient. Using secure multiparty computation, a group of size $\Theta(\log n)$ can generate $O(\log n)$ random bits with $O(\log^3 n)$ message complexity \cite{awerbuch2006towards}.  However,  Awerbuch and Scheideler \cite{awerbuch:random} are able to improve on this situation by designing a distributed random number generator (RNG) ---called a round-robin RNG---that generates $\Theta(\log n)$ random bits with a small bias, but requires only $O(\log^2 n)$ message complexity. The authors then show how, despite this bias, a revised version of the cuckoo rule---called the {\it de Bruijn cuckoo rule}---can use the  $O(\log n)$ random bits per join to maintain the balancing and majority conditions \whp~over a polynomial number of join operations. 


Another aspect of group-based defenses that has received is reducing the communication costs. While all-to-all communication (as illustrated in Figure \ref{fig:P2P}) suffices, each communication between adjacent groups  yields a message complexity of $\Theta(\log^2 n)$, given that groups are each of size $\Theta(\log n)$. Early work by Saia and Young \cite{saia2008reducing}  reduces the message complexity to $O(\log n)$ in expectation, as well as reducing the bit complexity in the case where large messages are sent. Subsequent work by Young et al. \cite{young:towards,young:practical} uses cryptographic methods to give a deterministic method with $O(\log n)$ message complexity, and a randomized method with $O(1)$ expected message complexity, for communication between two adjacent groups. Notably, the application of this latter result in a DHT where each lookup requires \whp~ $O(\log n)$ hops (e.g., the Chord DHT), results in $O(\log n)$ hops in expectation. Thus,  this performance is comparable to that of non-group-based communication, demonstrating that group-based approaches can achieve near-equivalent efficiency in terms of hop count.


Yet another optimization is explored with respect to the group size, where  Jaiyeola et al. \cite{JaiyeolaPSYZ18} show how PoW can be used to reduce the group size to $O(\log\log n)$. Despite shrinking the groups size substantially, the authors show that the system can withsand a Byzantine adversary that controls a constant fraction of the peers. The use of PoW  incurs an additional computational cost that is absent from prior work in this area. Additionally, some small $o(1)$-fraction of the ID space in the DHT cannot facilitate lookups reliably; however, \whp~the vast majority of lookups will succeed. 


Saad and Saia \cite{saad:self-healing2} consider a model where an adversary controls $t\leq (1/4-\epsilon)n$ nodes in an $n$ node network, where $\epsilon>0$ is a (small) constant. In this setting, the authors explore how to perform reliable multiparty communication (MPC), where each node has an input, and the nodes wish to jointly compute a function $f$ over these inputs. Again, the proposed approach makes use of $\Theta(\log n)$-sized groups, where each group contains at most $1/4$-fraction of  faulty nodes. These groups are used as gates to form a general circuit. An interesting aspect of the authors' approach is that, once a corrupted message is detected, the corresponding  bad peer can be  ``deleted'' from the network.  This deletion is effected by having good peers ignore messages from this bad peer, thus effectively quarantining the bad peer; in this way the algorithm is self-healing. This algorithm accomplishes MPC with  $O (m + n \log n)$ messages,   $O(m + n \log n)$ computation, and $O(l)$ latency, where $l$ is the depth of the circuit that computes $f$. Moreover, the authors show that the adversary can cause only $O (t  (\log^* m)^2 )$ corruptions before all of its bad peers are deleted. Simulation results in \cite{saad2017theoretical} illustrate that this algorithm greatly reduces the message cost compared to the naive approach.


An earlier paper by Knockel et al. \cite{knockel2013self} proposes a less general result in a similar setting where, instead of computing a circuit, the authors show how to maintain communication between all good peers with near optimal communication cost and latency. Another similar paper by Boshrooyeh and Ozkasap \cite{boshrooyeh} proposes a security mechanism for skip graphs, making use of cryptographic signatures to isolate badly behaving peers. Likewise, Ismail et al. \cite{Ismail_a,Ismail_b} propose a method for cleaning up bad peers in groups. Informally, their heuristic works by having good peers issue requests to peers they suspect of being bad, based on corrupted responses. When a sufficient number of peers agree that a peer is bad, this peer is evicted from the system (i.e., removed from the routing tables of good peers that made this diagnosis).

The Fireflies system by Johansen et al. \cite{johansen2006fireflies} shares some common ground with the above approaches in that it  makes use of ``accusations''. Specifically, with the identifier space represented by a ring, a predecessor of peer $u$ can accuse $u$ of being Byzantine. Since positions are (pseudo)random due to hashing, it is unlikely that any good peer will have a large number of Byzantine predecessors. Thus, the value of $t$ can be chosen to reduce the probability to some tolerable value, ensuring that at least $t+1$ (i.e., a majority of its) predecessors are good. This property also underpins the authors' claim that the good peers remain connected within the Fireflies topology. Peers periodically broadcast their accusations to the network and update their view based on broadcasts they receive from other peers. If an accusation is made by peer $v$ and the accused peer does not respond within a bounded amount of time, then $v$ removes the accused peer from its own view of the network. By itself, Fireflies does not provide secure routing. The authors argue that a single-hop DHT can be built on top of Fireflies, but this requires that each good peer has a large forwarding table. A construction for DHTs where peers have only a partial view of the network is not provided. The authors experiment with their construction on PlanetLab. With up to $20$\% of the peers being Byzantine, Fireflies gave good performance; specifically, in the cases where Byzantine peers sent spurious accusations or refused to participate in forwarding accusations.


Guerraoui et al. \cite{guerraoui2013highly} consider a model where the system size is always at least $n_0$, but allows for a polynomial change in $n_0$ over time. This stands in contrast to many other results in the static setting, where $n$ does not change by more than a constant factor (recall Section \ref{sec:definitions-terminology}). In this setting, the authors propose P2P system that can tolerate a Byzantine adversary who controls strictly less than a $1/3$-fraction of the peers. A key ingredient is the familiar use of groups of $\Theta(\log n)$ peers in order to majority filter on decisions, where $n$ is the current systems size. Groups may merge or split, depending upon whether a group finds itself to be too small or too large, respectively, as the system size changes. The groups link together to form a network topology which has good expansion. A group can be chosen uniformly at random by performing a random walk on this expander graph, which is important for placing newly-joined peers into the system. Overall, the authors show that secure lookup is possible with polylogarithmic-in-$n$ latency and message complexity.

Another work that deals with system churn is by Kuhn et al. \cite{kuhn:towards} present a construction that tolerates worst-case churn using groups organized into a hypercube topology. The authors assume a model where time proceeds in rounds, and an attacker uses peer insertions and deletions in each round in an attempt to partition the network (e.g., see \cite{heck} for an empirical exploration of such an attack). The authors prove that, even with $O(\log n)$ peers being inserted or removed per round, it is possible to preserve connectivity and all network content, while maintaining logarithmic degree and optimal diameter.

Aradhya et al. \cite{aradhya:robust-blockchain} also address the design of a robust DHT in  a dynamic setting. Here,  the number joins and departures obeys a half life (i.e.,  given $H$ peers currently in the system, over the half-life time, at most $H/2$ new peers can join or at most $H/2$ peers can depart). Their approach makes use of $\Theta(\log n)$-sized groups (called committees) that are linked in a hypercubic topology; however, a novel  ingredient used in this DHT construction is a blockchain. The authors define a \defn{directory}, which consists of the most recent $\Theta(\sqrt{n}/\log n)$ blocks that store information on the most recent $\tilde{\Theta}(\sqrt{n})$ peers. Each of these associated directory peers is responsible for maintaining information on a subset of the committees. This directory allows recovery even in the event that a constant fraction of the committees fail (i.e., lost their respective good majority). Furthermore, the construction can quickly recover from failures where a large fraction of directory peers and committees fail, so long as a large fraction of honest peers can still run the blockchain protocol, and the process by which new peers join the system is still functional. Additionally, the authors provide an argument demonstrating that the half-life their system can tolerate is asymptotically optimal up to logarithmic factors.


Finally, more-recent work by Augustine et al. \cite{augustine2022fully} addresses a P2P setting where up to $\frac{n}{\mathrm{polylog}(n)}$ nodes may be Byzantine, where $n$ is the system size, which does not change; that is, the static setting. The approach assumes that the underlying network is an $n$-node expander graph of constant degree, and every peer has a constant-factor approximation to $n$. The main goal is to build a robust DHT by creating a virtual hypercube structure on top of the expander graph. This structure uses groups of size $\Theta(\log^3 n)$, which facilitate Byzantine agreement and random sampling, in addition to secure routing with $\poly{\log n}$ latency and communication complexity. 
Similar to Guerraoui et al. \cite{guerraoui2013highly}, the membership of each group is formed by using random walks on the expander graph in order to select peers (nearly) uniformly at random from the system. Shuffling is a familiar ingredient that the authors employ to maintain the good majority in each group.

\subsection{Constructions that Tolerate High Churn}\label{high_churn}

As discussed in Section \ref{sec:overview-security}, high churn can be viewed as a form of DoS attack. In recent years, several results have been proposed by the research community aimed at tolerating high churn under adversarial settings.

Zhang et al. \cite{zhang} argue that many defenses against the eclipse attack have high communication cost. To address this issue, the authors propose the use of computational puzzles in order for each ID to join the network. While this is not the first approach to employ computational puzzles in a P2P system (e.g., Castro et al. \cite{castro:secure}, Li et al. \cite{li2012sybilcontrol}, Rowaihy et al. \cite{Rowaihy:2007}), the authors highlight that they have the advantage of not relying on a certification authority, and that their approach induces little churn. 
In terms of shortcomings, the authors note that their defense only imposes a prohibitive cost on the attacker when the DHT is sufficiently large, and that legitimate peers who suffer intermittent connectivity, may find themselves incurring significant costs to join and rejoin the network.

Anceaume et al. \cite{anceaume2008peercube} propose PeerCube, which is designed to thwart eclipse attacks in a high churn setting. Peers organize themselves into ``clusters'' (i.e., groups), which are sets of peers of size $O(\log n)$, and these clusters form a hypercubic topology. Clusters facilitate key operations, such as storage and retrieval, by using a subset of their peers known as ``core'' members. These core members are tightly connected and can employ Byzantine consensus protocols to overcome malicious interference. By not using the entire membership of a cluster, the authors argue that the impact of churn is reduced. Similar to Guerraoui et al. \cite{guerraoui2013highly}, clusters must split or merge whenever they become to large or small. The authors claim that each cluster can maintain a good majority by using a randomized join procedure, and ultimately they argue that routing in Peercube between any two peers requires $O(\log n)$ messages and $O(\log n )$ hops. PeerSim \cite{peersim} simulations indicate that Peercube can indeed mitigate attacks; for example, with 15\% of the system being malicious, 98\% of requests are successful.  

Going beyond just detecting the presence of an attack, the approach by Xie and Zhu \cite{Xie} also identifies  malicious nodes. Two random sampling approaches are used to determine whether an attack is underway; if any peer advertises the loss of a broadcast message, then this indicates malicious behavior. A key assumption is that there is a special entity, known as a group controller, that has information on the network topology and can sample peers. Peers are diagnosed as malicious  via the use of a reputation system. Backtracking from leaf peer with missing messages towards the root, the authors define a method for calculating a ``suspicion'' value per peer. Simulation results indicate that the methods work well, achieving a false positive rate of less than $2\%$, even in highly dynamic settings where message loss may also occur due to network churn.

Augustine et al. \cite{augustine2013fast} tackle Byzantine agreement in highly dynamic networks with frequent peer and link changes due to adversarial churn, modeled as a regular constant-degree expander graph (recall the notion of expansion in Section \ref{sec:graph-based}). They propose two randomized algorithms to achieve almost-everywhere Byzantine agreement with high probability despite significant churn and Byzantine nodes. The first handles $O(\sqrt{n}/\poly{\log n})$ churn per round and a topology-oblivious adversary controlling $O(\sqrt{n}/\poly{\log n})$ Byzantine peers. It uses random walks, where each peer initiates $\Theta(\log n)$ tokens (containing ID, counter, and $O(\log n)$ bits). Tokens traverse random neighbors, decrementing the counter until zero. After $\poly{\log n}$ iterations, all but $O(\sqrt{n}/\poly{\log n})$ good nodes agree on a value from a good node, requiring $\poly{\log n}$ bits per round per node. 

Their second algorithm tolerates $O(\sqrt{n}/\poly{\log n})$ churn against a topology-aware adversary controlling $O(\sqrt{n})$ Byzantine peers. Peers communicate directly using IDs and initially assign values based on adversarial input. After a fixed number of rounds, each non-faulty peer runs a {\it support estimation} algorithm to determine how many peers have a current value of $``0"$ or $``1"$. If the support for one value dominates by   $\Theta(\sqrt{n})$ from $n/2$, the peers deterministically update their values in favor of the dominant value.  The authors show that after $O(\log^{3}n)$ rounds, a large imbalance is established with high probability, and all nodes except $O(\sqrt{n})$ agree on a value. However, here a polynomial in $n$ number of bits must be transmitted/processed by each node per round. 

Continuing on in this model, 
Augustine et al. \cite{augustine2015fast} consider the classic distributed problem of leader election under Byzantine faults.  Specifically, the underlying topology is a sparse, bounded-degree expander graph. The number of Byzantine nodes is O ($n^{1/2-\epsilon}$), for some (small) constant $\epsilon>0$, and the amount of churn per round is $O ( \sqrt{n} / \poly{\log n}$, where $n$ is the stable
network size.

The intuition for the authors' randomized algorithm is as follows. The nodes are given the choice randomly to become a candidate participating in the election with a probability of $\Theta(\log n/\sqrt{n})$. Candidate nodes generate ``tickets'' to participate in the “leader lottery”. Each ticket is represented as a private random bit string of length $\Theta(\log n)$ bits. The faulty nodes can also participate in this “leader lottery”. A store mechanism is implemented to make sure that the information stays in the network irrespective of the topology changes and high node churn. The ticket of some honest node $u$  is stored in ${\tilde{\Theta}}(\sqrt{n})$ nodes, which can act to verify that this is indeed $u$'s ticket (and reject claims by Byzantine nodes).  Intuitively, given that the adversary can inflict $O( \sqrt{n} / \poly{\log n}$ churn per round, $u$'s witnesses are safe for $\poly{\log n}$ rounds. During this time, an almost-everywhere consensus algorithm is run that allows a (slightly biased) random coin flip to generate the winning ticket over $\poly{\log n}$ rounds. Then, in another $\poly{\log n}$ rounds, a sampling algorithm can be used to verify $u$'s claim and propagate this almost the entire set of honest nodes using random walks. 
Note that if multiple honest node have the same ticket, then the node with the smallest ID becomes the leader. 


\subsection{Summary of Results}

The following table summarizes the results surveyed in Section \ref{DA_Routing_Storage}.

\input{Tables/eclipse-storage_routing-defense-table}

\subsection{Discussion and Future Work}

The defense mechanisms surveyed in Sections \ref{redundant_routing} to \ref{high_churn} address redundant routing, group-based approaches, and solutions in  high-churn settings. Each method relies on distinct model assumptions, balancing efficiency and security differently, as we now discuss.

Redundant routing is very helpful against the eclipse and storage attacks. Mechanisms such as auxiliary node lists in Chord~\cite{rottondi} or predecessor-based verification in HALO~\cite{kapadia2008halo}, enhance resilience by diversifying paths. However, while these methods improve robustness, they can increase communication overhead and latency. For example, Kapadia and Triandopoulo~\cite{kapadia2008halo} demonstrate that HALO's reliance on redundant predecessors ensures secure routing, even with $25\%$ Byzantine peers, but this comes at the cost of logarithmic message complexity. Similarly, Li et al.~\cite{li} show that Kademlia modifications reduce eclipse vulnerability, but require additional verification steps, which may slow down lookups. A key challenge lies in optimizing redundancy to balance security guarantees with performance.

Group-based defenses (e.g., \cite{saia2008reducing,awerbuch2006towards}), leverage collective decision-making to limit the influence of Byzantine nodes. These methods are  effective at preserving an honest majority within each group; however, their scalability remains a concern. For instance, Guerraoui et al. \cite{guerraoui2013highly} achieve secure lookup in polylogarithmic time using groups of size $O(\log{n})$, but dynamic group splitting/merging introduces significant overhead; for example, splitting incurs $O(\log^4 n)$ rounds of communication between involved peers. Another example is the cuckoo rule proposed by Awerbuch and Scheideler \cite{awerbuch:random,awerbuch2006towards,awerbuch:towards2,awerbuch:cuckoo-journal}, along with the commensal cuckoo rule by Sen and Freedman \cite{sen2012commensal}, which gives strong guarantees on maintaining a good majority in every group, but requires periodic shuffling even in the absence of any bad peers.

Under high churn, we sometimes see a group-based approach (e.g.,\cite{kuhn:towards,anceaume2008peercube}) paired with techniques like resource burning (e.g., \cite{zhang,castro:secure,li2012sybilcontrol,Rowaihy:2007}) or random walks on expander graphs (\cite{augustine2015fast,augustine2013fast}). In practice,  these techniques can be  expensive, consuming network resources such as computational power and bandwidth, along with increasing latency relative to the setting where there are no attacks. 

In terms of future work, there are many directions to pursue regarding the above approaches for defending
against attacks on routing. First, are there any benefit to using RB to mitigate routing problems. For example, Castro et al. \cite{castro:secure} consider the idea of a monetary cost to purchase locations in the ID space. RB would appear to be a natural substitute for money, and the challenges of issuing RB puzzles and verification has been examined by Li et al. \cite{li2012sybilcontrol}. Similarly, perhaps RB may be combined with the use of groups to inflict a cost on an adversary who seeks to cause many joins and departure events.  In both cases, it might be worth exploring whether solutions exist where the RB cost to good peers grows slowly with the cost to the Byzantine adversary, thus providing a resource-competitive result \cite{gupta2020resource}.

The notion of shuffling (e.g., the cuckoo rule and its variants) for group-based defenses can yield strong security guarantees. However, shuffling can be needlessly expensive in cases where there is no attack or the attack is minimal. For example, even if only good peers are joining a DHT, the cuckoo rule still requires a (small) amount of IDs be relocated. In practice, this means terminating and establishing new communication links between machines, which incurs a latency and state cost.  Is there a way to reduce such costs in order to avoid this overhead in the absence of an aggressive attack? One approach would be to have a ``light-weight'' mechanism for detecting whether an attack is underway and then responding with a more ``heavy-weight'' shuffling operation. 

Aradhya et al. \cite{aradhya:robust-blockchain} demonstrate that incorporating blockchain technology into the design of a DHT can enhance resilience against severe failures, including Sybil attacks. It would be valuable to investigate additional advantages of blockchain integration, particularly in reputation systems, where peers could efficiently audit the actions of others to ensure accuracy and trustworthiness.

Along similar lines, the use of ML in diagnosing bad peers may lower the cost of failed lookups. Performance would then be parameterized by the accuracy of an ML black box (see \cite{mitzenmacher:predictions}); for example, this has been done recently for denial-of-service attacks in the traditional client-server model \cite{chakraborty:bankrupting}. Here, route selection could be based on feedback from this black box, allowing us to lower message complexity by reducing the amount of replication typically used in secure DHT construction (e.g., groups or redundant routes).

Another promising avenue for future research is the application of agentic artificial intelligence (AI) to enhance network security and resilience. Agentic AI is characterized by autonomous goal-directed behavior, persistent memory, and adaptive decision-making. It offers the potential to operate as a decentralized enforcement and monitoring layer within P2P networks. These agents might be employed to observe node behavior over time, detect deviations from protocol norms, and autonomously respond to suspected malicious activity. 
Additionally, in response to adversarial conditions, agentic AI could facilitate distributed coordination among nodes to share reputational data and adapt routing strategies. 
It is an open question as to whether such an approach would provide robust security guarantees similar to the results surveyed here.

\section{Final Thoughts}
This survey reviews the landscape of security threats in P2P networks and discusses the defense strategies proposed in the recent literature. We focused on two broad classes of attacks: Sybil attacks, and attacks on routing. We observe from the literature that the lack of a central authority, and an inability to verify identities, poses common challenges that are dealt with in a variety of ways. 
In the context of the Sybil attack, we examined a range of countermeasures, including graph-structured defenses, ML-based classifiers, and resource-burning techniques. For attacks on routing, we explored defenses based on redundant routing, group-based approaches, and solutions that can tolerate high churn. 

We feel that many interesting  challenges remain. In our discussion sections, we have attempted to highlight promising avenues for future work. We also note that, while the surveyed papers each focus on a narrow security issue, we believe that in many cases, these techniques may be deployed in tandem. For example, ML approaches might complement group-based techniques, and yield stronger resilience than using either as a standalone solution. 

Our survey, together with Urdaneta et al. \cite{urdaneta2011survey}, highlights the ongoing challenges in designing secure  P2P systems. We anticipate sustained research activity in this field, and we eagerly await a future survey paper within another decade. 

\begin{acks}
This work is supported by NSF award CNS-2210300. 
\end{acks}

\bibliographystyle{ACM-Reference-Format}
\bibliography{rb}

\end{document}

%% file: Tables/sybil-defense-table.tex
\begin{longtable}{|p{2.8cm}|p{3.7cm}|p{8.3cm}|}
    \hline
    \rowcolor{LightCyan} {\bf Type of Defense} &  {\bf Name(s)} &  {\bf Brief Summary}\\
    \hline
    \endfirsthead 

    \hline
    \rowcolor{LightCyan} {\bf Type of Defense} &  {\bf Name(s)} &  {\bf Brief Summary}\\
    \hline
    \endhead 

    \hline
    \multirow{2}{*}{\makecell[tl]{Leveraging\\ Social Networks}} & Yu et al. \cite{yu_sybilguard} & SybilGuard introduces the use of mixing time properties in social graphs. Specifically, they assume that the good subgraph tends to be fast-mixing, while the bad subgraph is not. SybilGuard tolerates $O(\sqrt{n} \log n)$ bad nodes, provided that the number of attack edges remains $o(\sqrt{n}/\log n)$, where $n$ is the number of good nodes.\\
    \cline{2-3}
    &Yu et al. \cite{yu_sybillimit} & SybilLimit builds on SybilGuard's approach by reducing the number of bad nodes admitted per attack edge to $O(\log n)$ with high probability.\\
    \cline{2-3}
    &Lesniewski-Laas et al. \cite{lesniewski-laas:whanau} & Wh\~{a}nau is a Sybil-resistant DHT that leverages social connections to establish secure routing using routing tables of size  $O(\sqrt{n} \log n)$ entries per node, and tolerating up to $O(n/\log n)$ attack edges. \\
    \cline{2-3}
    &Cao et al. \cite{cao} & SybilRank ranks nodes by evaluating the length of random walks, assigning lower trust to nodes found at the tail-end of these walks. It incorporates trust propagation to other nodes and employs the Louvain method for community detection.\\
    \cline{2-3}
    &Wei et al. \cite{wei:sybildefender} & SybilDefender identifies communities of Sybil nodes using random walks. It achieves a lower false-negative rate and better scalability compared to SybilGuard and SybilLimit. It incorporates relationship and activity ratings to reduce attack edges for accuracy and run-time.\\
    \cline{2-3}
    &Danezis and Mittal \cite{danezis2009sybilinfer} & SybilInfer is a centralized algorithm that uses Bayesian inference to identify Sybil nodes by detecting reduced graph conductance.\\
    \cline{2-3}
     &Scheideler and Schmid \cite{scheideler2009distributed} & The SHELL distributed heap (min) enabling fast updates with $O(\log^2 n)$ network adjustments and $O(\log n)$ routing time. However, the system assumes timestamps are unforgeable, which is a strong requirement for resilience against churn and Sybil attacks. \\
    \cline{2-3}
    &Tran et al. \cite{GateKeeper} & GateKeeper leverages random expander graphs and a ticket-distribution scheme to limit bad nodes. This defense ensures that honest nodes are included while restricting bad nodes to $O(\log k)$ per attack edge, where $k$ is the number of attack edges. \\
    \hline
    \multirow{2}{*}{\makecell[tl]{Machine Learning\\ Methods}} & Gong et al. \cite{gong_sybilbelief} & 
        SybilBelief uses Markov Random Fields and Loopy Belief Propagation (LBP). It overcomes noise issues in previous methods, providing robustness and scaling linearly with network size.\\
    \cline{2-3}
    & Gao et al. \cite{gao_sybilframe} & 
        SybilFrame enhances SybilBelief by integrating local node classification with Markov Random Fields and LBP. It is particularly effective in environments with low trust and a high density of attack edges.\\
    \cline{2-3}
    & Gao et al. \cite{gao_sybilfuse} & 
        SybilFuse combines node and edge classifiers to compute local trust scores for both nodes and edges.\\
    \cline{2-3}
    & Boshmaf et al. \cite{boshmaf_botnet} & 
        Sybil adversaries can avoid detection by mixing Sybil IDs into the network topology, minimizing structural differences. They explore these concepts and their implications, particularly with respect to DHTs.\\
        \cline{2-3}
    & Haribabu et al. \cite{haribabu2010detecting} &  Employs a neural network approach to identify suspicious peers, and then issues each such peer a CAPTCHA for confirmation.\\
    \cline{2-3}
    & Cai and Rojas-Cessa \cite{cai2014containing} & 
        Propose a framework with local trust tables, $k$-means clustering, and transaction verification to protect trust schemes from Sybil attacks.\\
    \hline
    \multirow{2}{*}{\makecell[tl]{Resource Burning\\ Approaches}} & Li et al. \cite{li2012sybilcontrol} &
        SybilControl uses computational puzzles to restrict Sybil IDs. It maintains similar routing latency to the Chord DHT, even under a strong Sybil attack.\\
    \cline{2-3}
    & Gupta et al. \cite{gupta_proof} & 
        This work builds a secure distributed system by requiring PoW for entry and periodic purges to limit bad IDs. Good IDs incur a cost of $O(T + g)$, where $T$ is the adversary's cost and $g$ is the number of good IDs.\\
    \cline{2-3}
    & Gupta et al. \cite{gupta2019peace,gupta2023bankrupting} & 
        This work builds on \cite{gupta_proof}, but the difficulty of the entrance puzzle is tunable. By carefully tuning the puzzle's hardness, the algorithm incurs a cost of $O(\sqrt{TJ} + J )$, where now $T$ represents the RB rate of the adversary, and $J$ is the join rate of good IDs.\\
    \cline{2-3}
    & G\"{u}nther and Pietrzak  \cite{gunther:putting} & 
        Makes use of Proof of Space in DHTs to limit the number of Sybil peers. \\
    \cline{2-3}
    &  Li et al. \cite{Liu2020PoWSybil} & 
        Peers are periodically required to solve PoW puzzles issued to all identities simultaneously at fixed intervals, with difficult increasing for those peers that fail to respond in time.  \\
    \hline
    & Da Costa Cordeiro et al. \cite{Cordeiro2012AdaptivePuzzle} & 
        Frequent identity-creation requests by peers or associated peers results in that increasingly harder puzzles are assigned prior to fulfilling the request. \\
    \hline
\end{longtable}

%% file: Tables/eclipse-storage_routing-defense-table.tex
\begin{longtable}{|p{2.8cm}|p{3.7cm}|p{8.3cm}|}
    \hline
    \rowcolor{LightCyan} {\bf Type of Defense} &  {\bf Name(s)} &  {\bf Brief Summary}\\
    \hline
    \endfirsthead 
    
    \hline
    \rowcolor{LightCyan} {\bf Type of Defense} &  {\bf Name(s)} &  {\bf Brief Summary}\\
    \hline
    \endhead 
    
    \hline
    \multirow{2}{*}{Redundant Routing} & Li et al. \cite{li} & Implemented improvements to Kademila DHT to defend against eclipse attacks using peer verification and random searches. This method assumes at least $95\%$ correct peers and shows strong performance in a 10,000-peer network.\\
    \cline{2-3}
    &Germanus et al. \cite{germanus_susceptibility} & Explored topology-aware eclipse attacks in structured P2P systems and proposed heuristics to enhance routing resilience. Later work introduced a scalable mitigation strategy ensuring $90\%$ to $100\%$ lookup success, even with high churn in large-scale networks.\\
    \cline{2-3}
    &Heilman et al. \cite{heilman} & Monte Carlo simulations show that eclipse attacks on the Bitcoin network require a large number of IP addresses. Potential attackers include botnets, and ten countermeasures are proposed, three of which were adopted by Bitcoin in 2015.\\
    \cline{2-3}
    &Rottondi et al. \cite{rottondi} & Adding extra peers to nodes in Chord DHT can mitigate eclipse attacks, with centralized method being the most effective.\\
     \cline{2-3}
    & Kapadia and Triandopoulos \cite{kapadia2008halo}  &  HALO uses redundant routing through $\Theta (\log{n})$ predecessors to tolerate random Byzantine faults. It ensures secure routing with low failure rates even when up to $25\%$ of peers are faulty.\\
    \hline
    \multirow{2}{*}{\makecell[tl]{Group-Based \\ Approaches}} & Saia and Young \cite{saia2008reducing}  & Quorums of $\Theta(\log{n})$ peers improve communication efficiency in robust DHTs by reducing complexity to $\Theta(\log{n})$ while handling Byzantine faults. Binning functions and majority agreement ensure reliable message transmission and secure peer assignments.\\
    \hline
    & Sen and Freedman \cite{sen2012commensal} & The commensal cuckoo rule improves the original rule by moving only some peers based on group size when a new peer joins, preventing uneven peer distribution. Simulations with up to $8000$ peers show improved robustness to Byzantine peers in groups of size $64$.\\
    
    \cline{2-3}
    & Saad and Saia \cite{saad:self-healing2} & An adversary controls $(1/4-\epsilon)n$ nodes, and reliable multiparty communication (MPC) is achieved using $\Theta(\log{n})$-sized groups. The self-healing algorithm reduces message costs, quarantines malicious nodes upon detection, and achieves MPC with $O(m+n\log{n})$ messages, computation, and $O(l)$ latency.\\

    \cline{2-3}
    & Knockel et al. \cite{knockel2013self} &  Demonstrating that all good peers can maintain communication with near-optimal cost and latency, rather than computing a circuit.\\
    \cline{2-3}
    & Boshrooyeh and Ozkasap \cite{boshrooyeh} & A security framework for skip graphs that leverages cryptographic signatures to detect and isolate adversarial or misbehaving peers.  \\
    \cline{2-3}
    & Ismail et al. \cite{Ismail_a,Ismail_b} & A heuristic for removing bad peers, in which good peers test suspected peers and remove them once misbehavior is sufficiently verified.   \\
    \cline{2-3}
    & Guerraoui et al. \cite{guerraoui2013highly} & A dynamic P2P system tolerating Byzantine adversaries ($<1/3$ of peers) adapts to system size changes by splitting or merging groups of $\Theta (\log{n})$ peers. Using an expander graph topology, it achieves a secure search with polylogarithmic latency and message complexity.\\
    
    \cline{2-3}
    & Kuhn et al. \cite{kuhn:towards} &  A dynamic network model ensures functionality under worst-case churn attacks by dynamically distributing tokens and aggregating information to adjust network size. It prioritizes data availability and low overhead during peer reshuffling.\\ 
    \cline{2-3}
    & Nambiar and Wright \cite{nambiar2006salsa} & The Salsa DHT organizes peers into groups using IP hash-based binary trees, enabling efficient and secure lookups. It ensures $96\%$ search success even with $20\%$ compromised nodes. \\ 
    \cline{2-3}
    & Johansen et al. \cite{johansen2006fireflies} &  Fireflies is an intrusion-tolerant overlay using gossip communication, a certificate authority, and hash-based IDs to ensure Byzantine nodes cannot manipulate their positions. It uses accusations and updates to stay connected and performs well up to $20\%$ Byzantine nodes.\\ 
    \hline
    & Augustine et al. \cite{augustine2022fully} & A robust DHT is built over a static expander graph, using $\Theta(\log^3 n)$-sized committees to tolerate up to $\frac{n}{\mathrm{polylog}(n)}$ Byzantine nodes. Random walks from committees, and periodic shuffling maintain good majorities with $\poly{\log n}$ overhead.\\ 
    \cline{2-3}
    & Aradhya et al. \cite{aradhya:robust-blockchain} &  A robust DHT is constructed using $\Theta(\log{n})$-sized committees linked in a hypercubic topology, addressing a dynamic setting where joins and departures follow a half-life constraint. A blockchain-based directory of the most recent $\Theta (\sqrt n/\log{n})$ blocks enables recovery from committee failures while preserving robustness under near-optimal half-life conditions.  \\
    \hline
    \multirow{1}{*}{\makecell[tl]{Constructions that\\ Tolerate High Churn}} & Zhang et al. \cite{zhang} & Message-specific computational puzzles reduce eclipse attacks by requiring IDs to solve puzzles before joining, eliminating the need for certification authorities. This approach is effective only in large DHTs and can be costly for peers with unstable connections.\\
    \cline{2-3}
    & Anceaume et al. \cite{anceaume2008peercube} & PeerCube resists eclipse attacks in high churn by organizing peers into hypercubic clusters of size $O(\log n)$, using Byzantine consensus among core members for secure operations. Simulations show $98\%$ success even with $15\%$ malicious peers. \\
    \cline{2-3}
    & Kuhn et. al \cite{kuhn:towards} & The DHT construction handles a few adversarial peer additions and deletions, ensuring routing in $O(\log{n})$ hops. It maintains a group size of $\Theta(\log{n})$ while each peer has $O(\log{^2n})$ degree. Adapting a hypercube and splitting/merging groups yields well-distributed membership. \\
    \cline{2-3}
    & Xie and Zhu \cite{Xie} & Malicious nodes in a tree-structured overlay are identified using random sampling and a group controller. Tracing message loss from leaves to the root assigns suspicion values, achieving under $2\%$ false positive even in dynamic conditions. \\
    \cline{2-3}
    
    & Augustine et al. \cite{augustine2013fast} &  Two randomized algorithms achieve Byzantine agreement in dynamic networks under adversarial churn. The first one tolerates $O\left(\frac{\sqrt{n}}{\text{poly}(\log n)}\right)$ churn and a topology-oblivious adversary controlling the same number of Byzantine peers, while the second algorithm tolerates $O\left(\frac{\sqrt{n}}{\text{poly}(\log n)}\right)$ churn with an adversary aware of the network topology and controlling  $O(\sqrt{n})$ Byzantine peers, where $n$ is the network size.\\
     \hline
    & Augustine et al. \cite{augustine2015fast} & Consider the classic distributed problem of leader election under Byzantine faults in sparse, bounded-degree expander graphs with up to O ($n^{1/2-\epsilon}$) Byzantine nodes. The system tolerates churn of $O ( \sqrt{n} / \poly{(\log n)}$ per round, where $n$ is the stable network size. \\
    \hline
\end{longtable}

%% file: survey.bbl

\begin{thebibliography}{161}


\ifx \showCODEN    \undefined \def \showCODEN     #1{\unskip}     \fi
\ifx \showISBNx    \undefined \def \showISBNx     #1{\unskip}     \fi
\ifx \showISBNxiii \undefined \def \showISBNxiii  #1{\unskip}     \fi
\ifx \showISSN     \undefined \def \showISSN      #1{\unskip}     \fi
\ifx \showLCCN     \undefined \def \showLCCN      #1{\unskip}     \fi
\ifx \shownote     \undefined \def \shownote      #1{#1}          \fi
\ifx \showarticletitle \undefined \def \showarticletitle #1{#1}   \fi
\ifx \showURL      \undefined \def \showURL       {\relax}        \fi
\providecommand\bibfield[2]{#2}
\providecommand\bibinfo[2]{#2}
\providecommand\natexlab[1]{#1}
\providecommand\showeprint[2][]{arXiv:#2}

\bibitem[Ahmed et~al\mbox{.}(2016)]%
        {ahmed2016survey}
\bibfield{author}{\bibinfo{person}{Mohiuddin Ahmed}, \bibinfo{person}{Abdun Naser~Mahmood}, {and} \bibinfo{person}{Jiankun Hu}.} \bibinfo{year}{2016}\natexlab{}.
\newblock \showarticletitle{A survey of network anomaly detection techniques}.
\newblock \bibinfo{journal}{\emph{J. Netw. Comput. Appl.}} \bibinfo{volume}{60}, \bibinfo{number}{C} (\bibinfo{date}{Jan.} \bibinfo{year}{2016}), \bibinfo{pages}{19–31}.
\newblock
\showISSN{1084-8045}
\href{https://doi.org/10.1016/j.jnca.2015.11.016}{doi:\nolinkurl{10.1016/j.jnca.2015.11.016}}


\bibitem[Alvisi et~al\mbox{.}(2013)]%
        {alvisi2013sok}
\bibfield{author}{\bibinfo{person}{Lorenzo Alvisi}, \bibinfo{person}{Allen Clement}, \bibinfo{person}{Alessandro Epasto}, \bibinfo{person}{Silvio Lattanzi}, {and} \bibinfo{person}{Alessandro Panconesi}.} \bibinfo{year}{2013}\natexlab{}.
\newblock \showarticletitle{Sok: The evolution of sybil defense via social networks}. In \bibinfo{booktitle}{\emph{2013 IEEE symposium on security and privacy}}. \bibinfo{pages}{382--396}.
\newblock


\bibitem[Anceaume et~al\mbox{.}(2008)]%
        {anceaume2008peercube}
\bibfield{author}{\bibinfo{person}{Emmanuelle Anceaume}, \bibinfo{person}{Romaric Ludinard}, \bibinfo{person}{Aina Ravoaja}, {and} \bibinfo{person}{F Brasileiro}.} \bibinfo{year}{2008}\natexlab{}.
\newblock \showarticletitle{Peercube: A hypercube-based p2p overlay robust against collusion and churn}. In \bibinfo{booktitle}{\emph{Proceedings of the 2008 Second IEEE International Conference on Self-Adaptive and Self-Organizing Systems}}. IEEE, \bibinfo{pages}{15--24}.
\newblock


\bibitem[Aradhya et~al\mbox{.}(2023)]%
        {aradhya:robust-blockchain}
\bibfield{author}{\bibinfo{person}{Vijeth Aradhya}, \bibinfo{person}{Seth Gilbert}, {and} \bibinfo{person}{Aquinas Hobor}.} \bibinfo{year}{2023}\natexlab{}.
\newblock \showarticletitle{Robust Overlays Meet Blockchains: On Handling High Churn and Catastrophic Failures}. In \bibinfo{booktitle}{\emph{Proceedings of the 25th International Symposium Stabilization, Safety, and Security of Distributed Systems (SSS)}}. \bibinfo{publisher}{Springer-Verlag}, \bibinfo{address}{Berlin, Heidelberg}, \bibinfo{pages}{191–206}.
\newblock
\href{https://doi.org/10.1007/978-3-031-44274-2_15}{doi:\nolinkurl{10.1007/978-3-031-44274-2_15}}


\bibitem[Arge et~al\mbox{.}(2005)]%
        {skip-webs:arge}
\bibfield{author}{\bibinfo{person}{Lars Arge}, \bibinfo{person}{David Eppstein}, {and} \bibinfo{person}{Michael~T. Goodrich}.} \bibinfo{year}{2005}\natexlab{}.
\newblock \showarticletitle{Skip-webs: efficient distributed data structures for multi-dimensional data sets}. In \bibinfo{booktitle}{\emph{Proceedings of the Twenty-Fourth Annual ACM Symposium on Principles of Distributed Computing}} (Las Vegas, NV, USA) \emph{(\bibinfo{series}{PODC '05})}. \bibinfo{publisher}{Association for Computing Machinery}, \bibinfo{address}{New York, NY, USA}, \bibinfo{pages}{69–76}.
\newblock
\showISBNx{1581139942}
\href{https://doi.org/10.1145/1073814.1073827}{doi:\nolinkurl{10.1145/1073814.1073827}}


\bibitem[Aspnes and Shah(2007)]%
        {aspnes2003skip}
\bibfield{author}{\bibinfo{person}{James Aspnes} {and} \bibinfo{person}{Gauri Shah}.} \bibinfo{year}{2007}\natexlab{}.
\newblock \showarticletitle{Skip graphs}.
\newblock \bibinfo{journal}{\emph{ACM Trans. Algorithms}} \bibinfo{volume}{3}, \bibinfo{number}{4} (\bibinfo{date}{Nov.} \bibinfo{year}{2007}), \bibinfo{pages}{37–es}.
\newblock
\showISSN{1549-6325}
\href{https://doi.org/10.1145/1290672.1290674}{doi:\nolinkurl{10.1145/1290672.1290674}}


\bibitem[Augustine et~al\mbox{.}(2022)]%
        {augustine2022fully}
\bibfield{author}{\bibinfo{person}{John Augustine}, \bibinfo{person}{Soumyottam Chatterjee}, {and} \bibinfo{person}{Gopal Pandurangan}.} \bibinfo{year}{2022}\natexlab{}.
\newblock \showarticletitle{A Fully-Distributed Scalable Peer-to-Peer Protocol for Byzantine-Resilient Distributed Hash Tables}. In \bibinfo{booktitle}{\emph{Proceedings of the 34th ACM Symposium on Parallelism in Algorithms and Architectures}} (Philadelphia, PA, USA) \emph{(\bibinfo{series}{SPAA '22})}. \bibinfo{publisher}{Association for Computing Machinery}, \bibinfo{address}{New York, NY, USA}, \bibinfo{pages}{87–98}.
\newblock
\showISBNx{9781450391467}
\href{https://doi.org/10.1145/3490148.3538588}{doi:\nolinkurl{10.1145/3490148.3538588}}


\bibitem[Augustine et~al\mbox{.}(2013)]%
        {augustine2013fast}
\bibfield{author}{\bibinfo{person}{John Augustine}, \bibinfo{person}{Gopal Pandurangan}, {and} \bibinfo{person}{Peter Robinson}.} \bibinfo{year}{2013}\natexlab{}.
\newblock \showarticletitle{Fast byzantine agreement in dynamic networks}. In \bibinfo{booktitle}{\emph{Proceedings of the 2013 ACM symposium on Principles of distributed computing}}. \bibinfo{pages}{74--83}.
\newblock


\bibitem[Augustine et~al\mbox{.}(2015)]%
        {augustine2015fast}
\bibfield{author}{\bibinfo{person}{John Augustine}, \bibinfo{person}{Gopal Pandurangan}, {and} \bibinfo{person}{Peter Robinson}.} \bibinfo{year}{2015}\natexlab{}.
\newblock \showarticletitle{Fast byzantine leader election in dynamic networks}. In \bibinfo{booktitle}{\emph{International Symposium on Distributed Computing}}. Springer, \bibinfo{pages}{276--291}.
\newblock


\bibitem[Awerbuch and Scheideler(2004a)]%
        {awerbuch_scheideler:group}
\bibfield{author}{\bibinfo{person}{Baruch Awerbuch} {and} \bibinfo{person}{Christian Scheideler}.} \bibinfo{year}{2004}\natexlab{a}.
\newblock \showarticletitle{Group Spreading: A Protocol for Provably Secure Distributed Name Service}. In \bibinfo{booktitle}{\emph{Automata, Languages and Programming}}, \bibfield{editor}{\bibinfo{person}{Josep D{\'i}az}, \bibinfo{person}{Juhani Karhum{\"a}ki}, \bibinfo{person}{Arto Lepist{\"o}}, {and} \bibinfo{person}{Donald Sannella}} (Eds.). \bibinfo{publisher}{Springer Berlin Heidelberg}, \bibinfo{address}{Berlin, Heidelberg}, \bibinfo{pages}{183--195}.
\newblock
\showISBNx{978-3-540-27836-8}


\bibitem[Awerbuch and Scheideler(2004b)]%
        {awerbuch:hyperring}
\bibfield{author}{\bibinfo{person}{Baruch Awerbuch} {and} \bibinfo{person}{Christian Scheideler}.} \bibinfo{year}{2004}\natexlab{b}.
\newblock \showarticletitle{The Hyperring: a Low-Congestion Deterministic Data Structure for Distributed Environments}. In \bibinfo{booktitle}{\emph{Proceedings of the $15th$ Annual ACM-SIAM Symposium on Discrete Algorithms (SODA)}}. \bibinfo{pages}{318--327}.
\newblock


\bibitem[Awerbuch and Scheideler(2006a)]%
        {awerbuch:random}
\bibfield{author}{\bibinfo{person}{Baruch Awerbuch} {and} \bibinfo{person}{Christian Scheideler}.} \bibinfo{year}{2006}\natexlab{a}.
\newblock \showarticletitle{Robust Random Number Generation for Peer-to-peer Systems}. In \bibinfo{booktitle}{\emph{Proceedings of the $10th$ International Conference On Principles of Distributed Systems (OPODIS)}}. \bibinfo{pages}{275--289}.
\newblock


\bibitem[Awerbuch and Scheideler(2006b)]%
        {awerbuch2006towards}
\bibfield{author}{\bibinfo{person}{Baruch Awerbuch} {and} \bibinfo{person}{Christian Scheideler}.} \bibinfo{year}{2006}\natexlab{b}.
\newblock \showarticletitle{Towards a scalable and robust {DHT}}. In \bibinfo{booktitle}{\emph{Proceedings of the Eighteenth Annual ACM Symposium on Parallelism in Algorithms and Architectures (SPAA)}}. \bibinfo{pages}{318--327}.
\newblock


\bibitem[Awerbuch and Scheideler(2007)]%
        {awerbuch:towards2}
\bibfield{author}{\bibinfo{person}{Baruch Awerbuch} {and} \bibinfo{person}{Christian Scheideler}.} \bibinfo{year}{2007}\natexlab{}.
\newblock \showarticletitle{Towards Scalable and Robust Overlay Networks}. In \bibinfo{booktitle}{\emph{Proceedings of the $6^{th}$ International Workshop on Peer-to-Peer Systems (IPTPS)}}. \bibinfo{pages}{n. pag.}
\newblock


\bibitem[Awerbuch and Scheideler(2008)]%
        {awerbuch:cuckoo-journal}
\bibfield{author}{\bibinfo{person}{Baruch Awerbuch} {and} \bibinfo{person}{Christian Scheideler}.} \bibinfo{year}{2008}\natexlab{}.
\newblock \showarticletitle{Towards a Scalable and Robust {DHT}}.
\newblock \bibinfo{journal}{\emph{Theory of Computing Systems}}  \bibinfo{volume}{45} (\bibinfo{year}{2008}), \bibinfo{pages}{234--260}.
\newblock


\bibitem[Baird et~al\mbox{.}(2005)]%
        {baird2005scattertype}
\bibfield{author}{\bibinfo{person}{Henry~S Baird}, \bibinfo{person}{Michael~A Moll}, {and} \bibinfo{person}{Sui-Yu Wang}.} \bibinfo{year}{2005}\natexlab{}.
\newblock \showarticletitle{ScatterType: A legible but hard-to-segment {CAPTCHA}}. In \bibinfo{booktitle}{\emph{Proceedings of the Eighth International Conference on Document Analysis and Recognition (ICDAR)}}. \bibinfo{pages}{935--939}.
\newblock


\bibitem[Ball et~al\mbox{.}(2018)]%
        {ball2018proofs}
\bibfield{author}{\bibinfo{person}{Marshall Ball}, \bibinfo{person}{Alon Rosen}, \bibinfo{person}{Manuel Sabin}, {and} \bibinfo{person}{Prashant~Nalini Vasudevan}.} \bibinfo{year}{2018}\natexlab{}.
\newblock \showarticletitle{Proofs of work from worst-case assumptions}. In \bibinfo{booktitle}{\emph{Proceedings of the Annual International Cryptology Conference (CRYPTO)}}. Springer, \bibinfo{pages}{789--819}.
\newblock


\bibitem[Benet and Greco(2018)]%
        {filecoin}
\bibfield{author}{\bibinfo{person}{Juan Benet} {and} \bibinfo{person}{Nicola Greco}.} \bibinfo{year}{2018}\natexlab{}.
\newblock \showarticletitle{Filecoin: A Decentralized Storage Network}.
\newblock \bibinfo{journal}{\emph{Protocol Labs}}  \bibinfo{volume}{1} (\bibinfo{date}{July} \bibinfo{year}{2018}), \bibinfo{pages}{1--36}.
\newblock
\urldef\tempurl%
\url{https://filecoin.io/filecoin.pdf}
\showURL{%
\tempurl}


\bibitem[Benevenuto et~al\mbox{.}(2010)]%
        {benevenuto}
\bibfield{author}{\bibinfo{person}{Fabricio Benevenuto}, \bibinfo{person}{Gabriel Magno}, \bibinfo{person}{Tiago Rodrigues}, {and} \bibinfo{person}{Virgilio Almeida}.} \bibinfo{year}{2010}\natexlab{}.
\newblock \showarticletitle{Detecting spammers on Twitter}. In \bibinfo{booktitle}{\emph{Proceedings of the Collaboration, electronic messaging, anti-abuse and spam conference (CEAS)}}, Vol.~\bibinfo{volume}{6}. \bibinfo{pages}{12}.
\newblock


\bibitem[Bilge et~al\mbox{.}(2009)]%
        {bilge:all}
\bibfield{author}{\bibinfo{person}{Leyla Bilge}, \bibinfo{person}{Thorsten Strufe}, \bibinfo{person}{Davide Balzarotti}, {and} \bibinfo{person}{Engin Kirda}.} \bibinfo{year}{2009}\natexlab{}.
\newblock \showarticletitle{All your contacts are belong to us: automated identity theft attacks on social networks}. In \bibinfo{booktitle}{\emph{Proceedings of the 18th International Conference on World Wide Web (WWW)}}. \bibinfo{publisher}{Association for Computing Machinery}, \bibinfo{address}{New York, NY, USA}, \bibinfo{pages}{551–560}.
\newblock
\showISBNx{9781605584874}
\href{https://doi.org/10.1145/1526709.1526784}{doi:\nolinkurl{10.1145/1526709.1526784}}


\bibitem[Blondel et~al\mbox{.}(2008)]%
        {louvain}
\bibfield{author}{\bibinfo{person}{Vincent~D Blondel}, \bibinfo{person}{Jean-Loup Guillaume}, \bibinfo{person}{Renaud Lambiotte}, {and} \bibinfo{person}{Etienne Lefebvre}.} \bibinfo{year}{2008}\natexlab{}.
\newblock \showarticletitle{Fast unfolding of communities in large networks}.
\newblock \bibinfo{journal}{\emph{Journal of Statistical Mechanics: Theory and Experiment}} \bibinfo{volume}{2008}, \bibinfo{number}{10} (\bibinfo{date}{Oct.} \bibinfo{year}{2008}), \bibinfo{pages}{P10008}.
\newblock
\showISSN{1742-5468}
\href{https://doi.org/10.1088/1742-5468/2008/10/p10008}{doi:\nolinkurl{10.1088/1742-5468/2008/10/p10008}}


\bibitem[Boshmaf et~al\mbox{.}(2013)]%
        {boshmaf_botnet}
\bibfield{author}{\bibinfo{person}{Yazan Boshmaf}, \bibinfo{person}{Ildar Muslukhov}, \bibinfo{person}{Konstantin Beznosov}, {and} \bibinfo{person}{Matei Ripeanu}.} \bibinfo{year}{2013}\natexlab{}.
\newblock \showarticletitle{Design and analysis of a social botnet}.
\newblock \bibinfo{journal}{\emph{Computer Networks}} \bibinfo{volume}{57}, \bibinfo{number}{2} (\bibinfo{year}{2013}), \bibinfo{pages}{556--578}.
\newblock


\bibitem[{Boshrooyeh} and {Ozkasap}(2017)]%
        {boshrooyeh}
\bibfield{author}{\bibinfo{person}{S.~T. {Boshrooyeh}} {and} \bibinfo{person}{O. {Ozkasap}}.} \bibinfo{year}{2017}\natexlab{}.
\newblock \showarticletitle{Guard: Secure routing in skip graph}. In \bibinfo{booktitle}{\emph{Proceedings of the 2017 IFIP Networking Conference (IFIP Networking) and Workshops}}. \bibinfo{pages}{1--2}.
\newblock
\showISSN{null}
\href{https://doi.org/10.23919/IFIPNetworking.2017.8264893}{doi:\nolinkurl{10.23919/IFIPNetworking.2017.8264893}}


\bibitem[Buczak and Guven(2015)]%
        {buczak2015survey}
\bibfield{author}{\bibinfo{person}{Anna~L Buczak} {and} \bibinfo{person}{Erhan Guven}.} \bibinfo{year}{2015}\natexlab{}.
\newblock \showarticletitle{A survey of data mining and machine learning methods for cyber security intrusion detection}.
\newblock \bibinfo{journal}{\emph{IEEE Communications Surveys \& Tutorials}} \bibinfo{volume}{18}, \bibinfo{number}{2} (\bibinfo{year}{2015}), \bibinfo{pages}{1153--1176}.
\newblock


\bibitem[Cai and Rojas-Cessa(2014)]%
        {cai2014containing}
\bibfield{author}{\bibinfo{person}{Lin Cai} {and} \bibinfo{person}{Roberto Rojas-Cessa}.} \bibinfo{year}{2014}\natexlab{}.
\newblock \showarticletitle{Containing sybil attacks on trust management schemes for peer-to-peer networks}. In \bibinfo{booktitle}{\emph{2014 IEEE International Conference on Communications (ICC)}}. IEEE, \bibinfo{pages}{841--846}.
\newblock


\bibitem[Cao et~al\mbox{.}(2012)]%
        {cao}
\bibfield{author}{\bibinfo{person}{Qiang Cao}, \bibinfo{person}{Michael Sirivianos}, \bibinfo{person}{Xiaowei Yang}, {and} \bibinfo{person}{Tiago Pregueiro}.} \bibinfo{year}{2012}\natexlab{}.
\newblock \showarticletitle{Aiding the detection of fake accounts in large scale social online services}. In \bibinfo{booktitle}{\emph{Proceedings of the Presented as part of the 9th $\{$USENIX$\}$ Symposium on Networked Systems Design and Implementation ($\{$NSDI$\}$ 12)}}. \bibinfo{pages}{197--210}.
\newblock


\bibitem[Casino et~al\mbox{.}(2018)]%
        {casino}
\bibfield{author}{\bibinfo{person}{Fran Casino}, \bibinfo{person}{Thomas Dasaklis}, {and} \bibinfo{person}{Constantinos Patsakis}.} \bibinfo{year}{2018}\natexlab{}.
\newblock \showarticletitle{A systematic literature review of blockchain-based applications: Current status, classification and open issues}.
\newblock \bibinfo{journal}{\emph{Telematics and Informatics}} (\bibinfo{date}{11} \bibinfo{year}{2018}).
\newblock
\href{https://doi.org/10.1016/j.tele.2018.11.006}{doi:\nolinkurl{10.1016/j.tele.2018.11.006}}


\bibitem[Castro et~al\mbox{.}(2002)]%
        {castro:secure}
\bibfield{author}{\bibinfo{person}{Miguel Castro}, \bibinfo{person}{Peter Druschel}, \bibinfo{person}{Ayalvadi Ganesh}, \bibinfo{person}{Antony Rowstron}, {and} \bibinfo{person}{Dan~S. Wallach}.} \bibinfo{year}{2002}\natexlab{}.
\newblock \showarticletitle{Secure Routing for Structured Peer-to-Peer Overlay Networks}. In \bibinfo{booktitle}{\emph{Proceedings of the $5^{th}$ Usenix Symposium on Operating Systems Design and Implementation (OSDI)}}. \bibinfo{pages}{299--314}.
\newblock


\bibitem[Castro and Liskov(2001)]%
        {castro:byzantine}
\bibfield{author}{\bibinfo{person}{Miguel Castro} {and} \bibinfo{person}{Barbara Liskov}.} \bibinfo{year}{2001}\natexlab{}.
\newblock \showarticletitle{{Byzantine Fault Tolerance Can Be Fast}}. In \bibinfo{booktitle}{\emph{Proceedings of the International Conference on Dependable Systems and Networks}}. \bibinfo{pages}{513--518}.
\newblock


\bibitem[Castro and Liskov(2002)]%
        {castro:practical}
\bibfield{author}{\bibinfo{person}{Miguel Castro} {and} \bibinfo{person}{Barbara Liskov}.} \bibinfo{year}{2002}\natexlab{}.
\newblock \showarticletitle{{Practical Byzantine Fault Tolerance and Proactive Recovery}}.
\newblock \bibinfo{journal}{\emph{ACM Transactions on Computer Systems}}  \bibinfo{volume}{20(4)} (\bibinfo{year}{2002}), \bibinfo{pages}{398--461}.
\newblock


\bibitem[Chakraborty et~al\mbox{.}(2025)]%
        {chakraborty:bankrupting}
\bibfield{author}{\bibinfo{person}{Trisha Chakraborty}, \bibinfo{person}{Abir Islam}, \bibinfo{person}{Valerie King}, \bibinfo{person}{Daniel Rayborn}, \bibinfo{person}{Jared Saia}, {and} \bibinfo{person}{Maxwell Young}.} \bibinfo{year}{2025}\natexlab{}.
\newblock \showarticletitle{Bankrupting DoS Attackers}. In \bibinfo{booktitle}{\emph{Proceedings of the 32nd International Colloquium On Structural Information and Communication Complexity (SIROCCO)}}. \bibinfo{pages}{no pagination}.
\newblock


\bibitem[Chandola et~al\mbox{.}(2009)]%
        {chandola2009anomaly}
\bibfield{author}{\bibinfo{person}{Varun Chandola}, \bibinfo{person}{Arindam Banerjee}, {and} \bibinfo{person}{Vipin Kumar}.} \bibinfo{year}{2009}\natexlab{}.
\newblock \showarticletitle{Anomaly detection: A survey}.
\newblock \bibinfo{journal}{\emph{ACM computing surveys (CSUR)}} \bibinfo{volume}{41}, \bibinfo{number}{3} (\bibinfo{year}{2009}), \bibinfo{pages}{1--58}.
\newblock


\bibitem[Chawathe et~al\mbox{.}(2003)]%
        {chawathe2003making}
\bibfield{author}{\bibinfo{person}{Yatin Chawathe}, \bibinfo{person}{Sylvia Ratnasamy}, \bibinfo{person}{Lee Breslau}, \bibinfo{person}{Nick Lanham}, {and} \bibinfo{person}{Scott Shenker}.} \bibinfo{year}{2003}\natexlab{}.
\newblock \showarticletitle{Making gnutella-like p2p systems scalable}. In \bibinfo{booktitle}{\emph{Proceedings of the 2003 conference on Applications, technologies, architectures, and protocols for computer communications}}. \bibinfo{pages}{407--418}.
\newblock


\bibitem[Christin et~al\mbox{.}(2005)]%
        {christin2005content}
\bibfield{author}{\bibinfo{person}{Nicolas Christin}, \bibinfo{person}{Andreas~S Weigend}, {and} \bibinfo{person}{John Chuang}.} \bibinfo{year}{2005}\natexlab{}.
\newblock \showarticletitle{Content availability, pollution and poisoning in file sharing peer-to-peer networks}. In \bibinfo{booktitle}{\emph{Proceedings of the 6th ACM conference on Electronic commerce}}. \bibinfo{pages}{68--77}.
\newblock


\bibitem[Clarke et~al\mbox{.}(2001)]%
        {clarke2001freenet}
\bibfield{author}{\bibinfo{person}{Ian Clarke}, \bibinfo{person}{Oskar Sandberg}, \bibinfo{person}{Brandon Wiley}, {and} \bibinfo{person}{Theodore~W Hong}.} \bibinfo{year}{2001}\natexlab{}.
\newblock \showarticletitle{Freenet: A distributed anonymous information storage and retrieval system}. In \bibinfo{booktitle}{\emph{Proceedings of the Designing privacy enhancing technologies: international workshop on design issues in anonymity and unobservability Berkeley, CA, USA, July 25--26, 2000 Proceedings}}. Springer, \bibinfo{pages}{46--66}.
\newblock


\bibitem[Clifford(1990)]%
        {clifford1990markov}
\bibfield{author}{\bibinfo{person}{Peter Clifford}.} \bibinfo{year}{1990}\natexlab{}.
\newblock \showarticletitle{Markov random fields in statistics}.
\newblock \bibinfo{journal}{\emph{Disorder in physical systems: A volume in honour of John M. Hammersley}} (\bibinfo{year}{1990}), \bibinfo{pages}{19--32}.
\newblock


\bibitem[Cohen(2003)]%
        {cohen2003incentives}
\bibfield{author}{\bibinfo{person}{Bram Cohen}.} \bibinfo{year}{2003}\natexlab{}.
\newblock \showarticletitle{Incentives build robustness in BitTorrent}. In \bibinfo{booktitle}{\emph{Proceedings of the Workshop on Economics of Peer-to-Peer systems}}, Vol.~\bibinfo{volume}{6}. Berkeley, CA, USA, \bibinfo{pages}{68--72}.
\newblock


\bibitem[Cramer and Fuhrmann(2005)]%
        {cramer:proximity}
\bibfield{author}{\bibinfo{person}{C. Cramer} {and} \bibinfo{person}{T. Fuhrmann}.} \bibinfo{year}{2005}\natexlab{}.
\newblock \showarticletitle{Proximity neighbor selection for a {DHT} in wireless multi-hop networks}. In \bibinfo{booktitle}{\emph{Proceedings of the Fifth IEEE International Conference on Peer-to-Peer Computing (P2P)}}. \bibinfo{pages}{3--10}.
\newblock


\bibitem[Croman et~al\mbox{.}(2016)]%
        {croman2016scaling}
\bibfield{author}{\bibinfo{person}{Kyle Croman}, \bibinfo{person}{Christian Decker}, \bibinfo{person}{Ittay Eyal}, \bibinfo{person}{Adem~Efe Gencer}, \bibinfo{person}{Ari Juels}, \bibinfo{person}{Ahmed Kosba}, \bibinfo{person}{Andrew Miller}, \bibinfo{person}{Prateek Saxena}, \bibinfo{person}{Elaine Shi}, \bibinfo{person}{Emin~G{\"u}n Sirer}, {et~al\mbox{.}}} \bibinfo{year}{2016}\natexlab{}.
\newblock \showarticletitle{On scaling decentralized blockchains}. In \bibinfo{booktitle}{\emph{Proceedings of the International Conference on Financial Cryptography and Data Security}}. Springer, \bibinfo{pages}{106--125}.
\newblock


\bibitem[Da~Costa~Cordeiro et~al\mbox{.}(2012)]%
        {Cordeiro2012AdaptivePuzzle}
\bibfield{author}{\bibinfo{person}{Weverton~Luis Da~Costa~Cordeiro}, \bibinfo{person}{Fl\'{a}Vio~Roberto Santos}, \bibinfo{person}{Gustavo~Huff Mauch}, \bibinfo{person}{Marinho~Pilla Barcelos}, {and} \bibinfo{person}{Luciano~Paschoal Gaspary}.} \bibinfo{year}{2012}\natexlab{}.
\newblock \showarticletitle{Identity management based on adaptive puzzles to protect P2P systems from Sybil attacks}.
\newblock \bibinfo{journal}{\emph{Comput. Netw.}} \bibinfo{volume}{56}, \bibinfo{number}{11} (\bibinfo{date}{July} \bibinfo{year}{2012}), \bibinfo{pages}{2569–2589}.
\newblock
\showISSN{1389-1286}
\href{https://doi.org/10.1016/j.comnet.2012.03.026}{doi:\nolinkurl{10.1016/j.comnet.2012.03.026}}


\bibitem[Damiani et~al\mbox{.}(2002)]%
        {damiani2002reputation}
\bibfield{author}{\bibinfo{person}{Ernesto Damiani}, \bibinfo{person}{De~Capitani di Vimercati}, \bibinfo{person}{Stefano Paraboschi}, \bibinfo{person}{Pierangela Samarati}, {and} \bibinfo{person}{Fabio Violante}.} \bibinfo{year}{2002}\natexlab{}.
\newblock \showarticletitle{A reputation-based approach for choosing reliable resources in peer-to-peer networks}. In \bibinfo{booktitle}{\emph{Proceedings of the 9th ACM conference on Computer and communications security}}. \bibinfo{pages}{207--216}.
\newblock


\bibitem[Danezis and Mittal(2009)]%
        {danezis2009sybilinfer}
\bibfield{author}{\bibinfo{person}{George Danezis} {and} \bibinfo{person}{Prateek Mittal}.} \bibinfo{year}{2009}\natexlab{}.
\newblock \showarticletitle{Sybilinfer: Detecting sybil nodes using social networks}. In \bibinfo{booktitle}{\emph{Proceedings of NDSS}}, Vol.~\bibinfo{volume}{9}. San Diego, CA, \bibinfo{pages}{1--15}.
\newblock


\bibitem[Douceur(2002)]%
        {douceur}
\bibfield{author}{\bibinfo{person}{John~R Douceur}.} \bibinfo{year}{2002}\natexlab{}.
\newblock \showarticletitle{The sybil attack}. In \bibinfo{booktitle}{\emph{Proceedings of International workshop on peer-to-peer systems}}. Springer, \bibinfo{pages}{251--260}.
\newblock


\bibitem[Dwork et~al\mbox{.}(2003)]%
        {dwork2003memory}
\bibfield{author}{\bibinfo{person}{Cynthia Dwork}, \bibinfo{person}{Andrew Goldberg}, {and} \bibinfo{person}{Moni Naor}.} \bibinfo{year}{2003}\natexlab{}.
\newblock \showarticletitle{On memory-bound functions for fighting spam}. In \bibinfo{booktitle}{\emph{Proceedings of the Annual International Cryptology Conference}}. Springer, \bibinfo{pages}{426--444}.
\newblock


\bibitem[Dwork and Naor(1993)]%
        {dwork:pricing}
\bibfield{author}{\bibinfo{person}{Cynthia Dwork} {and} \bibinfo{person}{Moni Naor}.} \bibinfo{year}{1993}\natexlab{}.
\newblock \showarticletitle{Pricing via Processing or Combatting Junk Mail}. In \bibinfo{booktitle}{\emph{Proceedings of the $12^{th}$ Annual International Cryptology Conference on Advances in Cryptology}}. \bibinfo{pages}{139--147}.
\newblock


\bibitem[Ethereum(2016)]%
        {ethereum}
\bibfield{author}{\bibinfo{person}{Ethereum}.} \bibinfo{year}{2016}\natexlab{}.
\newblock \bibinfo{title}{{Ethereum: Blockchain App. Program}}.
\newblock
\newblock
\shownote{\url{https://www.ethereum.org/}}.


\bibitem[Fraigniaud and Gauron(2006)]%
        {fraigniaud:d2b}
\bibfield{author}{\bibinfo{person}{Pierre Fraigniaud} {and} \bibinfo{person}{Philippe Gauron}.} \bibinfo{year}{2006}\natexlab{}.
\newblock \showarticletitle{{D2B}: A De {Bruijn} Based Content-addressable Network}.
\newblock \bibinfo{journal}{\emph{Theoretical Computer Science}} \bibinfo{volume}{355}, \bibinfo{number}{1} (\bibinfo{date}{April} \bibinfo{year}{2006}), \bibinfo{pages}{65--79}.
\newblock


\bibitem[Fujita(2011)]%
        {fujita:proximity-aware}
\bibfield{author}{\bibinfo{person}{Satoshi Fujita}.} \bibinfo{year}{2011}\natexlab{}.
\newblock \showarticletitle{Proximity-Aware {DHT} for Efficient Lookup Service in Peer-to-Peer Applications}. In \bibinfo{booktitle}{\emph{Proceedings of the 14th IEEE International Conference on Computational Science and Engineering}}. \bibinfo{pages}{464--470}.
\newblock


\bibitem[Gao et~al\mbox{.}(2015)]%
        {gao_sybilframe}
\bibfield{author}{\bibinfo{person}{Peng Gao}, \bibinfo{person}{Neil Gong}, \bibinfo{person}{Sanjeev Kulkarni}, \bibinfo{person}{Kurt Thomas}, {and} \bibinfo{person}{Prateek Mittal}.} \bibinfo{year}{2015}\natexlab{}.
\newblock \showarticletitle{Sybilframe: A defense-in-depth framework for structure-based sybil detection}.
\newblock \bibinfo{journal}{\emph{arXiv preprint arXiv:1503.02985}} (\bibinfo{date}{03} \bibinfo{year}{2015}).
\newblock


\bibitem[Gao et~al\mbox{.}(2018)]%
        {gao_sybilfuse}
\bibfield{author}{\bibinfo{person}{Peng Gao}, \bibinfo{person}{Binghui Wang}, \bibinfo{person}{Neil~Zhenqiang Gong}, \bibinfo{person}{Sanjeev~R. Kulkarni}, \bibinfo{person}{Kurt Thomas}, {and} \bibinfo{person}{Prateek Mittal}.} \bibinfo{year}{2018}\natexlab{}.
\newblock \showarticletitle{SYBILFUSE: Combining Local Attributes with Global Structure to Perform Robust Sybil Detection}. In \bibinfo{booktitle}{\emph{Proceedings of the 2018 IEEE Conference on Communications and Network Security (CNS)}}. \bibinfo{pages}{1--9}.
\newblock
\href{https://doi.org/10.1109/CNS.2018.8433147}{doi:\nolinkurl{10.1109/CNS.2018.8433147}}


\bibitem[Germanus et~al\mbox{.}(2015)]%
        {germanus_pass}
\bibfield{author}{\bibinfo{person}{Daniel Germanus}, \bibinfo{person}{Hatem Ismail}, {and} \bibinfo{person}{Neeraj Suri}.} \bibinfo{year}{2015}\natexlab{}.
\newblock \showarticletitle{Pass: an address space slicing framework for p2p eclipse attack mitigation}. In \bibinfo{booktitle}{\emph{Proceedings of the 2015 IEEE 34th Symposium on Reliable Distributed Systems (SRDS)}}. IEEE, \bibinfo{pages}{74--83}.
\newblock


\bibitem[Germanus et~al\mbox{.}(2012)]%
        {germanus_susceptibility}
\bibfield{author}{\bibinfo{person}{Daniel Germanus}, \bibinfo{person}{Robert Langenberg}, \bibinfo{person}{Abdelmajid Khelil}, {and} \bibinfo{person}{Neeraj Suri}.} \bibinfo{year}{2012}\natexlab{}.
\newblock \showarticletitle{Susceptibility analysis of structured p2p systems to localized eclipse attacks}. In \bibinfo{booktitle}{\emph{Proceedings of the 2012 IEEE 31st Symposium on Reliable Distributed Systems}}. IEEE, \bibinfo{pages}{11--20}.
\newblock


\bibitem[Germanus et~al\mbox{.}(2014)]%
        {germanus_mitigating}
\bibfield{author}{\bibinfo{person}{Daniel Germanus}, \bibinfo{person}{Stefanie Roos}, \bibinfo{person}{Thorsten Strufe}, {and} \bibinfo{person}{Neeraj Suri}.} \bibinfo{year}{2014}\natexlab{}.
\newblock \showarticletitle{Mitigating eclipse attacks in peer-to-peer networks}. In \bibinfo{booktitle}{\emph{Proceedings of the 2014 IEEE Conference on Communications and Network Security}}. IEEE, \bibinfo{pages}{400--408}.
\newblock


\bibitem[Gong et~al\mbox{.}(2014)]%
        {gong_sybilbelief}
\bibfield{author}{\bibinfo{person}{Neil~Zhenqiang Gong}, \bibinfo{person}{Mario Frank}, {and} \bibinfo{person}{Prateek Mittal}.} \bibinfo{year}{2014}\natexlab{}.
\newblock \showarticletitle{Sybilbelief: A semi-supervised learning approach for structure-based sybil detection}.
\newblock \bibinfo{journal}{\emph{IEEE Transactions on Information Forensics and Security}} \bibinfo{volume}{9}, \bibinfo{number}{6} (\bibinfo{year}{2014}), \bibinfo{pages}{976--987}.
\newblock


\bibitem[Goodrich et~al\mbox{.}(2006)]%
        {goodrich:rainbow}
\bibfield{author}{\bibinfo{person}{Michael~T. Goodrich}, \bibinfo{person}{Michael~J. Nelson}, {and} \bibinfo{person}{Jonathan~Z. Sun}.} \bibinfo{year}{2006}\natexlab{}.
\newblock \showarticletitle{The rainbow skip graph: a fault-tolerant constant-degree distributed data structure}. In \bibinfo{booktitle}{\emph{Proceedings of the Seventeenth Annual ACM-SIAM Symposium on Discrete Algorithm (SODA)}}. \bibinfo{pages}{384--393}.
\newblock


\bibitem[Gracia-Tinedo et~al\mbox{.}(2012)]%
        {gracia}
\bibfield{author}{\bibinfo{person}{Ra{\'u}l Gracia-Tinedo}, \bibinfo{person}{Pedro Garc{\'\i}a-L{\'o}pez}, {and} \bibinfo{person}{Marc S{\'a}nchez-Artigas}.} \bibinfo{year}{2012}\natexlab{}.
\newblock \showarticletitle{Sophia: A local trust system to secure key-based routing in non-deterministic {DHT}s}.
\newblock \bibinfo{journal}{\emph{J. Parallel and Distrib. Comput.}} \bibinfo{volume}{72}, \bibinfo{number}{12} (\bibinfo{year}{2012}), \bibinfo{pages}{1696--1712}.
\newblock


\bibitem[Guerraoui et~al\mbox{.}(2013)]%
        {guerraoui2013highly}
\bibfield{author}{\bibinfo{person}{Rachid Guerraoui}, \bibinfo{person}{Florian Huc}, {and} \bibinfo{person}{Anne-Marie Kermarrec}.} \bibinfo{year}{2013}\natexlab{}.
\newblock \showarticletitle{Highly dynamic distributed computing with byzantine failures}. In \bibinfo{booktitle}{\emph{Proceedings of the 2013 ACM Symposium on Principles of Distributed Computing (PODC)}}. \bibinfo{pages}{176--183}.
\newblock


\bibitem[G\"{u}nther and Pietrzak(2025)]%
        {gunther:putting}
\bibfield{author}{\bibinfo{person}{Christoph~U. G\"{u}nther} {and} \bibinfo{person}{Krzysztof Pietrzak}.} \bibinfo{year}{2025}\natexlab{}.
\newblock \bibinfo{title}{Putting Sybils on a Diet: Securing Distributed Hash Tables using Proofs of Space}.
\newblock \bibinfo{howpublished}{Workshop on Cryptographic Tools for Blockchains}.
\newblock


\bibitem[Guo and Yu(2022)]%
        {guo2022survey}
\bibfield{author}{\bibinfo{person}{Huaqun Guo} {and} \bibinfo{person}{Xingjie Yu}.} \bibinfo{year}{2022}\natexlab{}.
\newblock \showarticletitle{A survey on blockchain technology and its security}.
\newblock \bibinfo{journal}{\emph{Blockchain: research and applications}} \bibinfo{volume}{3}, \bibinfo{number}{2} (\bibinfo{year}{2022}), \bibinfo{pages}{100067}.
\newblock


\bibitem[Gupta et~al\mbox{.}(2018)]%
        {gupta_proof}
\bibfield{author}{\bibinfo{person}{Diksha Gupta}, \bibinfo{person}{Jared Saia}, {and} \bibinfo{person}{Maxwell Young}.} \bibinfo{year}{2018}\natexlab{}.
\newblock \showarticletitle{Proof of work without all the work}. In \bibinfo{booktitle}{\emph{Proceedings of the 19th International Conference on Distributed Computing and Networking}}. \bibinfo{pages}{1--10}.
\newblock


\bibitem[Gupta et~al\mbox{.}(2019)]%
        {gupta2019peace}
\bibfield{author}{\bibinfo{person}{Diksha Gupta}, \bibinfo{person}{Jared Saia}, {and} \bibinfo{person}{Maxwell Young}.} \bibinfo{year}{2019}\natexlab{}.
\newblock \showarticletitle{Peace through superior puzzling: An asymmetric sybil defense}. In \bibinfo{booktitle}{\emph{Proceedings of the 2019 IEEE International Parallel and Distributed Processing Symposium (IPDPS)}}. IEEE, \bibinfo{pages}{1083--1094}.
\newblock


\bibitem[Gupta et~al\mbox{.}(2020)]%
        {gupta2020resource}
\bibfield{author}{\bibinfo{person}{Diksha Gupta}, \bibinfo{person}{Jared Saia}, {and} \bibinfo{person}{Maxwell Young}.} \bibinfo{year}{2020}\natexlab{}.
\newblock \showarticletitle{Resource burning for permissionless systems}. In \bibinfo{booktitle}{\emph{Proceedings of the International Colloquium on Structural Information and Communication Complexity}}. Springer, \bibinfo{pages}{19--44}.
\newblock


\bibitem[Gupta et~al\mbox{.}(2023)]%
        {gupta2023bankrupting}
\bibfield{author}{\bibinfo{person}{Diksha Gupta}, \bibinfo{person}{Jared Saia}, {and} \bibinfo{person}{Maxwell Young}.} \bibinfo{year}{2023}\natexlab{}.
\newblock \showarticletitle{Bankrupting Sybil despite churn}.
\newblock \bibinfo{journal}{\emph{J. Comput. System Sci.}}  \bibinfo{volume}{135} (\bibinfo{year}{2023}), \bibinfo{pages}{89--124}.
\newblock


\bibitem[Haribabu et~al\mbox{.}(2010)]%
        {haribabu2010detecting}
\bibfield{author}{\bibinfo{person}{Kandi Haribabu}, \bibinfo{person}{Dushyant Arora}, \bibinfo{person}{Bhavik Kothari}, {and} \bibinfo{person}{Chittaranjan Hota}.} \bibinfo{year}{2010}\natexlab{}.
\newblock \showarticletitle{Detecting sybils in peer-to-peer overlays using neural networks and {CAPTCHAs}}. In \bibinfo{booktitle}{\emph{2010 International Conference on Computational Intelligence and Communication Networks}}. IEEE, \bibinfo{pages}{154--161}.
\newblock


\bibitem[Harvey et~al\mbox{.}(2003)]%
        {HJSTW}
\bibfield{author}{\bibinfo{person}{Nicholar Harvey}, \bibinfo{person}{Michael Jones}, \bibinfo{person}{Stefan Saroiu}, \bibinfo{person}{Marvin Theimer}, {and} \bibinfo{person}{Alec Wolman}.} \bibinfo{year}{2003}\natexlab{}.
\newblock \showarticletitle{SkipNet: A Scalable Overlay Network with Practical Locality Properties}. In \bibinfo{booktitle}{\emph{Proceedings of the Fourth USENIX Symposium on Internet Technologies and Systems (USITS)}}.
\newblock


\bibitem[Harvey and Munro(2004)]%
        {harvey:deterministic}
\bibfield{author}{\bibinfo{person}{Nicholas J.~A. Harvey} {and} \bibinfo{person}{J.~Ian Munro}.} \bibinfo{year}{2004}\natexlab{}.
\newblock \showarticletitle{Deterministic SkipNet}.
\newblock \bibinfo{journal}{\emph{Inform. Process. Lett.}}  \bibinfo{volume}{90(4)} (\bibinfo{year}{2004}), \bibinfo{pages}{205--208}.
\newblock


\bibitem[Hassanzadeh-Nazarabadi et~al\mbox{.}(2021)]%
        {hassanzadeh2021dht}
\bibfield{author}{\bibinfo{person}{Yahya Hassanzadeh-Nazarabadi}, \bibinfo{person}{Sanaz Taheri-Boshrooyeh}, \bibinfo{person}{Safa Otoum}, \bibinfo{person}{Seyhan Ucar}, {and} \bibinfo{person}{{\"O}znur {\"O}zkasap}.} \bibinfo{year}{2021}\natexlab{}.
\newblock \showarticletitle{{DHT}-based communications survey: architectures and use cases}.
\newblock \bibinfo{journal}{\emph{arXiv preprint arXiv:2109.10787}} (\bibinfo{year}{2021}).
\newblock


\bibitem[Heck et~al\mbox{.}(2017)]%
        {heck}
\bibfield{author}{\bibinfo{person}{Henner Heck}, \bibinfo{person}{Olga Kieselmann}, {and} \bibinfo{person}{Arno Wacker}.} \bibinfo{year}{2017}\natexlab{}.
\newblock \showarticletitle{Evaluating connection resilience for the overlay network Kademlia}. In \bibinfo{booktitle}{\emph{Proceedings of the 37th IEEE International Conference on Distributed Computing Systems (ICDCS)}}. IEEE, \bibinfo{pages}{2581--2584}.
\newblock


\bibitem[Heilman et~al\mbox{.}(2015)]%
        {heilman}
\bibfield{author}{\bibinfo{person}{Ethan Heilman}, \bibinfo{person}{Alison Kendler}, \bibinfo{person}{Aviv Zohar}, {and} \bibinfo{person}{Sharon Goldberg}.} \bibinfo{year}{2015}\natexlab{}.
\newblock \showarticletitle{Eclipse attacks on bitcoin’s peer-to-peer network}. In \bibinfo{booktitle}{\emph{Proceedings of the 24th $\{$USENIX$\}$ Security Symposium ($\{$USENIX$\}$ Security 15)}}. \bibinfo{pages}{129--144}.
\newblock


\bibitem[Hoffman et~al\mbox{.}(2009)]%
        {hoffman2009survey}
\bibfield{author}{\bibinfo{person}{Kevin Hoffman}, \bibinfo{person}{David Zage}, {and} \bibinfo{person}{Cristina Nita-Rotaru}.} \bibinfo{year}{2009}\natexlab{}.
\newblock \showarticletitle{A survey of attack and defense techniques for reputation systems}.
\newblock \bibinfo{journal}{\emph{ACM Computing Surveys (CSUR)}} \bibinfo{volume}{42}, \bibinfo{number}{1} (\bibinfo{year}{2009}), \bibinfo{pages}{1--31}.
\newblock


\bibitem[Honigsberg(2001)]%
        {honigsberg2001evolution}
\bibfield{author}{\bibinfo{person}{Peter~Jan Honigsberg}.} \bibinfo{year}{2001}\natexlab{}.
\newblock \showarticletitle{The evolution and revolution of Napster}.
\newblock \bibinfo{journal}{\emph{University of San Francisco Law Review}}  \bibinfo{volume}{36} (\bibinfo{year}{2001}), \bibinfo{pages}{473}.
\newblock


\bibitem[{Ismail} et~al\mbox{.}(2016)]%
        {Ismail_a}
\bibfield{author}{\bibinfo{person}{H. {Ismail}}, \bibinfo{person}{D. {Germanus}}, {and} \bibinfo{person}{N. {Suri}}.} \bibinfo{year}{2016}\natexlab{}.
\newblock \showarticletitle{Malicious peers eviction for P2P overlays}. In \bibinfo{booktitle}{\emph{Proceedings of the 2016 IEEE Conference on Communications and Network Security (CNS)}}. \bibinfo{pages}{216--224}.
\newblock


\bibitem[Ismail et~al\mbox{.}(2017)]%
        {Ismail_b}
\bibfield{author}{\bibinfo{person}{Hatem Ismail}, \bibinfo{person}{Daniel Germanus}, {and} \bibinfo{person}{Neeraj Suri}.} \bibinfo{year}{2017}\natexlab{}.
\newblock \showarticletitle{P2P routing table poisoning: A quorum-based sanitizing approach}.
\newblock \bibinfo{journal}{\emph{Computers \& Security}}  \bibinfo{volume}{65} (\bibinfo{year}{2017}), \bibinfo{pages}{283--299}.
\newblock


\bibitem[Jagadish et~al\mbox{.}(2005)]%
        {jagadish:baton}
\bibfield{author}{\bibinfo{person}{H.V. Jagadish}, \bibinfo{person}{Beng~Chin Ooi}, {and} \bibinfo{person}{Quang~Hieu Vu}.} \bibinfo{year}{2005}\natexlab{}.
\newblock \showarticletitle{\textsc{BATON}: A Balanced Tree Structure for Peer-to-Peer Networks}. In \bibinfo{booktitle}{\emph{Proceedings of the $31^{st}$ International conference on Very Large Data Bases (VLDB)}}. \bibinfo{pages}{661--672}.
\newblock


\bibitem[Jaiyeola et~al\mbox{.}(2018)]%
        {JaiyeolaPSYZ18}
\bibfield{author}{\bibinfo{person}{Mercy~O. Jaiyeola}, \bibinfo{person}{Kyle Patron}, \bibinfo{person}{Jared Saia}, \bibinfo{person}{Maxwell Young}, {and} \bibinfo{person}{Qian~M. Zhou}.} \bibinfo{year}{2018}\natexlab{}.
\newblock \showarticletitle{Tiny Groups Tackle Byzantine Adversaries}. In \bibinfo{booktitle}{\emph{Proceedings of the {IEEE} International Parallel and Distributed Processing Symposium, {IPDPS}}}. \bibinfo{pages}{1030--1039}.
\newblock


\bibitem[Jelasity et~al\mbox{.}({[n.\,d.]})]%
        {peersim}
\bibfield{author}{\bibinfo{person}{M{\'a}rk Jelasity}, \bibinfo{person}{Alberto Montresor}, \bibinfo{person}{Gian~Paolo Jesi}, {and} \bibinfo{person}{Spyros Voulgaris}.} \bibinfo{year}{[n.\,d.]}\natexlab{}.
\newblock \bibinfo{title}{The {Peersim} Simulator}.
\newblock
\newblock
\shownote{{\tt http://peersim.sf.net}}.


\bibitem[Jennings et~al\mbox{.}(2014)]%
        {jennings}
\bibfield{author}{\bibinfo{person}{C. Jennings}, \bibinfo{person}{B. Lowekamp}, \bibinfo{person}{E. Rescorla}, \bibinfo{person}{S. Baset}, {and} \bibinfo{person}{H. Schulzrinne}.} \bibinfo{year}{2014}\natexlab{}.
\newblock \showarticletitle{RFC 6940: REsource LOcation And Discovery (RELOAD) Base Protocol}.
\newblock  (\bibinfo{year}{2014}).
\newblock


\bibitem[Jethava and Rao(2022)]%
        {JETHAVA2022:hybrid-approach-sybil-detection}
\bibfield{author}{\bibinfo{person}{Gordhan Jethava} {and} \bibinfo{person}{Udai~Pratap Rao}.} \bibinfo{year}{2022}\natexlab{}.
\newblock \showarticletitle{User behavior-based and graph-based hybrid approach for detection of Sybil Attack in online social networks}.
\newblock \bibinfo{journal}{\emph{Computers and Electrical Engineering}}  \bibinfo{volume}{99} (\bibinfo{year}{2022}), \bibinfo{pages}{107753}.
\newblock
\showISSN{0045-7906}
\href{https://doi.org/10.1016/j.compeleceng.2022.107753}{doi:\nolinkurl{10.1016/j.compeleceng.2022.107753}}


\bibitem[Johansen et~al\mbox{.}(2006)]%
        {johansen2006fireflies}
\bibfield{author}{\bibinfo{person}{H{\aa}vard Johansen}, \bibinfo{person}{Andr{\'e} Allavena}, {and} \bibinfo{person}{Robbert Van~Renesse}.} \bibinfo{year}{2006}\natexlab{}.
\newblock \showarticletitle{Fireflies: scalable support for intrusion-tolerant network overlays}.
\newblock \bibinfo{journal}{\emph{ACM SIGOPS Operating Systems Review}} \bibinfo{volume}{40}, \bibinfo{number}{4} (\bibinfo{year}{2006}), \bibinfo{pages}{3--13}.
\newblock


\bibitem[Kaashoek and Karger(2003)]%
        {kaashoek2003koorde}
\bibfield{author}{\bibinfo{person}{M~Frans Kaashoek} {and} \bibinfo{person}{David~R Karger}.} \bibinfo{year}{2003}\natexlab{}.
\newblock \showarticletitle{Koorde: A simple degree-optimal distributed hash table}. In \bibinfo{booktitle}{\emph{Proceedings of the Peer-to-Peer Systems II: Second International Workshop, IPTPS 2003, Berkeley, CA, USA, February 21-22, 2003. Revised Papers 2}}. Springer, \bibinfo{pages}{98--107}.
\newblock


\bibitem[Kapadia and Triandopoulos(2008)]%
        {kapadia2008halo}
\bibfield{author}{\bibinfo{person}{Apu Kapadia} {and} \bibinfo{person}{Nikos Triandopoulos}.} \bibinfo{year}{2008}\natexlab{}.
\newblock \showarticletitle{Halo: High-Assurance Locate for Distributed Hash Tables.}. In \bibinfo{booktitle}{\emph{Proceedings of the NDSS}}, Vol.~\bibinfo{volume}{8}. Citeseer, \bibinfo{pages}{142}.
\newblock


\bibitem[King et~al\mbox{.}(2007)]%
        {king2007choosing}
\bibfield{author}{\bibinfo{person}{Valerie King}, \bibinfo{person}{Scott Lewis}, \bibinfo{person}{Jared Saia}, {and} \bibinfo{person}{Maxwell Young}.} \bibinfo{year}{2007}\natexlab{}.
\newblock \showarticletitle{Choosing a random peer in chord}.
\newblock \bibinfo{journal}{\emph{Algorithmica}} \bibinfo{volume}{49}, \bibinfo{number}{2} (\bibinfo{year}{2007}), \bibinfo{pages}{147--169}.
\newblock


\bibitem[King and Saia(2004)]%
        {king:choosing}
\bibfield{author}{\bibinfo{person}{Valerie King} {and} \bibinfo{person}{Jared Saia}.} \bibinfo{year}{2004}\natexlab{}.
\newblock \showarticletitle{{Choosing a Random Peer}}. In \bibinfo{booktitle}{\emph{Proceedings of the {$23^{rd}$ ACM Symposium on Principles of Distributed Computing (PODC)}}}. \bibinfo{pages}{125--130}.
\newblock


\bibitem[Knockel et~al\mbox{.}(2013)]%
        {knockel2013self}
\bibfield{author}{\bibinfo{person}{Jeffrey Knockel}, \bibinfo{person}{George Saad}, {and} \bibinfo{person}{Jared Saia}.} \bibinfo{year}{2013}\natexlab{}.
\newblock \showarticletitle{Self-healing of byzantine faults}. In \bibinfo{booktitle}{\emph{Proceedings of the Symposium on Self-Stabilizing Systems}}. Springer, \bibinfo{pages}{98--112}.
\newblock


\bibitem[Kohnen(2012a)]%
        {kohnen_b}
\bibfield{author}{\bibinfo{person}{Michael Kohnen}.} \bibinfo{year}{2012}\natexlab{a}.
\newblock \showarticletitle{Analysis and optimization of routing trust values in a Kademlia-based Distributed Hash Table in a malicious environment}. In \bibinfo{booktitle}{\emph{Proceedings of the 2012 2nd Baltic Congress on Future Internet Communications}}. IEEE, \bibinfo{pages}{252--259}.
\newblock


\bibitem[Kohnen(2012b)]%
        {kohnen_a}
\bibfield{author}{\bibinfo{person}{Michael Kohnen}.} \bibinfo{year}{2012}\natexlab{b}.
\newblock \showarticletitle{Applying trust and reputation mechanisms to a Kademlia-based Distributed Hash Table}. In \bibinfo{booktitle}{\emph{Proceedings of the 2012 IEEE International Conference on Communications (ICC)}}. IEEE, \bibinfo{pages}{1036--1041}.
\newblock


\bibitem[Kowalski(2019)]%
        {kowalski2019introduction}
\bibfield{author}{\bibinfo{person}{Emmanuel Kowalski}.} \bibinfo{year}{2019}\natexlab{}.
\newblock \bibinfo{booktitle}{\emph{An introduction to expander graphs}}.
\newblock \bibinfo{publisher}{Soci{\'e}t{\'e} math{\'e}matique de France Paris}.
\newblock


\bibitem[Kuhn et~al\mbox{.}(2010)]%
        {kuhn:towards}
\bibfield{author}{\bibinfo{person}{Fabian Kuhn}, \bibinfo{person}{Stefan Schmid}, {and} \bibinfo{person}{Roger Wattenhofer}.} \bibinfo{year}{2010}\natexlab{}.
\newblock \showarticletitle{Towards worst-case churn resistant peer-to-peer systems}.
\newblock \bibinfo{journal}{\emph{Distributed Computing}} \bibinfo{volume}{22}, \bibinfo{number}{5} (\bibinfo{year}{2010}), \bibinfo{pages}{249--267}.
\newblock
Issue 4.
\href{https://doi.org/10.1007/s00446-010-0099-z}{doi:\nolinkurl{10.1007/s00446-010-0099-z}}


\bibitem[Lamport et~al\mbox{.}(1982)]%
        {Lamport}
\bibfield{author}{\bibinfo{person}{Leslie Lamport}, \bibinfo{person}{Robert Shostak}, {and} \bibinfo{person}{Marshall Pease}.} \bibinfo{year}{1982}\natexlab{}.
\newblock \showarticletitle{The Byzantine Generals Problem}.
\newblock \bibinfo{journal}{\emph{ACM Trans. Program. Lang. Syst.}} \bibinfo{volume}{4}, \bibinfo{number}{3} (\bibinfo{date}{July} \bibinfo{year}{1982}), \bibinfo{pages}{382–401}.
\newblock
\showISSN{0164-0925}
\href{https://doi.org/10.1145/357172.357176}{doi:\nolinkurl{10.1145/357172.357176}}


\bibitem[Langville and Meyer(2004)]%
        {langville2004deeper}
\bibfield{author}{\bibinfo{person}{Amy~N. Langville} {and} \bibinfo{person}{Carl~D. Meyer}.} \bibinfo{year}{2004}\natexlab{}.
\newblock \showarticletitle{Deeper Inside PageRank}.
\newblock \bibinfo{journal}{\emph{Internet Mathematics}}  \bibinfo{volume}{1} (\bibinfo{year}{2004}), \bibinfo{pages}{2004}.
\newblock


\bibitem[Laurie and Clayton(2004)]%
        {laurie-proof}
\bibfield{author}{\bibinfo{person}{Ben Laurie} {and} \bibinfo{person}{Richard Clayton}.} \bibinfo{year}{2004}\natexlab{}.
\newblock \showarticletitle{"{Proof}-of-Work" Proves Not to Work}. In \bibinfo{booktitle}{\emph{Proceedings of the $3^{rd}$ Annual Workshop on Economics and Information Security (WEIS)}}.
\newblock


\bibitem[Lesniewski-Laas and Kaashoek(2010)]%
        {lesniewski-laas:whanau}
\bibfield{author}{\bibinfo{person}{Chris Lesniewski-Laas} {and} \bibinfo{person}{M.~Frans Kaashoek}.} \bibinfo{year}{2010}\natexlab{}.
\newblock \showarticletitle{Whanau: A {Sybil}-Proof Distributed Hash Table}. In \bibinfo{booktitle}{\emph{Proceedings of the $7^{th}$ USENIX Conference on Networked Systems Design and Implementation}} (San Jose, California) \emph{(\bibinfo{series}{NSDI'10})}. \bibinfo{pages}{8--8}.
\newblock


\bibitem[Li et~al\mbox{.}(2020)]%
        {Liu2020PoWSybil}
\bibfield{author}{\bibinfo{person}{Biaoqi Li}, \bibinfo{person}{Xiaodong Fu}, \bibinfo{person}{Kun Yue}, \bibinfo{person}{Li Liu}, \bibinfo{person}{Lijun Liu}, {and} \bibinfo{person}{Yong Feng}.} \bibinfo{year}{2020}\natexlab{}.
\newblock \showarticletitle{{PoW}-Based Sybil Attack Resistant Model for P2P Reputation Systems}. In \bibinfo{booktitle}{\emph{Blockchain and Trustworthy Systems}}, \bibfield{editor}{\bibinfo{person}{Zibin Zheng}, \bibinfo{person}{Hong-Ning Dai}, \bibinfo{person}{Xiaodong Fu}, {and} \bibinfo{person}{Benhui Chen}} (Eds.). \bibinfo{publisher}{Springer Singapore}, \bibinfo{address}{Singapore}, \bibinfo{pages}{166--177}.
\newblock


\bibitem[Li et~al\mbox{.}(2012)]%
        {li2012sybilcontrol}
\bibfield{author}{\bibinfo{person}{Frank Li}, \bibinfo{person}{Prateek Mittal}, \bibinfo{person}{Matthew Caesar}, {and} \bibinfo{person}{Nikita Borisov}.} \bibinfo{year}{2012}\natexlab{}.
\newblock \showarticletitle{SybilControl: Practical Sybil defense with computational puzzles}. In \bibinfo{booktitle}{\emph{Proceedings of the seventh ACM workshop on Scalable trusted computing}}. \bibinfo{pages}{67--78}.
\newblock


\bibitem[Li et~al\mbox{.}(2014)]%
        {li}
\bibfield{author}{\bibinfo{person}{Qiang Li}, \bibinfo{person}{Jie Yu}, {and} \bibinfo{person}{Zhoujun Li}.} \bibinfo{year}{2014}\natexlab{}.
\newblock \showarticletitle{An Enhanced Kad Protocol Resistant to Eclipse Attacks}. In \bibinfo{booktitle}{\emph{Proceedings of the 2014 9th IEEE International Conference on Networking, Architecture, and Storage}}. \bibinfo{pages}{83--87}.
\newblock
\href{https://doi.org/10.1109/NAS.2014.19}{doi:\nolinkurl{10.1109/NAS.2014.19}}


\bibitem[Liang and Kumar(2005)]%
        {liang:pollution}
\bibfield{author}{\bibinfo{person}{Jian Liang} {and} \bibinfo{person}{Rakesh Kumar}.} \bibinfo{year}{2005}\natexlab{}.
\newblock \showarticletitle{{Pollution in \textsc{P2P} file sharing systems}}. In \bibinfo{booktitle}{\emph{Proceedings of the International Conference on Computer Communications (INFOCOM)}}.
\newblock


\bibitem[Liang et~al\mbox{.}(2006a)]%
        {liang2006fasttrack}
\bibfield{author}{\bibinfo{person}{Jian Liang}, \bibinfo{person}{Rakesh Kumar}, {and} \bibinfo{person}{Keith~W Ross}.} \bibinfo{year}{2006}\natexlab{a}.
\newblock \showarticletitle{The FastTrack overlay: A measurement study}.
\newblock \bibinfo{journal}{\emph{Computer Networks}} \bibinfo{volume}{50}, \bibinfo{number}{6} (\bibinfo{year}{2006}), \bibinfo{pages}{842--858}.
\newblock


\bibitem[Liang et~al\mbox{.}(2006b)]%
        {liang:poison}
\bibfield{author}{\bibinfo{person}{Jian Liang}, \bibinfo{person}{Naoum Naoumov}, {and} \bibinfo{person}{Keith~W. Ross}.} \bibinfo{year}{2006}\natexlab{b}.
\newblock \showarticletitle{The Index Poisoning Attack in \textsc{P2P} File Sharing Systems}. In \bibinfo{booktitle}{\emph{Proceedings of the INFOCOM}}. \bibinfo{pages}{1--12}.
\newblock


\bibitem[Liu and Camp(2006)]%
        {liu:proof}
\bibfield{author}{\bibinfo{person}{Debin Liu} {and} \bibinfo{person}{L.~Jean Camp}.} \bibinfo{year}{2006}\natexlab{}.
\newblock \showarticletitle{Proof of Work Can Work}. In \bibinfo{booktitle}{\emph{Proceedings of the $5^{th}$ Workshop on the Economics of Information Security (WEIS)}}.
\newblock


\bibitem[Liu et~al\mbox{.}(2010)]%
        {liu}
\bibfield{author}{\bibinfo{person}{Yaqiong Liu}, \bibinfo{person}{Weilian Xue}, \bibinfo{person}{Keqiu Li}, \bibinfo{person}{Zhongxian Chi}, \bibinfo{person}{Geyong Min}, {and} \bibinfo{person}{Wenyu Qu}.} \bibinfo{year}{2010}\natexlab{}.
\newblock \showarticletitle{{DHT}rust: A robust and distributed reputation system for trusted peer-to-peer networks}. In \bibinfo{booktitle}{\emph{Proceedings of the 2010 IEEE Global Telecommunications Conference GLOBECOM 2010}}. IEEE, \bibinfo{pages}{1--6}.
\newblock


\bibitem[Lua et~al\mbox{.}(2005)]%
        {P2P:network-overlay-survey}
\bibfield{author}{\bibinfo{person}{Eng~Keong Lua}, \bibinfo{person}{J. Crowcroft}, \bibinfo{person}{M. Pias}, \bibinfo{person}{R. Sharma}, {and} \bibinfo{person}{S. Lim}.} \bibinfo{year}{2005}\natexlab{}.
\newblock \showarticletitle{A survey and comparison of peer-to-peer overlay network schemes}.
\newblock \bibinfo{journal}{\emph{IEEE Communications Surveys \& Tutorials}} \bibinfo{volume}{7}, \bibinfo{number}{2} (\bibinfo{year}{2005}), \bibinfo{pages}{72--93}.
\newblock
\href{https://doi.org/10.1109/COMST.2005.1610546}{doi:\nolinkurl{10.1109/COMST.2005.1610546}}


\bibitem[Malatras(2015)]%
        {malatras:survey}
\bibfield{author}{\bibinfo{person}{Apostolos Malatras}.} \bibinfo{year}{2015}\natexlab{}.
\newblock \showarticletitle{State-of-the-art survey on P2P overlay networks in pervasive computing environments}.
\newblock \bibinfo{journal}{\emph{Journal of Network and Computer Applications}}  \bibinfo{volume}{55} (\bibinfo{year}{2015}), \bibinfo{pages}{1--23}.
\newblock
\showISSN{1084-8045}
\href{https://doi.org/10.1016/j.jnca.2015.04.014}{doi:\nolinkurl{10.1016/j.jnca.2015.04.014}}


\bibitem[Malkhi et~al\mbox{.}(2002)]%
        {malkhi2002viceroy}
\bibfield{author}{\bibinfo{person}{Dahlia Malkhi}, \bibinfo{person}{Moni Naor}, {and} \bibinfo{person}{David Ratajczak}.} \bibinfo{year}{2002}\natexlab{}.
\newblock \showarticletitle{Viceroy: A scalable and dynamic emulation of the butterfly}. In \bibinfo{booktitle}{\emph{Proceedings of the twenty-first annual symposium on Principles of distributed computing}}. \bibinfo{pages}{183--192}.
\newblock


\bibitem[Manshaei et~al\mbox{.}(2013)]%
        {manshaei:game-theory}
\bibfield{author}{\bibinfo{person}{Mohammad~Hossein Manshaei}, \bibinfo{person}{Quanyan Zhu}, \bibinfo{person}{Tansu Alpcan}, \bibinfo{person}{Tamer Bac\c{s}ar}, {and} \bibinfo{person}{Jean-Pierre Hubaux}.} \bibinfo{year}{2013}\natexlab{}.
\newblock \showarticletitle{Game theory meets network security and privacy}.
\newblock \bibinfo{journal}{\emph{Comput. Surveys}} \bibinfo{volume}{45}, \bibinfo{number}{3}, Article \bibinfo{articleno}{25} (\bibinfo{date}{July} \bibinfo{year}{2013}), \bibinfo{numpages}{39}~pages.
\newblock
\showISSN{0360-0300}
\href{https://doi.org/10.1145/2480741.2480742}{doi:\nolinkurl{10.1145/2480741.2480742}}


\bibitem[Maymounkov and Mazières(2002)]%
        {maymounkov}
\bibfield{author}{\bibinfo{person}{Petar Maymounkov} {and} \bibinfo{person}{David Mazières}.} \bibinfo{year}{2002}\natexlab{}.
\newblock \showarticletitle{Kademlia: A Peer-to-peer Information System Based on the XOR Metric}. In \bibinfo{booktitle}{\emph{Proceedings of Peer-to-Peer Systems}}, Vol.~\bibinfo{volume}{2429}. \bibinfo{publisher}{Springer Berlin Heidelberg}, \bibinfo{address}{Berlin, Heidelberg}, \bibinfo{pages}{53--65}.
\newblock
\showISBNx{978-3-540-45748-0}
\href{https://doi.org/10.1007/3-540-45748-8_5}{doi:\nolinkurl{10.1007/3-540-45748-8_5}}


\bibitem[Mitzenmacher and Vassilvitskii(2022)]%
        {mitzenmacher:predictions}
\bibfield{author}{\bibinfo{person}{Michael Mitzenmacher} {and} \bibinfo{person}{Sergei Vassilvitskii}.} \bibinfo{year}{2022}\natexlab{}.
\newblock \showarticletitle{Algorithms with predictions}.
\newblock \bibinfo{journal}{\emph{Commun. ACM}} \bibinfo{volume}{65}, \bibinfo{number}{7} (\bibinfo{date}{June} \bibinfo{year}{2022}), \bibinfo{pages}{33–35}.
\newblock


\bibitem[Mohaisen and Hollenbeck(2014)]%
        {mohaisen2014improving}
\bibfield{author}{\bibinfo{person}{Aziz Mohaisen} {and} \bibinfo{person}{Scott Hollenbeck}.} \bibinfo{year}{2014}\natexlab{}.
\newblock \showarticletitle{Improving social network-based sybil defenses by rewiring and augmenting social graphs}. In \bibinfo{booktitle}{\emph{Proceedings of the 4th International Workshop Information on Security Applications (WISA)}}. Springer, \bibinfo{pages}{65--80}.
\newblock


\bibitem[Monrat et~al\mbox{.}(2019)]%
        {monrat2019survey}
\bibfield{author}{\bibinfo{person}{Ahmed~Afif Monrat}, \bibinfo{person}{Olov Schel{\'e}n}, {and} \bibinfo{person}{Karl Andersson}.} \bibinfo{year}{2019}\natexlab{}.
\newblock \showarticletitle{A survey of blockchain from the perspectives of applications, challenges, and opportunities}.
\newblock \bibinfo{journal}{\emph{IEEE Access}}  \bibinfo{volume}{7} (\bibinfo{year}{2019}), \bibinfo{pages}{117134--117151}.
\newblock


\bibitem[Moradi and Keyvanpour(2015)]%
        {moradi2015captcha}
\bibfield{author}{\bibinfo{person}{Mohammad Moradi} {and} \bibinfo{person}{MohammadReza Keyvanpour}.} \bibinfo{year}{2015}\natexlab{}.
\newblock \showarticletitle{{CAPTCHA and its alternatives: A review}}.
\newblock \bibinfo{journal}{\emph{Security and Communication Networks}} \bibinfo{volume}{8}, \bibinfo{number}{12} (\bibinfo{year}{2015}), \bibinfo{pages}{2135--2156}.
\newblock


\bibitem[Moran and Orlov(2019)]%
        {moran2019simple}
\bibfield{author}{\bibinfo{person}{Tal Moran} {and} \bibinfo{person}{Ilan Orlov}.} \bibinfo{year}{2019}\natexlab{}.
\newblock \showarticletitle{Simple proofs of space-time and rational proofs of storage}. In \bibinfo{booktitle}{\emph{Annual International Cryptology Conference}}. Springer, \bibinfo{pages}{381--409}.
\newblock


\bibitem[Murphy et~al\mbox{.}(1999)]%
        {murphy2013loopy}
\bibfield{author}{\bibinfo{person}{Kevin~P. Murphy}, \bibinfo{person}{Yair Weiss}, {and} \bibinfo{person}{Michael~I. Jordan}.} \bibinfo{year}{1999}\natexlab{}.
\newblock \showarticletitle{Loopy belief propagation for approximate inference: an empirical study}. In \bibinfo{booktitle}{\emph{Proceedings of the Fifteenth Conference on Uncertainty in Artificial Intelligence}} (Stockholm, Sweden) \emph{(\bibinfo{series}{UAI'99})}. \bibinfo{publisher}{Morgan Kaufmann Publishers Inc.}, \bibinfo{address}{San Francisco, CA, USA}, \bibinfo{pages}{467–475}.
\newblock
\showISBNx{1558606149}


\bibitem[Nakamoto(2008)]%
        {nakamoto:bitcoin}
\bibfield{author}{\bibinfo{person}{Satoshi Nakamoto}.} \bibinfo{year}{2008}\natexlab{}.
\newblock \bibinfo{title}{{Bitcoin: A Peer-to-Peer Electronic Cash System}}.
\newblock
\newblock
\shownote{\url{http://bitcoin.org/bitcoin.pdf}}.


\bibitem[Nambiar and Wright(2006)]%
        {nambiar2006salsa}
\bibfield{author}{\bibinfo{person}{Arjun Nambiar} {and} \bibinfo{person}{Matthew Wright}.} \bibinfo{year}{2006}\natexlab{}.
\newblock \showarticletitle{Salsa: a structured approach to large-scale anonymity}. In \bibinfo{booktitle}{\emph{Proceedings of the 13th ACM conference on Computer and communications security}}. \bibinfo{pages}{17--26}.
\newblock


\bibitem[Neudecker(2017)]%
        {bitcoin-sybil}
\bibfield{author}{\bibinfo{person}{Till Neudecker}.} \bibinfo{year}{2017}\natexlab{}.
\newblock \bibinfo{title}{Bitcoin Cash ({BCH}) {Sybil} Nodes on the {Bitcoin} Peer-to-Peer Network}.
\newblock
\newblock
\shownote{\url{http://dsn.tm.kit.edu/publications/files/332/bch_sybil.pdf}}.


\bibitem[on~Risks to the Public~in Computers and Systems(1993)]%
        {detweiler:pseudospoofing}
\bibfield{author}{\bibinfo{person}{The Risks Digest:~Forum on~Risks to the Public~in Computers} {and} \bibinfo{person}{Related Systems}.} \bibinfo{year}{1993}\natexlab{}.
\newblock \bibinfo{title}{Volume 15, Issue 27, Tuesday 17, November 1993}.
\newblock \bibinfo{howpublished}{\url{http://catless.ncl.ac.uk/Risks/15.27.html\#subj1.10}}.
\newblock
\newblock
\shownote{Forum messages concerning L. Detweiler's claims of pseudospoofing}.


\bibitem[Pandurangan et~al\mbox{.}(2016)]%
        {pandurangan2016dex}
\bibfield{author}{\bibinfo{person}{Gopal Pandurangan}, \bibinfo{person}{Peter Robinson}, {and} \bibinfo{person}{Amitabh Trehan}.} \bibinfo{year}{2016}\natexlab{}.
\newblock \showarticletitle{DEX: self-healing expanders}.
\newblock \bibinfo{journal}{\emph{Distributed Computing}} \bibinfo{volume}{29}, \bibinfo{number}{3} (\bibinfo{year}{2016}), \bibinfo{pages}{163--185}.
\newblock


\bibitem[Pease et~al\mbox{.}(1980)]%
        {pease}
\bibfield{author}{\bibinfo{person}{Marshall~C. Pease}, \bibinfo{person}{Robert~E. Shostak}, {and} \bibinfo{person}{Leslie Lamport}.} \bibinfo{year}{1980}\natexlab{}.
\newblock \showarticletitle{Reaching Agreement in the Presence of Faults}.
\newblock \bibinfo{journal}{\emph{J. ACM}}  \bibinfo{volume}{27(2)} (\bibinfo{year}{1980}), \bibinfo{pages}{228--234}.
\newblock


\bibitem[Pourebrahimi et~al\mbox{.}(2005)]%
        {pourebrahimi2005survey}
\bibfield{author}{\bibinfo{person}{B Pourebrahimi}, \bibinfo{person}{K Bertels}, {and} \bibinfo{person}{S Vassiliadis}.} \bibinfo{year}{2005}\natexlab{}.
\newblock \showarticletitle{A survey of peer-to-peer networks}. In \bibinfo{booktitle}{\emph{Proceedings of the 16th annual workshop on Circuits, Systems and Signal Processing}}. \bibinfo{pages}{570--577}.
\newblock


\bibitem[Pr{\"{u}}nster et~al\mbox{.}(2022)]%
        {prunster:total}
\bibfield{author}{\bibinfo{person}{Bernd Pr{\"{u}}nster}, \bibinfo{person}{Alexander Marsalek}, {and} \bibinfo{person}{Thomas Zefferer}.} \bibinfo{year}{2022}\natexlab{}.
\newblock \showarticletitle{Total Eclipse of the Heart - Disrupting the InterPlanetary File System}. In \bibinfo{booktitle}{\emph{Proceedings of the 31st {USENIX} Security Symposium (USENIX)}}, \bibfield{editor}{\bibinfo{person}{Kevin R.~B. Butler} {and} \bibinfo{person}{Kurt Thomas}} (Eds.). \bibinfo{pages}{3735--3752}.
\newblock


\bibitem[Pugh(1990)]%
        {pugh1990skip}
\bibfield{author}{\bibinfo{person}{William Pugh}.} \bibinfo{year}{1990}\natexlab{}.
\newblock \showarticletitle{Skip lists: a probabilistic alternative to balanced trees}.
\newblock \bibinfo{journal}{\emph{Commun. ACM}} \bibinfo{volume}{33}, \bibinfo{number}{6} (\bibinfo{year}{1990}), \bibinfo{pages}{668--676}.
\newblock


\bibitem[Ratnasamy et~al\mbox{.}(2001)]%
        {ratnasamy2001scalable}
\bibfield{author}{\bibinfo{person}{Sylvia Ratnasamy}, \bibinfo{person}{Paul Francis}, \bibinfo{person}{Mark Handley}, \bibinfo{person}{Richard Karp}, {and} \bibinfo{person}{Scott Shenker}.} \bibinfo{year}{2001}\natexlab{}.
\newblock \showarticletitle{A scalable content-addressable network}. In \bibinfo{booktitle}{\emph{Proceedings of the 2001 conference on Applications, technologies, architectures, and protocols for computer communications}}. \bibinfo{pages}{161--172}.
\newblock


\bibitem[Ripeanu et~al\mbox{.}(2010)]%
        {ripeanu2010search}
\bibfield{author}{\bibinfo{person}{Matei Ripeanu}, \bibinfo{person}{Adriana Iamnitchi}, \bibinfo{person}{Ian Foster}, {and} \bibinfo{person}{Anne Rogers}.} \bibinfo{year}{2010}\natexlab{}.
\newblock \showarticletitle{In search of simplicity: a self-organizing group communication overlay}.
\newblock \bibinfo{journal}{\emph{Concurrency and Computation: Practice and Experience}} \bibinfo{volume}{22}, \bibinfo{number}{7} (\bibinfo{year}{2010}), \bibinfo{pages}{788--815}.
\newblock


\bibitem[Rottondi et~al\mbox{.}(2014)]%
        {rottondi}
\bibfield{author}{\bibinfo{person}{Cristina Rottondi}, \bibinfo{person}{Alessandro Panzeri}, \bibinfo{person}{Constantin Yagne}, {and} \bibinfo{person}{Giacomo Verticale}.} \bibinfo{year}{2014}\natexlab{}.
\newblock \showarticletitle{Mitigation of the eclipse attack in chord overlays}.
\newblock \bibinfo{journal}{\emph{Procedia Computer Science}}  \bibinfo{volume}{32} (\bibinfo{year}{2014}), \bibinfo{pages}{1115--1120}.
\newblock


\bibitem[Rowaihy et~al\mbox{.}(2007)]%
        {Rowaihy:2007}
\bibfield{author}{\bibinfo{person}{H. Rowaihy}, \bibinfo{person}{W. Enck}, \bibinfo{person}{P. McDaniel}, {and} \bibinfo{person}{T. La~Porta}.} \bibinfo{year}{2007}\natexlab{}.
\newblock \showarticletitle{Limiting {Sybil} Attacks in Structured {P2P} Networks}. In \bibinfo{booktitle}{\emph{Proceedings of the 26th IEEE International Conference on Computer Communications (INFOCOM)}}. \bibinfo{pages}{2596--2600}.
\newblock


\bibitem[Rowstron and Druschel(2001)]%
        {rowstron2001pastry}
\bibfield{author}{\bibinfo{person}{Antony Rowstron} {and} \bibinfo{person}{Peter Druschel}.} \bibinfo{year}{2001}\natexlab{}.
\newblock \showarticletitle{Pastry: Scalable, decentralized object location, and routing for large-scale peer-to-peer systems}. In \bibinfo{booktitle}{\emph{Proceedings of the Middleware 2001: IFIP/ACM International Conference on Distributed Systems Platforms Heidelberg, Germany, November 12--16, 2001 Proceedings 2}}. Springer, \bibinfo{pages}{329--350}.
\newblock


\bibitem[Saad and Saia(2014)]%
        {saad:self-healing2}
\bibfield{author}{\bibinfo{person}{George Saad} {and} \bibinfo{person}{Jared Saia}.} \bibinfo{year}{2014}\natexlab{}.
\newblock \showarticletitle{Self-Healing Computation}. In \bibinfo{booktitle}{\emph{Proceedings of the International Symposium on Stabilization, Safety, and Security of Distributed Systems (SSS)}}. \bibinfo{pages}{195--210}.
\newblock


\bibitem[Saad and Saia(2017)]%
        {saad2017theoretical}
\bibfield{author}{\bibinfo{person}{George Saad} {and} \bibinfo{person}{Jared Saia}.} \bibinfo{year}{2017}\natexlab{}.
\newblock \showarticletitle{A theoretical and empirical evaluation of an algorithm for self-healing computation}.
\newblock \bibinfo{journal}{\emph{Distributed Computing}} \bibinfo{volume}{30}, \bibinfo{number}{6} (\bibinfo{year}{2017}), \bibinfo{pages}{391--412}.
\newblock


\bibitem[Saia and Young(2008)]%
        {saia2008reducing}
\bibfield{author}{\bibinfo{person}{Jared Saia} {and} \bibinfo{person}{Maxwell Young}.} \bibinfo{year}{2008}\natexlab{}.
\newblock \showarticletitle{Reducing communication costs in robust peer-to-peer networks}.
\newblock \bibinfo{journal}{\emph{Inform. Process. Lett.}} \bibinfo{volume}{106}, \bibinfo{number}{4} (\bibinfo{year}{2008}), \bibinfo{pages}{152--158}.
\newblock


\bibitem[Saroiu et~al\mbox{.}(2003)]%
        {saroiu2003measuring}
\bibfield{author}{\bibinfo{person}{Stefan Saroiu}, \bibinfo{person}{Krishna~P Gummadi}, {and} \bibinfo{person}{Steven~D Gribble}.} \bibinfo{year}{2003}\natexlab{}.
\newblock \showarticletitle{Measuring and analyzing the characteristics of Napster and Gnutella hosts}.
\newblock \bibinfo{journal}{\emph{Multimedia systems}} \bibinfo{volume}{9}, \bibinfo{number}{2} (\bibinfo{year}{2003}), \bibinfo{pages}{170--184}.
\newblock


\bibitem[Scheideler and Schmid(2009)]%
        {scheideler2009distributed}
\bibfield{author}{\bibinfo{person}{Christian Scheideler} {and} \bibinfo{person}{Stefan Schmid}.} \bibinfo{year}{2009}\natexlab{}.
\newblock \showarticletitle{A distributed and oblivious heap}. In \bibinfo{booktitle}{\emph{Proceedings of the International Colloquium on Automata, Languages, and Programming}}. Springer, \bibinfo{pages}{571--582}.
\newblock


\bibitem[Sen and Freedman(2012)]%
        {sen2012commensal}
\bibfield{author}{\bibinfo{person}{Siddhartha Sen} {and} \bibinfo{person}{Michael~J Freedman}.} \bibinfo{year}{2012}\natexlab{}.
\newblock \showarticletitle{Commensal cuckoo: Secure group partitioning for large-scale services}.
\newblock \bibinfo{journal}{\emph{ACM SIGOPS Operating Systems Review}} \bibinfo{volume}{46}, \bibinfo{number}{1} (\bibinfo{year}{2012}), \bibinfo{pages}{33--39}.
\newblock


\bibitem[Shane and Isaac(2017)]%
        {shane:facebook}
\bibfield{author}{\bibinfo{person}{Scott Shane} {and} \bibinfo{person}{Mike Isaac}.} \bibinfo{year}{2017}\natexlab{}.
\newblock \showarticletitle{Facebook says it’s policing fake accounts. But they’re still easy to spot}.
\newblock \bibinfo{journal}{\emph{New York Times}}  \bibinfo{volume}{3} (\bibinfo{year}{2017}).
\newblock


\bibitem[Shaukat et~al\mbox{.}(2020)]%
        {shaukat2020survey}
\bibfield{author}{\bibinfo{person}{Kamran Shaukat}, \bibinfo{person}{Suhuai Luo}, \bibinfo{person}{Vijay Varadharajan}, \bibinfo{person}{Ibrahim~A Hameed}, {and} \bibinfo{person}{Min Xu}.} \bibinfo{year}{2020}\natexlab{}.
\newblock \showarticletitle{A survey on machine learning techniques for cyber security in the last decade}.
\newblock \bibinfo{journal}{\emph{IEEE access}}  \bibinfo{volume}{8} (\bibinfo{year}{2020}), \bibinfo{pages}{222310--222354}.
\newblock


\bibitem[Shen et~al\mbox{.}(2004)]%
        {shen:cycloid}
\bibfield{author}{\bibinfo{person}{Haiying Shen}, \bibinfo{person}{Cheng-Zhong Xu}, {and} \bibinfo{person}{Guihai Chen}.} \bibinfo{year}{2004}\natexlab{}.
\newblock \showarticletitle{Cycloid: a constant-degree and lookup-efficient {P2P} overlay network}. In \bibinfo{booktitle}{\emph{Proceedings of the 18th International Parallel and Distributed Processing Symposium (IPDPS), 2004}}. \bibinfo{pages}{26--35}.
\newblock


\bibitem[Shoker(2017)]%
        {shoker2017sustainable}
\bibfield{author}{\bibinfo{person}{Ali Shoker}.} \bibinfo{year}{2017}\natexlab{}.
\newblock \showarticletitle{Sustainable blockchain through proof of exercise}. In \bibinfo{booktitle}{\emph{2017 IEEE 16th International Symposium on Network Computing and Applications (NCA)}}. IEEE, \bibinfo{pages}{1--9}.
\newblock


\bibitem[Singh et~al\mbox{.}(2006)]%
        {singh2006eclipse}
\bibfield{author}{\bibinfo{person}{Atul Singh}, \bibinfo{person}{Tsuen-Wan Ngan}, \bibinfo{person}{Peter Druschel}, \bibinfo{person}{Dan~S Wallach}, {et~al\mbox{.}}} \bibinfo{year}{2006}\natexlab{}.
\newblock \showarticletitle{Eclipse attacks on overlay networks: Threats and defenses}.
\newblock  (\bibinfo{year}{2006}).
\newblock


\bibitem[Sridhar et~al\mbox{.}(2024)]%
        {sridhar:content}
\bibfield{author}{\bibinfo{person}{Srivatsan Sridhar}, \bibinfo{person}{Onur Ascigil}, \bibinfo{person}{Navin~V. Keizer}, \bibinfo{person}{Fran{\c{c}}ois Genon}, \bibinfo{person}{S{\'{e}}bastien Pierre}, \bibinfo{person}{Yiannis Psaras}, \bibinfo{person}{Etienne Rivi{\`{e}}re}, {and} \bibinfo{person}{Michal Kr{\'{o}}l}.} \bibinfo{year}{2024}\natexlab{}.
\newblock \showarticletitle{Content Censorship in the InterPlanetary File System}. In \bibinfo{booktitle}{\emph{Proceedings of the 31st Annual Network and Distributed System Security Symposium, (NDSS)}}.
\newblock


\bibitem[Stoica et~al\mbox{.}(2001)]%
        {stoica2001chord}
\bibfield{author}{\bibinfo{person}{Ion Stoica}, \bibinfo{person}{Robert Morris}, \bibinfo{person}{David Karger}, \bibinfo{person}{M~Frans Kaashoek}, {and} \bibinfo{person}{Hari Balakrishnan}.} \bibinfo{year}{2001}\natexlab{}.
\newblock \showarticletitle{Chord: A scalable peer-to-peer lookup service for internet applications}.
\newblock \bibinfo{journal}{\emph{ACM SIGCOMM computer communication review}} \bibinfo{volume}{31}, \bibinfo{number}{4} (\bibinfo{year}{2001}), \bibinfo{pages}{149--160}.
\newblock


\bibitem[Stutzbach and Rejaie(2006)]%
        {stutzbach2006improving}
\bibfield{author}{\bibinfo{person}{Daniel Stutzbach} {and} \bibinfo{person}{Reza Rejaie}.} \bibinfo{year}{2006}\natexlab{}.
\newblock \showarticletitle{Improving lookup performance over a widely-deployed {DHT}}. In \bibinfo{booktitle}{\emph{Proceedings of the 25TH IEEE International Conference on Computer Communications (INFOCOM)}}. IEEE, \bibinfo{pages}{1--12}.
\newblock


\bibitem[Sun et~al\mbox{.}(2020)]%
        {Yue2020SybilTrustGCN}
\bibfield{author}{\bibinfo{person}{Yue Sun}, \bibinfo{person}{Zhi Yang}, {and} \bibinfo{person}{Yafei Dai}.} \bibinfo{year}{2020}\natexlab{}.
\newblock \showarticletitle{TrustGCN: Enabling Graph Convolutional Network for Robust Sybil Detection in OSNs}. In \bibinfo{booktitle}{\emph{2020 IEEE/ACM International Conference on Advances in Social Networks Analysis and Mining (ASONAM)}}. \bibinfo{pages}{1--7}.
\newblock
\href{https://doi.org/10.1109/ASONAM49781.2020.9381325}{doi:\nolinkurl{10.1109/ASONAM49781.2020.9381325}}


\bibitem[Tochner and Zohar(2020)]%
        {tochner2020gametheoretic}
\bibfield{author}{\bibinfo{person}{Saar Tochner} {and} \bibinfo{person}{Aviv Zohar}.} \bibinfo{year}{2020}\natexlab{}.
\newblock \showarticletitle{How to Pick Your Friends A Game Theoretic Approach to P2P Overlay Construction}. In \bibinfo{booktitle}{\emph{Proceedings of the 2nd ACM Conference on Advances in Financial Technologies}} (New York, NY, USA) \emph{(\bibinfo{series}{AFT '20})}. \bibinfo{publisher}{Association for Computing Machinery}, \bibinfo{address}{New York, NY, USA}, \bibinfo{pages}{37–45}.
\newblock
\showISBNx{9781450381390}
\href{https://doi.org/10.1145/3419614.3423252}{doi:\nolinkurl{10.1145/3419614.3423252}}


\bibitem[Tran et~al\mbox{.}(2011)]%
        {GateKeeper}
\bibfield{author}{\bibinfo{person}{Nguyen Tran}, \bibinfo{person}{Jinyang Li}, \bibinfo{person}{Lakshminarayanan Subramanian}, {and} \bibinfo{person}{Sherman~S.M. Chow}.} \bibinfo{year}{2011}\natexlab{}.
\newblock \showarticletitle{Optimal Sybil-resilient node admission control}. In \bibinfo{booktitle}{\emph{Proceedings of IEEE INFOCOM}}. \bibinfo{pages}{3218--3226}.
\newblock
\href{https://doi.org/10.1109/INFCOM.2011.5935171}{doi:\nolinkurl{10.1109/INFCOM.2011.5935171}}


\bibitem[Tran et~al\mbox{.}(2009)]%
        {Tran:2009:SOC:1558977.1558979}
\bibfield{author}{\bibinfo{person}{Nguyen Tran}, \bibinfo{person}{Bonan Min}, \bibinfo{person}{Jinyang Li}, {and} \bibinfo{person}{Lakshminarayanan Subramanian}.} \bibinfo{year}{2009}\natexlab{}.
\newblock \showarticletitle{Sybil-Resilient Online Content Voting}. In \bibinfo{booktitle}{\emph{Proceedings of the 6th USENIX Symposium on Networked Systems Design and Implementation (NSDI)}} (Boston, Massachusetts). \bibinfo{pages}{15--28}.
\newblock


\bibitem[Tuenti(2006)]%
        {tuenti}
\bibfield{author}{\bibinfo{person}{Tuenti}.} \bibinfo{year}{2006}\natexlab{}.
\newblock \bibinfo{title}{Tuenti: A Private, Invitation-only Social Platform Used by Millions of People to Communicate and Share Every Day}.
\newblock \bibinfo{howpublished}{\url{http://www.tuenti.com}}.
\newblock
\newblock
\shownote{Accessed: 2025-05-27}.


\bibitem[Urdaneta et~al\mbox{.}(2011)]%
        {urdaneta2011survey}
\bibfield{author}{\bibinfo{person}{Guido Urdaneta}, \bibinfo{person}{Guillaume Pierre}, {and} \bibinfo{person}{Maarten~Van Steen}.} \bibinfo{year}{2011}\natexlab{}.
\newblock \showarticletitle{A survey of {DHT} security techniques}.
\newblock \bibinfo{journal}{\emph{ACM Computing Surveys (CSUR)}} \bibinfo{volume}{43}, \bibinfo{number}{2} (\bibinfo{year}{2011}), \bibinfo{pages}{1--49}.
\newblock


\bibitem[{Uruena} et~al\mbox{.}(2013)]%
        {uruena}
\bibfield{author}{\bibinfo{person}{M. {Uruena}}, \bibinfo{person}{R. {Cuevas}}, \bibinfo{person}{A. {Cuevas}}, {and} \bibinfo{person}{A. {Banchs}}.} \bibinfo{year}{2013}\natexlab{}.
\newblock \showarticletitle{A Model to Quantify the Success of a Sybil Attack Targeting RELOAD/Chord Resources}.
\newblock \bibinfo{journal}{\emph{IEEE Communications Letters}} \bibinfo{volume}{17}, \bibinfo{number}{2} (\bibinfo{date}{February} \bibinfo{year}{2013}), \bibinfo{pages}{428--431}.
\newblock


\bibitem[Viswanath et~al\mbox{.}(2012)]%
        {viswanath2012exploring}
\bibfield{author}{\bibinfo{person}{Bimal Viswanath}, \bibinfo{person}{Mainack Mondal}, \bibinfo{person}{Allen Clement}, \bibinfo{person}{Peter Druschel}, \bibinfo{person}{Krishna~P Gummadi}, \bibinfo{person}{Alan Mislove}, {and} \bibinfo{person}{Ansley Post}.} \bibinfo{year}{2012}\natexlab{}.
\newblock \showarticletitle{Exploring the design space of social network-based sybil defenses}. In \bibinfo{booktitle}{\emph{Proceedings of the 2012 Fourth International Conference on Communication Systems and Networks (COMSNETS 2012)}}. IEEE, \bibinfo{pages}{1--8}.
\newblock


\bibitem[Von~Ahn et~al\mbox{.}(2003)]%
        {von2003captcha}
\bibfield{author}{\bibinfo{person}{Luis Von~Ahn}, \bibinfo{person}{Manuel Blum}, \bibinfo{person}{Nicholas~J Hopper}, {and} \bibinfo{person}{John Langford}.} \bibinfo{year}{2003}\natexlab{}.
\newblock \showarticletitle{{CAPTCHA}: Using hard {AI} problems for security}. In \bibinfo{booktitle}{\emph{Proceedings of the International Conference on the Theory and Applications of Cryptographic Techniques}}. Springer, \bibinfo{pages}{294--311}.
\newblock


\bibitem[Walfish et~al\mbox{.}(2006)]%
        {walfish:ddos}
\bibfield{author}{\bibinfo{person}{Michael Walfish}, \bibinfo{person}{Mythili Vutukuru}, \bibinfo{person}{Hari Balakrishnan}, \bibinfo{person}{David Karger}, {and} \bibinfo{person}{Scott Shenker}.} \bibinfo{year}{2006}\natexlab{}.
\newblock \showarticletitle{{DDoS} Defense by Offense}. In \bibinfo{booktitle}{\emph{Proceedings of the 2006 Conference on Applications, Technologies, Architectures, and Protocols for Computer Communications (SIGCOMM)}}. \bibinfo{pages}{303--314}.
\newblock


\bibitem[Wang and Kangasharju(2012)]%
        {wang}
\bibfield{author}{\bibinfo{person}{Liang Wang} {and} \bibinfo{person}{Jussi Kangasharju}.} \bibinfo{year}{2012}\natexlab{}.
\newblock \showarticletitle{Real-world sybil attacks in BitTorrent mainline {DHT}}. In \bibinfo{booktitle}{\emph{Proceedings of the 2012 IEEE Global Communications Conference (GLOBECOM)}}. IEEE, \bibinfo{pages}{826--832}.
\newblock


\bibitem[Wei et~al\mbox{.}(2013)]%
        {wei:sybildefender}
\bibfield{author}{\bibinfo{person}{Wei Wei}, \bibinfo{person}{Fengyuan Xu}, \bibinfo{person}{Chiu~C. Tan}, {and} \bibinfo{person}{Qun Li}.} \bibinfo{year}{2013}\natexlab{}.
\newblock \showarticletitle{{SybilDefender}: A Defense Mechanism for {Sybil} Attacks in Large Social Networks}.
\newblock \bibinfo{journal}{\emph{IEEE Transactions on Parallel \& Distributed Systems}} \bibinfo{volume}{24}, \bibinfo{number}{12} (\bibinfo{year}{2013}), \bibinfo{pages}{2492--2502}.
\newblock
\showISSN{1045-9219}


\bibitem[Wouhaybi and Campbell(2004)]%
        {wouhaybi2004phenix}
\bibfield{author}{\bibinfo{person}{Rita~H Wouhaybi} {and} \bibinfo{person}{Andrew~T Campbell}.} \bibinfo{year}{2004}\natexlab{}.
\newblock \showarticletitle{Phenix: Supporting resilient low-diameter peer-to-peer topologies}. In \bibinfo{booktitle}{\emph{Proceedings of the IEEE INFOCOM 2004}}, Vol.~\bibinfo{volume}{1}. IEEE.
\newblock


\bibitem[Xie and Zhu(2008)]%
        {Xie}
\bibfield{author}{\bibinfo{person}{Liang Xie} {and} \bibinfo{person}{Sencun Zhu}.} \bibinfo{year}{2008}\natexlab{}.
\newblock \showarticletitle{Message Dropping Attacks in Overlay Networks: Attack Detection and Attacker Identification}.
\newblock \bibinfo{journal}{\emph{ACM Trans. Inf. Syst. Secur.}} \bibinfo{volume}{11}, \bibinfo{number}{3}, Article \bibinfo{articleno}{15} (\bibinfo{date}{March} \bibinfo{year}{2008}), \bibinfo{numpages}{30}~pages.
\newblock
\showISSN{1094-9224}
\href{https://doi.org/10.1145/1341731.1341736}{doi:\nolinkurl{10.1145/1341731.1341736}}


\bibitem[Yang et~al\mbox{.}(2011)]%
        {Yang:2011:USN:2068816.2068841}
\bibfield{author}{\bibinfo{person}{Zhi Yang}, \bibinfo{person}{Christo Wilson}, \bibinfo{person}{Xiao Wang}, \bibinfo{person}{Tingting Gao}, \bibinfo{person}{Ben~Y. Zhao}, {and} \bibinfo{person}{Yafei Dai}.} \bibinfo{year}{2011}\natexlab{}.
\newblock \showarticletitle{Uncovering Social Network Sybils in the Wild}. In \bibinfo{booktitle}{\emph{Proceedings of the 2011 ACM SIGCOMM Conference on Internet Measurement Conference (IMC)}}. \bibinfo{pages}{259--268}.
\newblock


\bibitem[{Yingying Ma} and {Fengjun Li}(2012)]%
        {ma:detecting}
\bibfield{author}{\bibinfo{person}{{Yingying Ma}} {and} \bibinfo{person}{{Fengjun Li}}.} \bibinfo{year}{2012}\natexlab{}.
\newblock \showarticletitle{Detecting review spam: Challenges and opportunities}. In \bibinfo{booktitle}{\emph{Proceedings of the 8th International Conference on Collaborative Computing: Networking, Applications and Worksharing (CollaborateCom)}}. \bibinfo{pages}{651--654}.
\newblock


\bibitem[Young et~al\mbox{.}(2010)]%
        {young:practical}
\bibfield{author}{\bibinfo{person}{Maxwell Young}, \bibinfo{person}{Aniket Kate}, \bibinfo{person}{Ian Goldberg}, {and} \bibinfo{person}{Martin Karsten}.} \bibinfo{year}{2010}\natexlab{}.
\newblock \showarticletitle{Practical Robust Communication in DHTs Tolerating a Byzantine Adversary}. In \bibinfo{booktitle}{\emph{Proceedings of the $30^{th}$ IEEE International Conference on Distributed Computing Systems (ICDCS)}}. \bibinfo{pages}{263--272}.
\newblock


\bibitem[Young et~al\mbox{.}(2013)]%
        {young:towards}
\bibfield{author}{\bibinfo{person}{Maxwell Young}, \bibinfo{person}{Aniket Kate}, \bibinfo{person}{Ian Goldberg}, {and} \bibinfo{person}{Martin Karsten}.} \bibinfo{year}{2013}\natexlab{}.
\newblock \showarticletitle{Towards Practical Communication in {Byzantine}-Resistant {DHTs}}.
\newblock \bibinfo{journal}{\emph{IEEE/ACM Transactions on Networking}} \bibinfo{volume}{21}, \bibinfo{number}{1} (\bibinfo{date}{Feb.} \bibinfo{year}{2013}), \bibinfo{pages}{190--203}.
\newblock


\bibitem[Yu et~al\mbox{.}(2008)]%
        {yu_sybillimit}
\bibfield{author}{\bibinfo{person}{Haifeng Yu}, \bibinfo{person}{Phillip~B Gibbons}, \bibinfo{person}{Michael Kaminsky}, {and} \bibinfo{person}{Feng Xiao}.} \bibinfo{year}{2008}\natexlab{}.
\newblock \showarticletitle{Sybillimit: A near-optimal social network defense against sybil attacks}. In \bibinfo{booktitle}{\emph{Proceedings of the 2008 IEEE Symposium on Security and Privacy (sp 2008)}}. IEEE, \bibinfo{pages}{3--17}.
\newblock


\bibitem[Yu et~al\mbox{.}(2006)]%
        {yu_sybilguard}
\bibfield{author}{\bibinfo{person}{Haifeng Yu}, \bibinfo{person}{Michael Kaminsky}, \bibinfo{person}{Phillip~B Gibbons}, {and} \bibinfo{person}{Abraham Flaxman}.} \bibinfo{year}{2006}\natexlab{}.
\newblock \showarticletitle{Sybilguard: defending against sybil attacks via social networks}. In \bibinfo{booktitle}{\emph{Proceedings of the 2006 Conference on Applications, Technologies, Architectures, and Protocols for Computer Communications (SIGCOMM)}}. \bibinfo{pages}{267--278}.
\newblock


\bibitem[Zhang et~al\mbox{.}(2013)]%
        {dht:TheoryPlatformAndApplication}
\bibfield{author}{\bibinfo{person}{Hao Zhang}, \bibinfo{person}{Yonggang Wen}, \bibinfo{person}{Haiyong Xie}, {and} \bibinfo{person}{Nenghai Yu}.} \bibinfo{year}{2013}\natexlab{}.
\newblock \bibinfo{booktitle}{\emph{Distributed Hash Table: Theory, Platforms and Applications}}.
\newblock \bibinfo{publisher}{Springer Publishing Company, Incorporated}.
\newblock
\showISBNx{1461490073}


\bibitem[Zhang et~al\mbox{.}(2011)]%
        {zhang}
\bibfield{author}{\bibinfo{person}{Ren Zhang}, \bibinfo{person}{Jianyu Zhang}, \bibinfo{person}{Yu Chen}, \bibinfo{person}{Nanhao Qin}, \bibinfo{person}{Bingshuang Liu}, {and} \bibinfo{person}{Yuan Zhang}.} \bibinfo{year}{2011}\natexlab{}.
\newblock \showarticletitle{Making eclipse attacks computationally infeasible in large-scale {DHT}s}. In \bibinfo{booktitle}{\emph{Proceedings of the 30th IEEE International Performance Computing and Communications Conference}}. IEEE, \bibinfo{pages}{1--8}.
\newblock


\end{thebibliography}
